\documentclass[trackchanges,twocolumn,resetfootnote]{aastex701}

\newcommand\eos{\textit{Eos}}

\usepackage{comment}
\usepackage{multirow}
\usepackage[flushleft]{threeparttable}
\usepackage{graphicx}
\usepackage{amsmath}
\usepackage{breqn}
\usepackage{enumitem}
\graphicspath{{./}{figures/}}
\begin{document}

\title{VENUS: an ultra-faint galaxy hosting the metal-poor type II supernova at \boldmath $z=5.13$ \\
Witnessing the initial metal enrichment with extremely frequent core-collapse supernovae?
}

\author[orcid=0000-0003-3983-5438,sname=Asada,gname=Yoshihisa]{Yoshihisa Asada}
\altaffiliation{Dunlap Fellow}
\affiliation{Dunlap Institute for Astronomy and Astrophysics, 50 St. George Street, Toronto ON M5S 3H4, Canada}
\email[show]{yoshi.asada@utoronto.ca}
\correspondingauthor{Yoshihisa Asada}

\author[0000-0001-7201-5066]{Seiji Fujimoto}
\affiliation{Department of Astronomy, The University of Texas at Austin, Austin, TX 78712, USA}
\affiliation{David A. Dunlap Department of Astronomy and Astrophysics, University of Toronto, 50 St. George Street, Toronto ON, M5S 3H4, Canada}
\email{seiji.fujimoto@utoronto.ca}

\author[0000-0003-2718-8640]{Joseph F. V. Allingham}
\affiliation{Department of Physics, Ben-Gurion University of the Negev, P.O. Box 653, Be'er-Sheva 84105, Israel}
\email{allingha@post.bgu.ac.il}

\author[orcid=0000-0003-4263-2228]{David~A.~Coulter}
\affiliation{Physics and Astronomy Department, Johns Hopkins University, Baltimore, MD 21218, USA}
\affiliation{Space Telescope Science Institute (STScI), 3700 San Martin Drive, Baltimore, MD 21218, USA}
\email{dcoulter@stsci.edu}

\author[orcid=0000-0003-2037-4619]{Conor~Larison}
\affiliation{Space Telescope Science Institute (STScI), 3700 San Martin Drive, Baltimore, MD 21218, USA}
\email{conorjlarison@gmail.com}

\author[0000-0003-2445-3891]{Matthew~R.~Siebert}
\affiliation{Space Telescope Science Institute (STScI), 3700 San Martin Drive, Baltimore, MD 21218, USA}
\email{msiebert@stsci.edu}

%%%%%%%%%
%% Survey builders + major contribution
%%%%%%%%%

\author[0000-0003-2680-005X]{Gabriel Brammer}
\affiliation{Cosmic Dawn Center (DAWN), Jagtvej 128, DK2200 Copenhagen N, Denmark}
\affiliation{Niels Bohr Institute, University of Copenhagen, Jagtvej 128, DK-2200 Copenhagen N, Denmark}
\email{gabriel.brammer@nbi.ku.dk}

\author[0000-0001-7410-7669]{Dan Coe}
\affiliation{Space Telescope Science Institute (STScI), 3700 San Martin Drive, Baltimore, MD 21218, USA}
\affiliation{Center for Astrophysical Sciences, Department of Physics and Astronomy, The Johns Hopkins University, 3400 N Charles St. Baltimore, MD 21218, USA}
\affiliation{Association of Universities for Research in Astronomy (AURA), Inc.~for the European Space Agency (ESA)}
\email{dcoe@stsci.edu}

\author[0000-0001-8460-1564]{Pratika Dayal}        \affiliation{Canadian Institute for Theoretical Astrophysics, 60 St George St, University of Toronto, Toronto, ON M5S 3H8, Canada}
\affiliation{David A. Dunlap Department of Astronomy and Astrophysics, University of Toronto, 50 St. George Street, Toronto ON, M5S 3H4, Canada}
\affiliation{Department of Physics, 60 St George St, University of Toronto, Toronto, ON M5S 3H8, Canada}
\email{pratika.dayal@utoronto.ca}

\author[0000-0001-7232-5355]{Qinyue Fei}
\affiliation{David A. Dunlap Department of Astronomy and Astrophysics, University of Toronto, 50 St. George Street, Toronto ON, M5S 3H4, Canada}
\email{qyfei.astro@gmail.com}

\author[orcid=0000-0001-6278-032X]{Lukas J.~Furtak} 
\affiliation{Department of Astronomy, The University of Texas at Austin, Austin, TX 78712, USA}
\affiliation{Cosmic Frontier Center, The University of Texas at Austin, Austin, TX 78712, USA}
\email{\url{furtak@utexas.edu}}

\author[0000-0002-5588-9156]{Vasily Kokorev}
\affiliation{Department of Astronomy, The University of Texas at Austin, Austin, TX 78712, USA}
\affiliation{Cosmic Frontier Center, The University of Texas at Austin, Austin, TX 78712, USA}
\email{vasily.kokorev.astro@gmail.com}

\author[0000-0003-2611-7269]{Keiichi Maeda}
\affiliation{Department of Astronomy, Kyoto University, Kitashirakawa-Oiwake-cho, Sakyo-ku, Kyoto, 606-8502, Japan}
\email{keiichi.maeda@kusastro.kyoto-u.ac.jp}

\author[0000-0002-9651-5716]{Richard Pan}
\affiliation{Department of Physics \& Astronomy, Tufts University, Medford, MA 02155, USA}
\email{richard.pan@tufts.edu}

\author[0000-0001-5492-1049]{Johan Richard}
\affiliation{Universit\'e Claude Bernard Lyon 1, CRAL UMR5574, ENS de Lyon, CNRS, Villeurbanne, F-69622, France}
\email{johan.richard@univ-lyon1.fr}

\author[0000-0002-4622-6617]{Fengwu Sun}
\affiliation{Center for Astrophysics $|$ Harvard \& Smithsonian, 60 Garden St., Cambridge, MA 02138, USA}
\email{fengwu.sun@cfa.harvard.edu}

%%%%%%%%%
%% Other co-authors
%%%%%%%%%
\author[0000-0002-5258-8761]{Abdurro'uf}
\affiliation{Department of Astronomy, Indiana University, 727 East Third Street, Bloomington, IN 47405, USA}
\email{fnuabdur@iu.edu}

\author[0000-0002-8686-8737]{Franz E.~Bauer}
\affiliation{Instituto de Alta Investigaci{\'{o}}n, Universidad de Tarapac{\'{a}}, Casilla 7D, Arica, 1010069, Chile}
\email{franz.e.bauer@gmail.com}

\author[0000-0001-5984-0395]{Maru\v{s}a Brada{\v c}}
\affiliation{Faculty of Mathematics and Physics, University of Ljubljana, Jadranska ulica 19, SI-1000 Ljubljana, Slovenia}
\affiliation{Department of Physics and Astronomy, University of California Davis, 1 Shields Avenue, Davis, CA 95616, USA}
\email{marusa.bradac@fmf.uni-lj.si}

\author[0000-0002-7908-9284]{Larry D.~Bradley}
\affiliation{Space Telescope Science Institute (STScI), 3700 San Martin Drive, Baltimore, MD 21218, USA}
\email{lbradley@stsci.edu}

\author[0000-0003-0212-2979]{Volker Bromm}
\affiliation{Department of Astronomy, The University of Texas at Austin, Austin, TX 78712, USA}
\affiliation{Cosmic Frontier Center, The University of Texas at Austin, Austin, TX 78712, USA}
\affiliation{Weinberg Institute for Theoretical Physics, University of Texas at Austin, Austin, TX 78712, USA}
\email{vbromm@astro.as.utexas.edu}

\author[0000-0002-0302-2577]{John Chisholm}
\affiliation{Department of Astronomy, The University of Texas at Austin, Austin, TX 78712, USA}
\affiliation{Cosmic Frontier Center, The University of Texas at Austin, Austin, TX 78712, USA}
\email{chisholm@austin.utexas.edu}

\author[0000-0003-1949-7638]{Christopher J.~Conselice}
\affiliation{Jodrell Bank Centre for Astrophysics, University of Manchester, Oxford Road, Manchester M13 9PL, UK}
\email{conselice@manchester.ac.uk}

\author[0000-0002-4781-9078]{Christa DeCoursey}
\affiliation{Steward Observatory, University of Arizona, 933 N. Cherry Ave, Tucson, AZ 85721, USA}
\email{cndecoursey@arizona.edu}

\author[0000-0002-9382-9832]{Andreas L.~Faisst}
\affiliation{IPAC, California Institute of Technology, 1200 E. California Blvd., Pasadena, CA 91125, USA}
\email{afaisst@caltech.edu}

\author[0000-0003-1625-8009]{Brenda Frye}
\affiliation{Steward Observatory, University of Arizona, 933 N. Cherry Ave., University of Arizona, Tucson, AZ  85721, USA}
\email{bfrye@arizona.edu}

\author[0000-0002-4837-1615]{Mauro González-Otero}
\affiliation{Instituto de Astrof\'isica de Andaluc\'ia--CSIC, Glorieta de la Astronom\'ia s/n, E--18008 Granada, Spain}
\email{mauromarago@gmail.com}

\author[0000-0002-6047-430X]{Yuichi Harikane}
\affiliation{Institute for Cosmic Ray Research, The University of Tokyo, 5-1-5 Kashiwanoha, Kashiwa, Chiba 277-8582, Japan}
\email{hari@icrr.u-tokyo.ac.jp}

\author[0000-0003-4512-8705]{Tiger Yu-Yang Hsiao}
\affiliation{Department of Astronomy, The University of Texas at Austin, Austin, TX 78712, USA}
\affiliation{Cosmic Frontier Center, The University of Texas at Austin, Austin, TX 78712, USA}
\email{tiger.hsiao@utexas.edu}

\author[orcid=0000-0001-9840-4959]{Kohei Inayoshi}
\affiliation{Kavli Institute for Astronomy and Astrophysics, Peking University, Beijing 100871, China}
\email{inayoshi@pku.edu.cn}

\author[0000-0002-6090-2853]{Yolanda Jim\'enez-Teja}
\affiliation{Instituto de Astrof\'isica de Andaluc\'ia--CSIC, Glorieta de la Astronom\'ia s/n, E--18008 Granada, Spain}
\affiliation{Observat\'orio Nacional, Rua General Jos\'e Cristino, 77 - Bairro Imperial de S\~ao Crist\'ov\~ao, Rio de Janeiro, 20921-400, Brazil}
\email{yojite@iaa.es}

\author[0000-0002-6610-2048]{Anton M.~Koekemoer}
\affiliation{Space Telescope Science Institute (STScI), 3700 San Martin Drive, Baltimore, MD 21218, USA}
\email{koekemoer@stsci.edu}

\author[0000-0002-4052-2394]{Kotaro Kohno}
\affiliation{Institute of Astronomy, Graduate School of Science, The University of Tokyo, 2-21-1 Osawa, Mitaka, Tokyo 181-0015, Japan}
\affiliation{Research Center for the Early Universe, Graduate School of Science, The University of Tokyo, 7-3-1 Hongo, Bunkyo-ku, Tokyo 113-0033, Japan}
\email{kkohno@ioa.s.u-tokyo.ac.jp}

\author[0000-0003-2540-7424]{Paulo A.~A.~Lopes} \affiliation{Observatório do Valongo, Universidade Federal do Rio de Janeiro, Ladeira do Pedro Antônio 43, Rio de Janeiro RJ 20080-090, Brazil}
\email{plopes@ov.ufrj.br}

\author[0000-0003-1581-7825]{Ray A.~Lucas}
\affiliation{Space Telescope Science Institute (STScI), 3700 San Martin Drive, Baltimore, MD 21218, USA}
\email{lucas@stsci.edu}

\author[0000-0002-4872-2294]{Georgios E.~Magdis} 
\affiliation{Cosmic Dawn Center (DAWN), Jagtvej 128, DK2200 Copenhagen N, Denmark}
\affiliation{DTU Space, Technical University of Denmark, Elektrovej 327, DK-2800 Kgs. Lyngby, Denmark}
\email{georma@space.dtu.dk}

\author[0000-0002-5694-6124]{Vladan Markov}
\affiliation{Faculty of Mathematics and Physics, University of Ljubljana, Jadranska ulica 19, SI-1000 Ljubljana, Slovenia}
\email{vladan.markov@fmf.uni-lj.si}

\author[0000-0003-3243-9969]{Nicholas Martis}
\affiliation{Faculty of Mathematics and Physics, University of Ljubljana, Jadranska ulica 19, SI-1000 Ljubljana, Slovenia}
\email{nicholas.martis@fmf.uni-lj.si}

\author[0000-0003-2871-127X]{Jorryt Matthee}
\affiliation{Institute of Science and Technology Austria (ISTA), Am Campus 1, 3400 Klosterneuburg, Austria}
\email{jorryt.matthee@ista.ac.at}

\author[0009-0000-1999-5472]{Minami Nakane}
\affiliation{Institute for Cosmic Ray Research, The University of Tokyo, 5-1-5 Kashiwanoha, Kashiwa, Chiba 277-8582, Japan}
\affiliation{Department of Physics, Graduate School of Science, The University of Tokyo, 7-3-1 Hongo, Bunkyo, Tokyo 113-0033, Japan}
\email{nakanem@icrr.u-tokyo.ac.jp}

\author[0000-0003-3729-1684]{Rohan P.~Naidu}
\affiliation{MIT Kavli Institute for Astrophysics and Space Research, 70 Vassar Street, Cambridge, MA 02139, USA}
\email{rnaidu@mit.edu}

\author[]{Ga\"{e}l Noirot} 
\affiliation{Space Telescope Science Institute (STScI), 3700 San Martin Drive, Baltimore, MD 21218, USA}
\email{gnoirot@stsci.edu}

\author[0000-0002-1049-6658]{Masami Ouchi}
\affiliation{National Astronomical Observatory of Japan, 2-21-1 Osawa, Mitaka, Tokyo 181-8588, Japan}
\affiliation{Institute for Cosmic Ray Research, The University of Tokyo, 5-1-5 Kashiwanoha, Kashiwa, Chiba 277-8582, Japan}
\affiliation{Department of Astronomical Science, SOKENDAI (The Graduate University for Advanced Studies), 2-21-1 Osawa, Mitaka, Tokyo, 181-8588, Japan}
\affiliation{Kavli Institute for the Physics and Mathematics of the Universe (WPI), The University of Tokyo, 5-1-5 Kashiwanoha, Kashiwa, Chiba 277-8583, Japan}
\email{ouchims@icrr.u-tokyo.ac.jp}

\author[0000-0002-4410-5387]{Armin Rest}
\affiliation{Space Telescope Science Institute (STScI), 3700 San Martin Drive, Baltimore, MD 21218, USA}
\affiliation{Physics and Astronomy Department, Johns Hopkins University, Baltimore, MD 21218, USA}
\email{arest@stsci.edu}

\author[0000-0003-4223-7324]{Massimo Ricotti}
\affiliation{Department of Astronomy, University of Maryland, College Park, MD 20742, USA}
\email{ricotti@umd.edu}

\author[0000-0002-7756-4440]{Louis-Gregory Strolger}
\affiliation{Space Telescope Science Institute (STScI), 3700 San Martin Drive, Baltimore, MD 21218, USA}
\email{strolger@stsci.edu}

\author[0000-0001-9317-2888]{Raffaella Schneider}
\affiliation{Department of Physics, Sapienza University of Rome, Pzz.le Aldo Moro 5, 00185 Rome, Italy}
\email{raffaella.schneider@uniroma1.it}

\author[0000-0001-6477-4011]{Francesco Valentino}
\affiliation{Cosmic Dawn Center (DAWN), Jagtvej 128, DK2200 Copenhagen N, Denmark}
\affiliation{DTU Space, Technical University of Denmark, Elektrovej 327, DK-2800 Kgs. Lyngby, Denmark}
\email{fmava@dtu.dk}

\author[0000-0002-5057-135X]{Eros Vanzella}
\affiliation{INAF -- OAS, Osservatorio di Astrofisica e Scienza dello Spazio di Bologna, via Gobetti 93/3, I-40129 Bologna, Italy}
\email{eros.vanzella@inaf.it}

\author[0000-0002-1681-0767]{Hayley Williams} 
\affiliation{School of Earth and Space Exploration, Arizona State University, Tempe, AZ 85287-6004, USA}
\email{hwill102@asu.edu}

\author[0000-0001-8156-6281]{Rogier A.~Windhorst}
\affiliation{School of Earth and Space Exploration, Arizona State University, Tempe, AZ 85287-6004, USA}
\email{Rogier.Windhorst@asu.edu}

\author[0000-0002-0350-4488]{Adi Zitrin}
\affiliation{Department of Physics, Ben-Gurion University of the Negev, P.O. Box 653, Be'er-Sheva 84105, Israel}
\email{zitrin@bgu.ac.il}

%\collaboration{all}{VENUS collaboration}

\begin{abstract}

We present the first characterization of the host galaxy of a recently discovered type IIP SN at $z=5.13$ (SN \eos).
SN \eos\ and its host galaxy are gravitationally lensed and multiply imaged. The total magnification $\mu\sim53$ enables spatially resolving the system, allowing us to localize the core-collapse supernova (CCSN) position and to characterize its local environment within an early galaxy. % for the first time.
Our observation reveals that the host is an ultra-faint ($M_{\rm UV}=-14.4\pm0.3$ mag) Lyman-$\alpha$ emitter with a very high equivalent width.
%The Ly$\alpha$ line profile is symmetric and close to the system redshift, and also is spatially offset from the stellar component position, possibly suggesting an anisotropic escape of the ionizing photons.
The host galaxy also shows very weak [O{\sc iii}]4959,5007 lines despite an H$\alpha$ line detection ([O{\sc iii}]5007/H$\beta <0.7$ with case B recombination).
Assuming that the weak [O{\sc iii}] is due to low gas-phase metallicity given the low-metallicity of SN \eos\ itself, SN \eos\ plausibly marks the formation and explosion of a metal-poor star in an extremely metal-poor environment ($<1\ \%\ Z_\odot$), facilitating the initial stages of the chemical enrichment of the host.
Finding the CCSN in such an ultra-faint galaxy at $z=5.13$ also indicates that the SN rate could be considerably higher in high-$z$, metal-poor environments, potentially implying e.g., a $Z$-dependent IMF, $Z$-dependent massive star explodability, or runaway stellar collisions in dense star clusters.
Without lensing, only SN \eos\ would be detectable and the host would be below the detection limit in any NIRCam surveys ever performed. The \eos\ host galaxy can thus be representative of the origin of {\it hostless} supernovae frequently found in \textit{JWST} blank field surveys.

\end{abstract}
%% !!! 249 words !!!



\section{Introduction} 
Characterizing the initial stages of early galaxy formation is one of the major science goals in extragalactic astronomy \citep[see, e.g.,][for reviews]{Bromm2011ARAA,Dayal2018PhR}.
With the advent of \textit{JWST}, it has been possible to derive the detailed physical properties of $z>3$ galaxies by enabling deep rest-frame optical observations of quantities such as metallicity, electron density, or star-formation burstiness \citep[e.g.,][]{Asada2024MNRAS,Curti2024AAP,Endsley2024MNRAS,Harikane2025ApJ}.
\textit{JWST} observations in gravitational lensing fields are particularly powerful, as they have provided opportunities to look into these physical parameters in faint-end galaxies otherwise impossible to detect \citep[e.g.,][]{Berg2025arXiv,Fujimoto2025NatAs,Asada2026arXiv,Zitrin2026arXiv}.
These observations have finally opened the window to study the fundamental physics behind early galaxy formation at high-$z$.

In theoretical studies, core-collapse supernovae (CCSNe) have been deemed to play major roles in almost all aspects of early galaxy evolution.
CCSNe are the main source of early metal enrichment \citep[e.g.,][]{Karlsson2013RvMP,Kobayashi2020ApJ} as well as dust production \citep[e.g.,][]{Schneider2024AARv,Shahbandeh2025ApJ}.
They also power galactic outflows \citep[e.g.,][]{Zhang2018Galax}, contributing to the baryon-cycle between the interstellar and circumstellar media \citep[e.g.,][]{Tollet2019MNRAS}.
Stellar feedback can create ionizing photon escape channels \citep[e.g.,][]{Katz2023MNRAS}, which could be a key trigger of the cosmic reionization.
Most importantly, these effects of CCSNe are thought to be more significant in low-mass galaxies; e.g., CCSNe-driven feedback and star-formation suppression are expected to work quite efficiently where the dark matter halo potentials are relatively shallow \citep[][]{Silk2012RAA,Nelson2019MNRAS}.

CCSNe in high-$z$ low-mass galaxies have thus been a long sought-after population in observational studies, although spectroscopic confirmation of high-$z$ CCSNe has been limited up to $z=3.9$ \citep{Cooke2012Natur} and only in bright galaxies.
This is due both to the difficulty in finding high-$z$ SNe themselves and also in identifying possible faint hosts.
After the launch of \textit{JWST}, deep multi-epoch NIRCam observations have found a large number of photometric candidates of CCSNe from time variability \citep[][]{DeCoursey2025aApJ,DeCoursey2025bApJ,Coulter2026ApJ}, but the spectroscopic confirmation of high-$z$ CCSNe has been challenging. It is even harder to characterize their host galaxies.
Indeed, non-negligible fractions of these \textit{JWST}-discovered photometric SN candidates do not have associated host identified ({\it hostless} SNe) even with NIRCam images \citep[][]{DeCoursey2025aApJ,Tee2025TNSTR}.
%\blue{Yoshi: COSMOS-3D TNS list has $\sim17 \%$ hostless candidates, but i'm not sure if they are all hostless or just lacks spec-z?}

\citet{Coulter2026arXiv} recently reported the spectroscopic confirmation of a type IIP SN at $z_{\rm spec}=5.13$ (SN~2024aijd; named ``SN \eos'') with \textit{JWST}/NIRCam and NIRSpec observations behind the gravitational lensing cluster MACS~J1931.8-2635 at $z_{\rm cluster}=0.35$ (MACS1931, hereafter).
SN \eos\ was first identified with the \textit{JWST}/NIRCam observation taken as part of the Vast Exploration for Nascent, Unexplored Sources program
\citep[VENUS, GO-6882;][]{Fujimoto_Go6882}.
SN \eos\ is predicted to be quintuply imaged, and SN~\eos\ was detected in the two highest-magnification images ($\mu\sim27$) at the epoch of the \textit{JWST} observation and is observed as a $\sim26$ mag bright source thanks to the lensing magnification.
A Director's Discretionary Time (DDT) observation was triggered to obtain a NIRSpec/PRISM spectrum, and \eos\ has been confirmed as a metal-poor ($Z_\star<10\ \% Z_\odot$) type IIP SN, at a late plateau phase ($+94$ days after the explosion).
The high magnification also enables the detection of the ultra-faint host galaxy, providing an unprecedented opportunity to characterize a galaxy hosting a metal-poor type IIP SN in the high-$z$ universe.

In this paper, we report the first characterization of the SN \eos\ host galaxy.
Although SN \eos\ is much brighter than the host at the epoch of the \textit{JWST} observations, we can derive the host information at a certain wavelength range (rest-frame UV) or by decomposing the host and \eos\ emissions (rest-frame optical).
We use the \textit{JWST} NIRCam and NIRSpec observations, decompose the SN \eos\ and the host galaxy component, and obtain the basic physical properties of the SN \eos\ host galaxy.
The paper is structured as follows.
%\blue{Describe the structure...}
We first briefly describe the data we use in Section \ref{sec:data} and analyses of the SN \eos\ host galaxy in Section \ref{sec:analysis}, including both photometric and spectroscopic data.
In Section \ref{sec:result}, we present the key characteristics of the host galaxy, and discuss the implication particularly related with the identification of SN \eos.
Throughout the paper, we assume a flat $\Lambda$-CDM cosmology with $\Omega_{m}=0.3$, $\Omega_{\Lambda}=0.7$, and $H_0=70\ {\rm km\ s^{-1}\ Mpc^{-1}}$, the Chabrier initial mass function \citep[IMF;][]{Chabrier2003PASP}, and quote all magnitudes in the AB system \citep{Oke1983ApJ}.

\section{Data}\label{sec:data}
We use \textit{JWST}/NIRCam images taken as part of the VENUS program and the NIRSpec/PRISM spectrum from a \textit{JWST} DDT program (PID: 9493; PI: D.~Coulter).
The NIRCam and NIRSpec data reduction processes are described in \citet{Coulter2026arXiv}.
From the NIRCam observations, we utilize F090W, F115W, F150W, F200W, F210M, F277W, F300M, F356W, F410M and F444W images, which have a typical 5$\sigma$ depth of $\sim28$ mag for point sources.
SN \eos\ stands out in the NIRCam images as it is doubly imaged, bright, and has a remarkably red color (Figure~\ref{fig:fig1} left).
The NIRSpec fixed-slit observations were carried out at the positions of both images of the SN \eos\ (image 1 and image 2; see white rectangles in Figure \ref{fig:fig1} left), each with 10,504 s exposure time. The spectra were reduced with the \texttt{msaexp} package \citep{Brammer2022zndo}, and confirmed that the SN \eos\ is a metal-poor type IIP SN at $z=5.13$ \citep{Coulter2026arXiv}.

We also utilize archival Very Large Telescope (VLT) MUSE observations.
The VLT/MUSE observations were taken on 2015 June and July as part of the CLASH-VLT survey \citep[095.A-0525; PI: Kneib,~J.-P.;][]{Rosati2014Msngr}, which is $\sim10$ years before the explosion of SN \eos\ is observed in the image 1 and image 2.
We retrieved the VLT/MUSE datacube from the ESO data archive \citep[see][for the detailed data processing]{Ciocan2021AAP}.

For gravitational lens modeling, we exploit the latest mass model developed by the VENUS collaboration \citep{Allingham2026arXiv}, which was also used in \citet{Coulter2026arXiv}.
The lens model was calibrated with 51 lensed images of 19 systems behind the cluster.
The model achieves the final reproduction RMS of $0.\!\!^{\prime\prime}44$.
The statistical errors are derived over a well-converged subset lens models. We compound them with systematics, conservatively taken to be 20\% for magnification values, based on lens models comparisons in past works \citep{Allingham2026arXiv}.

\section{Analysis of the \eos\ host}\label{sec:analysis}
\subsection{The \eos\ host in NIRCam images}\label{subsec:images}
\begin{figure*}[t]
\centering
\includegraphics[width=0.95\textwidth]{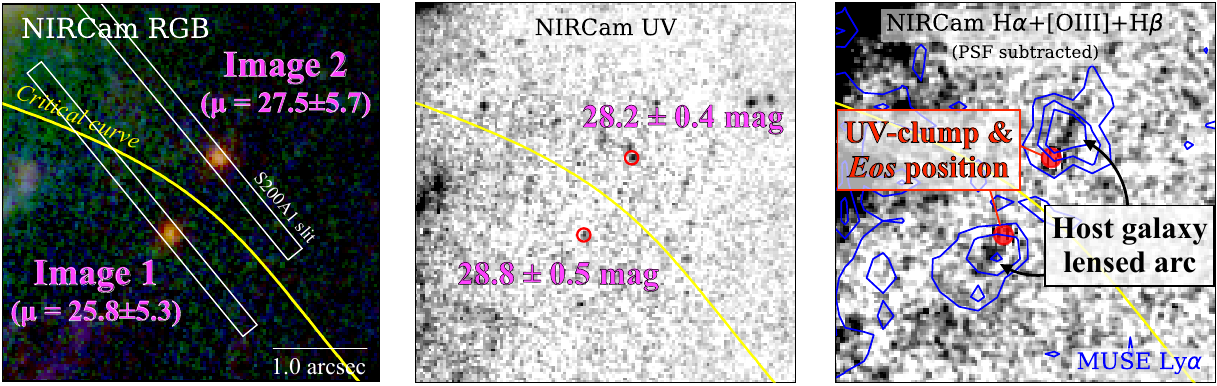}
\caption{\textit{JWST}/NIRCam images of the SN \eos\ host galaxy.
{\it Left}: the NIRCam RGB image (blue: F115W+F150W, green: F200W+F277W, red: F356W+F444W). The critical curve at $z=5.13$ is shown in yellow, obtained from the latest lens model \citep{Allingham2026arXiv}. SN \eos\ stands out as the bright and red, doubly-imaged source. The source is quintuply imaged in total, and the highest two magnification images (image 1 and image 2) near the critical line are magnified by $\mu\sim27$. The NIRSpec follow-up observation was taken with the S200A1 fixed-slit via the program DD-9493, and their slit positions are denoted by the white rectangles.
The cutout image is $4.\!^{\prime\prime}0$ wide.
{\it Middle}: NIRCam stacked image of the rest-frame UV (F090W+F115W+F150W).
The host galaxy is marginally detected at the same position of SN \eos. Red circles have a $r=0.\!^{\prime\prime}07$ radius.
{\it Right}: NIRCam stacked image of strong rest-optical emission line filters with point-source subtraction. With point-source subtraction, a lensed arc structure can be seen symmetrically imaged across the critical line. The blue contour presents the moment-zero map of the Ly$\alpha$ line taken with VLT/MUSE. The Ly$\alpha$ line emission is spatially offset from the position of the UV clump and SN \eos, and associated with the lensed arc structure.
}
\label{fig:fig1}
\end{figure*}

In contrast to SN \eos\ itself, the host galaxy is only barely detected in the individual NIRCam images.
In rest-frame UV, the emission from SN \eos\ has completely faded \citep[see Figure 5 in][]{Coulter2026arXiv} by the epoch of \textit{JWST}/NIRCam observations, and thus we stack the NIRCam images to increase the signal-to-noise ratios (S/Ns).
%where SN \eos\ has been already faded out at the epoch of \textit{JWST} observations, we stack the NIRCam images to increase the signal-to-noise ratios (S/Ns).
Given the redshift, we stack NIRCam F090W, F115W, and F150W to generate a deep NIRCam rest UV image of the \eos\ host.
We then perform fixed-aperture photometry in $0.\!^{\prime\prime}14$-diameter at the same position of \eos\ image 1 and image 2, applying a local-background subtraction and aperture correction to obtain the total flux.
The rest UV light of the host is marginally detected 
in the individual images, with $m=28.8\pm0.5$ mag and $28.4\pm0.4$ mag for image 1 and 2, respectively (Figure~\ref{fig:fig1} middle).
When combining the two images, which corresponds to a combined magnification of $\mu=53\pm8$, the rest UV light is detected at S/N=$3.0$ ($m=27.7\pm0.3$ mag).
%Given the lens magnification values obtained from the lens model, 
After correcting for the lens magnification, the \eos\ host galaxy is estimated to have an absolute UV magnitude of $M_{\rm UV}=-14.40\pm0.33$ mag.

In rest-frame optical, SN \eos\ is overwhelmingly bright and the host galaxy is outshined in the native NIRCam images. We thus perform point-source subtraction in each NIRCam image, and build the residual maps to look for the host (see Appendix \ref{apx:pssub} for details).
A spatially resolved, lensed arc component can be found in the residual maps of the F300M, F410M, and F444W filters.
Although the lensed arc component is only tentatively detected in the individual filters (S/N$\sim2$ per pixel), this becomes clearer and statistically significant when the three residual maps are combined (S/N$\sim4$-5 per pixel; Figure~\ref{fig:fig1} right). The arc morphology is also consistent with the shear map predicted by the lens model at this position, supporting the idear that the arc-like structure is a spatially resolved, lensed host galaxy component.
These filters are where the strong emission lines at $z=5.13$ fall ([O{\sc iii}]+H$\beta$ in F300M, and H$\alpha$ in F410M and F444W), we thus infer this arc structure as the lensed image of the extended nebular emission of the host galaxy.

\subsection{The \eos\ host in NIRSpec and MUSE spectra}\label{subsec:spec}

\subsubsection{NIRSpec spectrum}
\begin{figure*}[t]
\centering
\includegraphics[width=0.9\textwidth]{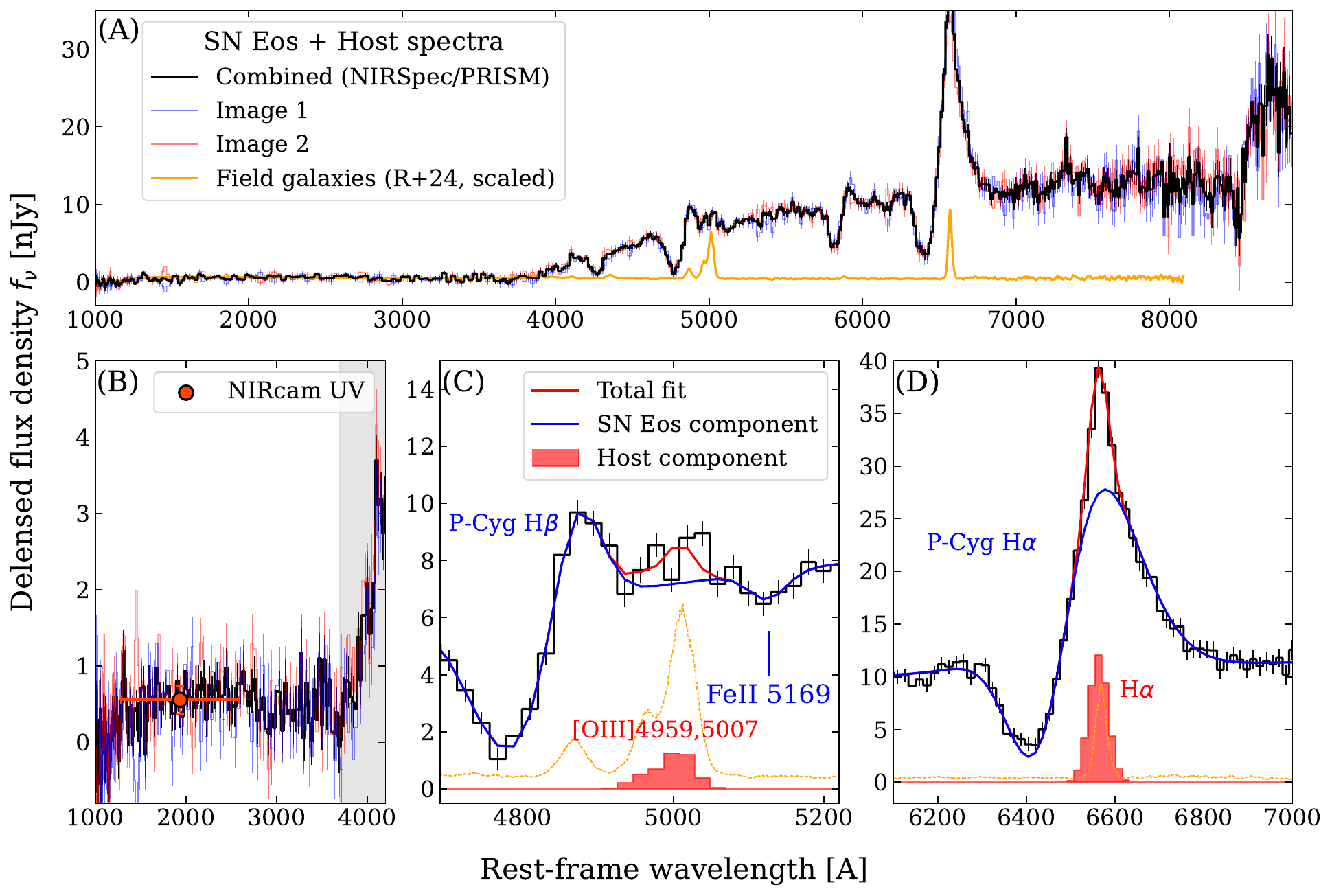}
\caption{\textit{JWST}/NIRSpec PRISM spectrum of SN \eos\ + its host galaxy.
(A) the full 1D spectrum of SN \eos\ + host. The spectrum in image 1 (blue), image 2 (red), and the composite spectrum (black) are shown. All spectra are corrected for the gravitational lensing effect. The orange curve shows the typical galaxy spectrum of faint galaxies at $z\sim5$ in the NIRSpec PRISM resolution \citep{Roberts-Borsani2024ApJ}, scaled to match the UV flux level of the SN \eos\ host galaxy.
(B) the rest-UV spectrum of the host. Although SN \eos\ outshines the host galaxy in rest-frame optical ($\lambda_{\rm rest}\gtrsim3700$ \AA; gray shaded), the host galaxy is clearly detected in rest UV, with no time variability between the epoch of the NIRCam observation (red filled circle) and the NIRSpec observation.
(C) NIRSpec spectral modeling around (rest-frame) 5000 \AA\ of the combined spectrum (black; image 1 + image 2).
The spectrum is well modeled with the type IIP SN component (black body continuum with a broad P-cygni H$\beta$ and the Fe {\sc ii}5169 absorption; blue curve), leaving only minimal residual for the host [O{\sc iii}]4959,5007 line (red filled step). The [O{\sc iii}] line is considerably weaker than the prediction by the scaled typical galaxy spectrum (orange curve).
(D) H$\alpha$ line profile modeling. Unlike [O{\sc iii}], the H$\alpha$ line profile is well modeled with the host galaxy (i.e., narrow) H$\alpha$ contribution, on top of the broad P-cyg H$\alpha$ line from the SN \eos.
The H$\alpha$ flux is roughly consistent with the prediction by the scaled reference spectrum (orange).
}
\label{fig:nirspec}
\end{figure*}

Although the NIRSpec/PRISM observation was intended to target SN \eos, the underlying host galaxy is inevitably included in the slit, enabling us to derive the host spectral properties.
As discussed in Section \ref{subsec:images}, SN \eos\ is dominating in rest-frame optical while already has faded out in rest-frame UV.
This can also be seen in the full 1D spectrum (Figure~\ref{fig:nirspec}A), 
and the rest UV continuum of the host galaxy is detected in both spectra of image 1 and image 2 (Figure~\ref{fig:nirspec}B).
The rest UV flux level in NIRSpec/PRISM spectrum is fully consistent with the rest UV flux measurement from the NIRCam composite image (Section~\ref{subsec:images}; red circle in Figure~\ref{fig:nirspec}B).
This stands in contrast to the SN \eos\ component, where the phase evolution is reported between the epochs of \textit{JWST}/NIRCam and NIRSpec observation ($\sim1$ month apart in the observer frame), confirming that the rest UV light is not time-variable and originates in the host galaxy.

Given the UV flux level of the host, strong emission lines such as H$\alpha$ and [O{\sc iii}]4959,5007 are also expected to be detectable on top of the SN \eos\ spectrum.
Orange curves in Figure~\ref{fig:nirspec} present the typical faint galaxy spectrum at $z\sim5$ in the NIRSpec/PRISM resolution \citep[$\langle M_{\rm UV}\rangle=-17.9$ mag;][]{Roberts-Borsani2024ApJ}, scaled to match the UV flux level in the \eos\ host.
The line peaks of H$\alpha$ and [O{\sc iii}]4959,5007 are expected to be considerably above the flux error levels of our NIRSpec/PRISM spectra, and comparable to the SN \eos\ continuum level.
We thus perform line profile modeling of these two lines including both SN \eos\ and the host galaxy component.
In the following, we utilized the python implementation of markov-chain monte-carlo (MCMC) sampling \citep[\texttt{emcee};][]{emcee} to obtain the posterior distributions of line profile parameters.

For the H$\alpha$ line, the SN \eos\ component (continuum + the broad P-cygni H$\alpha$) is modeled with two Gaussians, each representing the broad emission and absorption respectively, and a linear underlying continuum.
We first fit the pure-SN model to the NIRSpec/PRISM spectrum, and find that it leaves significant residuals in all spectra (image1, image2, and the composite).
This is considerably improved when a narrow Gaussian component is included in the model, representing the host-origin H$\alpha$ line (Figure~\ref{fig:nirspec}D).
The Bayesian Information Criteria (BIC) value differences between the pure-SN model and the host-included model ($\Delta{\rm BIC}={\rm BIC_{pureSN}} - {\rm BIC_{SN+host}}$) are 88.4, 26.7, and 123.7, in the image1, image2, and composite spectrum, respectively, which significantly favor host-included models in all spectra (see Appendix \ref{apx:NL_validation} for a more detailed discussion on the robustness of host-included models).
The derived H$\alpha$ line flux is fully consistent with the predicted H$\alpha$ level from the scaled typical faint galaxy spectrum (orange curve in panel D).
The inferred H$\alpha$ line flux gives an ionizing photon production efficiency $\xi_{\rm ion}$ of $\log(\xi_{\rm ion}/{\rm Hz\ erg^{-1}})=25.7\pm0.1$ for the SN Eos host galaxy, which is quite typical for faint LAEs \citep[e.g.,][]{Saxena2024AAP,Roberts-Borsani2024ApJ}.
We also confirm that the narrow-component detection at this level cannot be a false-positive due to the complex type IIP H$\alpha$ profile in the NIRSpec/PRISM resolution (Appendix \ref{apx:NL_validation}), and thus we conclude the narrow H$\alpha$ detected in the NIRSpec spectra should be the host-origin H$\alpha$.

%For [O{\sc iii}], on the other hand, the SN spectral profile around this wavelength needs more complex modeling.
The NIRSpec spectra around $\lambda_{\rm rest}\sim5000$ \AA\ are modeled including the [O{\sc iii}] line from the host and the photospheric continuum + multiple emission/absorption features of SN \eos.
Since the continuum shape around $\lambda_{\rm rest}\sim5000$ \AA\ cannot be well fit with a simple linear function, we first model the SN continuum with a blackbody masking wavelengths where strong emission or absorption are expected in type IIP spectra at this phase.
The best-fit blackbody temperature is $T_{\rm BB}=5500$ K, which is in excellent agreement with expected values for the photosphere at the late plateau phase of type IIP SNe \citep[e.g.,][]{Dastidar2018MNRAS,Faran2018MNRAS,Hillier2019AAP}.
Using the best-fit continuum model, we then model the line profiles around $\lambda_{\rm rest}\sim5000$ \AA\, including the broad P-Cygni H$\beta$ emission line and Fe{\sc ii} 5169 absorption line associated to the SN \eos, and [O{\sc iii}]4959,5007 emission lines from the host.
%The first two should be from SN \eos\ and the [O{\sc iii}] doublet is from the host.
We do not include the host-origin narrow H$\beta$ line given its insufficient S/N and resolution to decompose the SN \eos\ component and the narrow host component.
The NIRSpec spectrum is well modeled overall, and the [O{\sc iii}]4959,5007 line from the host is found remarkably weak (Figure~\ref{fig:nirspec}C).
Pure type IIP spectra at this phase should not emit O$^{++}$ lines and it should be much easier to detect the host-origin [O{\sc iii}]4959,5007 lines on top of the SN \eos\ spectrum if it exists, but there is almost no residual with the pure SN component (blue curve in the panel C).
Indeed, in contrast to the narrow H$\alpha$ component, the [O{\sc iii}] line component is not statistically required to fit the NIRSpec spectra according to the BIC; $\Delta{\rm BIC}={\rm BIC_{pureSN}} - {\rm BIC_{SN+[OIII]}}$ are only 3.0, 2.0, and 17.0 in the three spectra, and the host [O{\sc iii}]4959,5007 lines are hardly detected.
We thus treat the host [O{\sc iii}]5007 line as a non-detection, and adopt 97.72- (99.87-) percentile of the posterior distribution for the [O{\sc iii}]5007 flux as the 2-sigma (3-sigma) upper limit.
The weak [O{\sc iii}] lines lead to very low [O{\sc iii}]5007/H$\alpha$ ratios of the host galaxy in the all three spectra: $<0.25$, $<0.32$, and $<0.23$ for the image 1, image 2, and composite spectrum, respectively (at 2-sigma).
%Same results are obtained in all of the image 1, image 2, and the composite spectrum, and the [O{\sc iii}]5007/H$\alpha$ ratios in the host galaxy are estimated as $0.14\pm0.05$, $0.22\pm0.07$, and $0.17\pm0.03$, in the three spectra (image 1, 2, and composite).
The weak [O{\sc iii}] lines are further discussed in Section \ref{subsec:metallicity}.

%Pure type IIP spectra at this phase should not emit the O$^{++}$ lines and it is much easier to detect the host-origin [O{\sc iii}]4959,5007 lines on top of the SN \eos\ spectrum, but there is almost no residual with the pure SN component (blue curve in the panel C).
%Indeed, the best [O{\sc iii}]4959,5007 line flux is significantly below the expected value from the scaled typical galaxy spectrum (orange curve in the panel D).

\subsubsection{MUSE spectra}
We can obtain a cleaner picture of the host from the VLT/MUSE observations, since the MUSE data were taken $\sim10$ years before SN \eos\ explosion is observed in image 1 and 2.
As discussed in \citet{Coulter2026arXiv}, the Ly$\alpha$ emission line from the host is detected not only in image 1 and 2 but also from other multiple images (see Figure 3 in \citealt{Coulter2026arXiv} and Figure 2 in \citealt{Allingham2026arXiv}), confirming the host galaxy is a Lyman-alpha emitter (LAE).
%The total Ly$\alpha$ fluxes in image 1 and 2 are \blue{XXX, and }
We generate the Ly$\alpha$ moment-zero map from the $\lambda_{\rm obs}=7452-7458$ \AA\ cube, which is overlaid as the contour on the H$\alpha$+[O{\sc iii}]+H$\beta$ emission line NIRCam image in Figure~\ref{fig:fig1}, right.
The Ly$\alpha$ contour is obviously associated with the lensed arc component and offset from the \eos\ position in both images, which further supports the robustness of the lensed host detection in the NIRCam F300M+F410M+F444W image and host-origin of the Ly$\alpha$ emission.

We also extract the integrated 1D spectra of the Ly$\alpha$ emission, to measure the total line fluxes.
The spectra are extracted from the isophotal 9 spaxels in each image (corresponding to a 0.36 arcsec$^2$ aperture), and the line flux is measured by fitting a Gaussian profile with aperture corrections.
The resulting Ly$\alpha$ fluxes are $(3.6\pm0.3)\times10^{-18}$ and $(5.7\pm0.4)\times10^{-18}\ {\rm erg\ s^{-1}\ cm^{-2}}$ in image 1 and 2, respectively (before lens correction).
When we fit a Gaussian to the summed spectrum, we obtain $(10.1\pm0.6)\times10^{-18}\ {\rm erg\ s^{-1}\ cm^{-2}}$.
These values correspond to rest-frame Ly$\alpha$ equivalent widths (EWs) of $\sim140-150$\ \AA, placing the \eos\ host among the highest EW LAEs \citep[see, e.g.,][for general LAEs found with \textit{JWST}/NIRSpec]{Saxena2024AAP}.
%\blue{Yoshi: What do people usually do with MUSE aperture correction? Idk the PSF shape and PSF EE of the VLT/MUSE instrument, so just assume the 2D gaussian and estimate this value.}

\begin{deluxetable}{lcccccccc}
    \label{tab:host_phys}
    \tablecaption{Physical properties of the SN \eos\ host galaxy
  	}
    \tablewidth{0pt}
    \tablehead{
    \colhead{Image} & \colhead{$z_{\rm spec}$} & \colhead{$M_{\rm UV}$} & \colhead{${\textrm{EW}^0_{\textrm{Ly}\alpha}}$} & \colhead{$F_{\textrm{H}\alpha}$} & \colhead{$F_{\textrm{[OIII]5007}}$} & \colhead{$12+\log({\rm O/H})_{R3}$} & \colhead{log($M_\star$)} & \colhead{$\mu$} \\
    \colhead{} & \colhead{}& \colhead{ABmag} & \colhead{\AA} & \colhead{$10^{-20}$ cgs} & \colhead{$10^{-20}$ cgs} & \colhead{} & \colhead{$M_\odot$} & \colhead{}\\
    \colhead{(1)} & \colhead{(2)}& \colhead{(3)} & \colhead{(4)} & \colhead{(5)} & \colhead{(6)} & \colhead{(7)} & \colhead{(8)} & \colhead{(9)}
    }
    \startdata
    Image 1 & $5.133 \pm 0.001$ & $-14.15 \pm 0.54$ & $145 \pm 75$ & $9.2\pm2.3$ & $<2.3$ & $<6.68$ & $6.4^{+0.3}_{-0.3}$ & $25.79\pm5.31$ \\
    Image 2 & $5.131 \pm 0.002$ & $-14.60 \pm 0.38$ & $143 \pm 50$ & $5.8\pm1.7$ & $<1.8$ & $<6.76$ & $6.5^{+0.4}_{-0.4}$ & $27.49\pm5.67$ \\
    \hline
    Composite & $5.132 \pm 0.001$ & $-14.40 \pm 0.33$ & $156 \pm 48$ & $7.5 \pm 1.4$ & $<1.7$ & $<6.67$ & $6.5^{+0.4}_{-0.4}$ & $53.28\pm7.77$ \\
    \enddata
    \tablenotetext{}{(1) SN \eos\ host images. (2) System redshift, measured from the narrow H$\alpha$ line. (3) Lens-corrected absolute UV magnitude. (4) Rest-frame Ly$\alpha$ EW. (5) Lens-corrected H$\alpha$ line fluxes. (6) Lens-corrected [O{\sc iii}]5007 line fluxes. (7) Metallicity. (8) Stellar mass. (9) Lens magnification factor.
    }
    \tablenotetext{}{Upper limits are quoted at the 2-sigma level. Gravitational lens corrections are applied when needed, and the magnification factor uncertainties are propagated.}
\end{deluxetable}

\subsection{Host stellar mass estimation}
The stellar mass of the host galaxy is a key characterization, but a direct estimation with the standard approach of SED fitting cannot be applied to the \eos\ host given the limited information.
Instead, we estimate the stellar mass with two different approaches, each giving the lower and upper limit respectively, to obtain a rough estimation of the host stellar mass.
We first estimate the stellar mass from the UV and H$\alpha$ luminosities of the host galaxy.
Both H$\alpha$ and UV luminosities are known to trace the SFR but on different timescales \citep[$\sim5$-$10$ Myrs with H$\alpha$, $\sim100$ Myrs with UV; see e.g.,][]{Kennicutt1998ARAA}.
We can therefore roughly estimate the star-formation history (SFH) of the host galaxy in the recent 100 Myr with H$\alpha$ and UV luminosity. 
This enables us to estimate the stellar mass assuming the star-formation more than 100 Myrs ago is negligible, which likely gives a lower-limit estimation of the host galaxy stellar mass.

We model the SFH with a simple step function:
\begin{equation}
\textrm{SFR}(t) = 
\begin{cases} 
\psi_0\ [M_\odot\ {\rm yr}^{-1}] & (0<t<10\ {\rm Myr}) \\ 
\psi_1\ [M_\odot\ {\rm yr}^{-1}] & (10<t<100\ {\rm Myr})
\end{cases}
\end{equation}
where $t$ is the lookback time from the epoch of the observation.
Since the H$\alpha$- (UV-) based SFR is approximately 10 (100) Myr-averaged SFR, $\psi_0 \sim {\rm SFR}_{\textrm{H}\alpha}$ and $\psi_1 \sim (10{\rm SFR}_{\rm UV} - {\rm SFR}_{\textrm{H}\alpha})/9$.
The stellar mass can be then computed as:
\begin{eqnarray}
    M_\star &\sim& 10^7 (\psi_0 + 9R\psi_1) \notag\\
    &\sim& 10^7 (10R\ {\rm SFR}_{\rm UV} + (1-R)\ {\rm SFR}_{\textrm{H}\alpha})
\end{eqnarray}
where $R$ is the live mass fraction ($R\sim0.7$ with the Chabrier IMF at $\sim100$ Myr timescale).
We use the H$\alpha$ (UV) luminosity-to-SFR conversion factors given by \citet{Eldridge2017PASA} to estimate the ${\rm SFR}_{\textrm{H}\alpha}$ (${\rm SFR}_{\rm UV}$) values, and obtain the stellar mass as $M_\star\sim10^{6.0}$, $10^{6.2}$, and $10^{6.1}\ M_\odot$ from image 1, image 2, and the composite image, respectively.

We also estimate the stellar mass by scaling a typical faint SFG spectrum at the similar redshift to match the observed UV luminosity (the orange curve in Figure~\ref{fig:nirspec}).
%The SN \eos\ host galaxy is 22 times fainter than the stacked spectrum obtained by \citet{Roberts-Borsani2024ApJ}
Assuming the same mass-to-light ratio as the stacked spectrum by \citet{Roberts-Borsani2024ApJ}, the stellar mass of the \eos\ host is estimated as $M_\star\sim10^{6.8}\ M_\odot$.
The stellar mass estimation from the simplified SFH should be regarded as a lower limit due to the unknown contribution of the old stellar component ($t_{\rm age}>100$ Myr), while that from the scaled spectrum should be an upper limit because the mass-to-light ratio is likely lower in the \eos\ host galaxy than the stacked spectrum by \citet{Roberts-Borsani2024ApJ}: the host galaxy is $\sim20$ times fainter with a $\sim3$ times higher H$\alpha$-to-UV luminosity ratio than the stacked spectrum, which suggests the host galaxy has higher sSFR than the reference spectrum, leading to a lower mass-to-light ratio.
We thus take the mean value of the two estimations as fiducial, putting upper/lower errors from the two estimations.

Physical properties obtained in this section are listed in Table~\ref{tab:host_phys}.
%In the following, 
The physical properties estimated from different multiple images are in good agreement with each other.
Given the highest S/Ns, in the following, we use results from the composite image as the fiducial estimation to discuss the host galaxy properties.

\section{Results and Discussion}\label{sec:result}
\subsection{An ultra-faint LAE hosting SN Eos at $z=5.13$}\label{subsec:result_LAE}
\begin{figure}[t]
\centering
\includegraphics[width=0.95\linewidth]{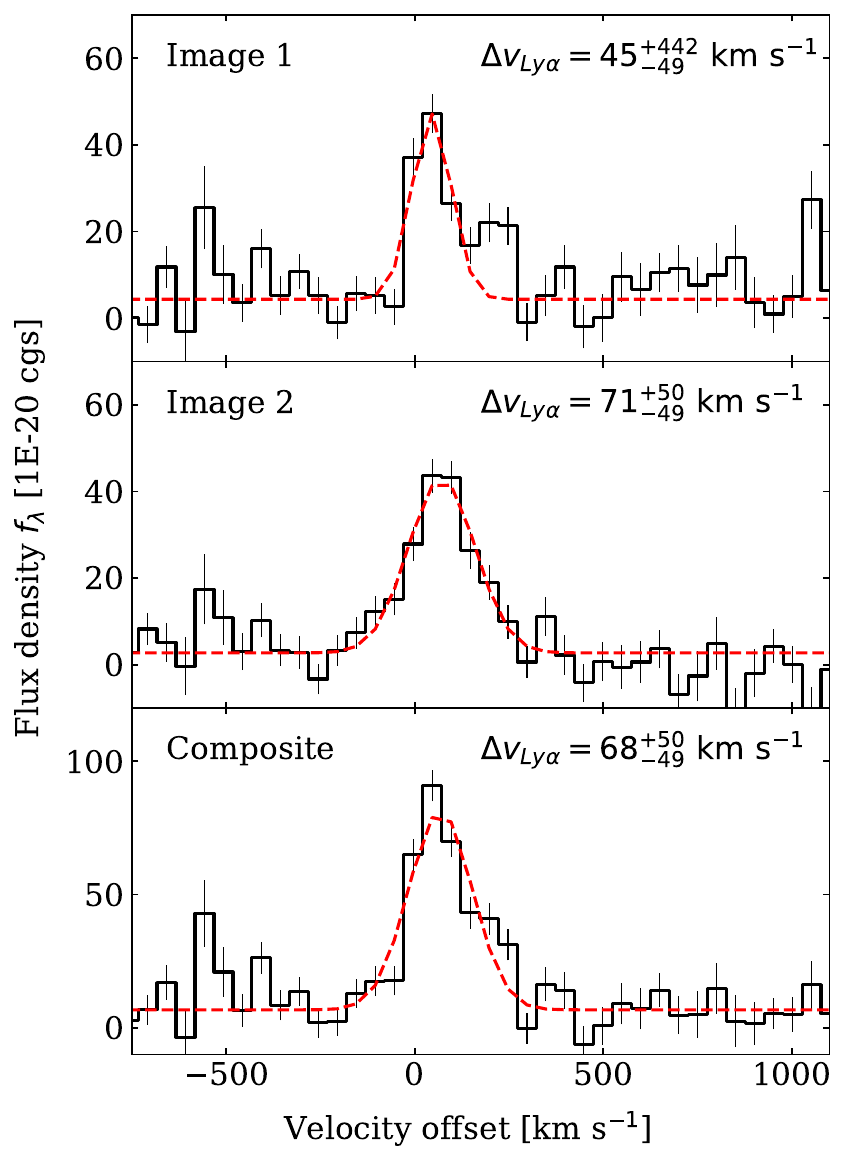}
\caption{The Ly$\alpha$ line profile taken with VLT/MUSE, extracted for image 1 (top), image 2 (middle), and the composite spectrum (bottom). 
The velocity offset is measured from the host H$\alpha$ line redshift ($z_{\rm sys}=5.13$).
Red dashed curves show the best-fit Gaussian, with which the line fluxes are measured.
}
\label{fig:LyA}
\end{figure}

The SN \eos\ host galaxy is revealed to be an ultra-faint LAE, $M_{\rm UV}=-14.4$ mag after gravitational lens correction.
The Ly$\alpha$ line profile of the SN \eos\ host is relatively symmetric and shows only negligible velocity offset from the system redshift ($\Delta v_{ \textrm{Ly}\alpha}<100\ {\rm km\ s^{-1}}$; Figure~\ref{fig:LyA}).
%The absolute UV magnitude and the high Ly$\alpha$ EW ($156\pm48$ \AA) places the \eos\ host as 
Such a high EW Ly$\alpha$ ($156\pm48$ \AA) line with a very small velocity offset is consistent with the (extrapolated) empirical relation \citep[e.g.,][]{Saxena2024AAP}, placing the SN \eos\ host as an extreme LAE at the faint-end \citep[see, e.g.,][for stacking analyses of MUSE-identified LAEs in a blank field]{Maseda2018ApJ}.

Moreover, the Ly$\alpha$ emission is spatially offset from the rest-frame UV clump position, where SN \eos\ exploded, and is rather associated with the lensed arc position found in the NIRCam emission-line filter images (Figure~\ref{fig:fig1} right).
%This suggests that the Ly$\alpha$ EW can be even higher locally.
On the other hand, the NIRSpec/PRISM slit was placed on the position of SN \eos, and we detect the H$\alpha$ line at the slit position as well (white rectangles in Figure \ref{fig:fig1} left).
These observations should suggest that the host is composed of a compact rest-frame UV clump, where \eos\ exploded, encompassed by an extended ionized nebular cloud emitting at least the H$\alpha$ line.
The Ly$\alpha$ line mainly arises from the extended nebular component, off-centered from the UV clump position.

\subsubsection{What gives rise to the spatially offset Ly$\alpha$ emission?}
Spatially offset Ly$\alpha$ emissions from the ionizing source (i.e., the UV clump) have been reported in literature, some of which are in gravitationally lensed fields \citep[e.g.,][]{Rauch2011MN,Lemaux2021MN,Claeyssens2022AAP,Markov2026AAP}.
They are often regarded as due to substructures within the star-forming region, a faint companion below the detection limit, Ly$\alpha$ resonant scattering, differential dust obscuration, or gas inflows/outflows.
In the case of the SN \eos\ host, the association of H$\alpha$ emitting gas and the symmetric line profile near the system redshift disfavor the resonant scattering and gas inflow/outflow scenarios.
Heavy dust obscuration is also less likely considering that Ly$\alpha$ mainly arises from the UV-dark region and the NIRSpec/PRISM spectra reveals the blue UV slope ($\beta_{\rm UV}=-2.2\pm0.2$).
The substructure scenario is generally taken as the predominant case when the offset is relatively small \citep[smaller than a few factors of the UV size; e.g.,][]{Claeyssens2022AAP}, while the Ly$\alpha$ emitting cloud in the \eos\ host is displaced by $\sim100$ pc, which is considerably larger than the UV clump size (unresolved with the NIRCam SW PSF with gravitational lens magnification of $\mu\sim27$; smaller than a few 10 pc).
The faint companion scenario cannot be fully ruled out, but in this case the Ly$\alpha$ EW of the companion needs to be even higher ($\gtrsim200$ \AA) and the detection of the SN \eos\ explosion at the position of the UV clump, not at the position of the companion, could be puzzling.

An alternative interpretation could be a combination of inhomogeneous geometry and a low column density of the neutral gas.
A low column density of the neutral gas is often invoked for high-EW LAEs with a low velocity offset \citep[e.g.,][]{Zackrisson2013ApJ,Verhamme2015AAP,Jaskot2019ApJ}, so that the Ly$\alpha$ photons can escape from the inter-stellar medium (ISM) with minimum resonant scattering.
This should be the case in the SN \eos\ host, at least in the line of sight toward the Ly$\alpha$ emitting region.
The associated H$\alpha$ emission indicates the Ly$\alpha$ line is directly coming from an extended H{\sc ii} region and not by scattering the resonant photons back, illuminated by the ionizing source in the host (the UV clump).
On the other hand, the Ly$\alpha$ emission in the line of sight towards the core of the host galaxy is attenuated, presumably due to a higher neutral gas density.

Such an anisotropic Ly$\alpha$ photon leakage has been suggested both from observations and simulations \citep[e.g.,][]{Heckman2011ApJ,Gronke2016ApJ,Kimm2019MNRAS,Smith2019MNRAS,Gazagnes2020AAP}, and we are possibly observing this in the ultra-faint galaxy at $z=5.13$.
The total Ly$\alpha$ flux (from MUSE/VLT) and H$\alpha$ flux (from \textit{JWST}/NIRSpec) gives the Ly$\alpha$/H$\alpha$ ratio of $2.54\pm0.34$, which corresponds to a Ly$\alpha$ photon escape fraction estimated $f_{\rm esc}({\rm Ly}\alpha)=29\pm4\ \%$ assuming case B recombination and no dust attenuation, and indeed a large fraction of Ly$\alpha$ photons are escaping from the ISM.
We again highlight that the VLT/MUSE observation was taken $\sim10$ years before the SN \eos\ explosion is observed in image 1 and 2.
This indicates that the escape of Ly$\alpha$ photons from this host galaxy is not directly triggered by the \eos\ explosion, although stellar feedback or prior SNe are expected to create the escaping channel \citep[e.g.,][]{Kimm2019MNRAS}.
Given the opacity in the line of sight to the SN \eos\ position at the epoch of the VLT/MUSE observation, the \eos\ explosion might be clearing the surrounding neutral gas and enabling Ly$\alpha$ photon escape in the future.

\subsection{Initial metal enrichment undergoing via CCSNe?}\label{subsec:metallicity}\label{subsec:result_metal}
The very low [O{\sc iii}]/H$\alpha$ ratio of the \eos\ host galaxy points to an extreme ISM condition of the host galaxy.
As shown in Figure~\ref{fig:nirspec}, both H$\alpha$ and [O{\sc iii}] lines from the host are expected to be detectable given its UV luminosity (the orange curve). The H$\alpha$ line is indeed identified roughly at a similar level to the prediction, while the [O{\sc iii}] line signal is clearly lacking and the upper limit on the [O{\sc iii}] line flux (red filled step in Figure~\ref{fig:nirspec}C) is considerably below the prediction.
It suggests the SN \eos\ host galaxy has a significantly lower [O{\sc iii}]/H$\alpha$ ratio than the stacked spectrum by \citet{Roberts-Borsani2024ApJ}.

A plausible interpretation comes with a very low gas-phase metallicity in the host galaxy H{\sc ii} region.
The $R3$ ($\equiv$[O{\sc iii}]5007/H$\beta$) line ratio has been widely used as an empirical estimator of the gas-phase metallicity in high-$z$ galaxies \citep[e.g.,][]{Nakajima2023ApJS,Curti2024AAP,Hsiao2025arXiv,Willott2025ApJ}.
Considering the blue color of the UV continuum of the \eos\ host (Figure~\ref{fig:nirspec}B), we assume that the dust attenuation in the \eos\ host is negligible and H$\alpha$/H$\beta$ ratio in the host galaxy is approximately case B prediction (2.86).
This results in an upper limit on [O{\sc iii}]/H$\beta$ of $R3<0.66$ at 2-sigma ($R3<0.81$ at 3-sigma).
Assuming an $R3$-metallicity conversion \citep{Nakajima2022ApJS}, the gas-phase metallicity is constrained as $12+\log({\rm O/H)}<6.67$ ($6.72$) at 2-sigma (3-sigma), which corresponds to $Z_{\rm gas}\lesssim1\ \%\ Z_\odot$.
This upper limit on the metallicity gets even lower if we use different conversions, such as \citet{Sanders2024ApJ,Sanders2025arXiv}.

\begin{figure*}[t]
\centering
\includegraphics[width=0.95\textwidth]{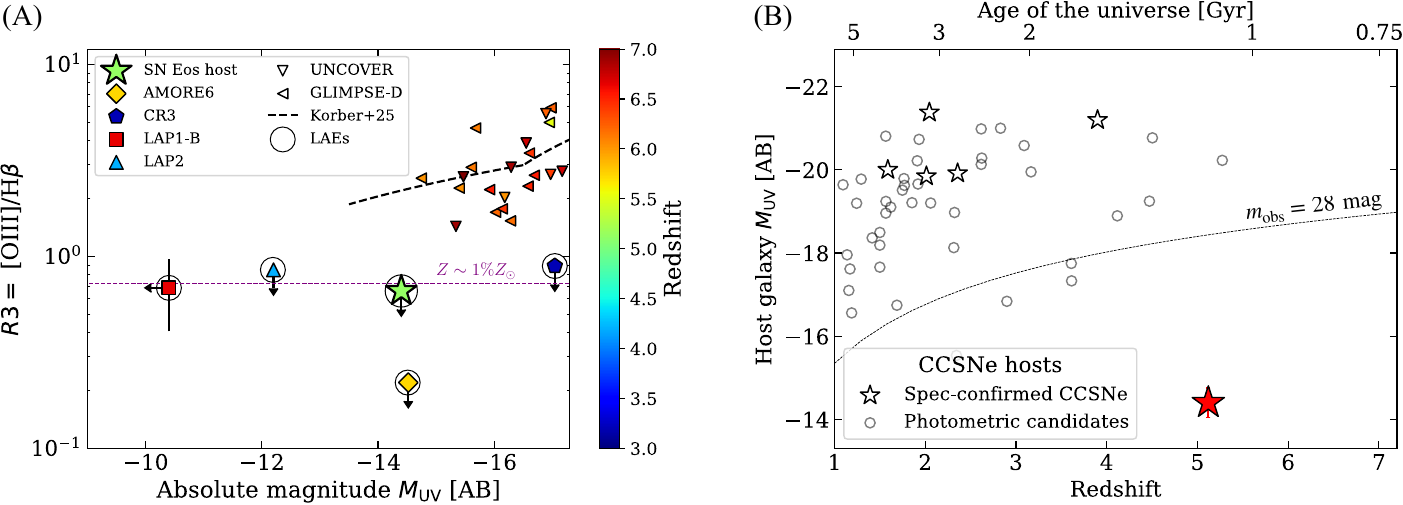}
\caption{Context of the SN \eos\ host.
(A) \eos\ host as a metal-poor galaxy candidate. The very low $R3$ value found in the \eos\ host implies that it is a member of extremely metal-poor galaxy candidates (CR3: \citealt{Cai2025ApJ}, AMORE6: \citealt{Morishita2025arXiv}, LAP1-B: \citealt{Vanzella2023AAP,Nakajima2025arXiv}, LAP2: \citealt{Vanzella2025arXiv}). They have considerably lower $R3$ values than other faint-end galaxies generally found in lensing fields (UNCOVER: \citealt{Chemerynska2024ApJ}, GLIMPSE-D: \citealt{Asada2026arXiv}, the black dashed line presents an empirical relation given by \citealt{Korber2026AAP}), and they are potentially extremely metal-poor galaxies assuming an $R3$-metallicity conversion ($\lesssim1\ \% Z_\odot$; purple dashed line given by \citealt{Nakajima2022ApJS} conversion).
(B) $M_{\rm UV}$-$z$ diagram of CCSN-host galaxies at $z>1$ found so far. Super Luminous SNe are also included.
%\eos\ host as an ultra-faint transient host galaxy at high-$z$ (the red star). %Only transients related to massive stars' explosion are plotted (CCSNe and GRBs). %SLSNe are also included. 
The \eos\ host (the red filled star) is the faintest galaxy host at the highest redshift among spectroscopically confirmed high-$z$ CCSN-host galaxies \citep[black open stars;][]{Cooke2009Natur,Cooke2012Natur,Schulze2018MNRAS}.
Hosts of photometric candidates of CCSNe found with NIRCam surveys are also plotted \citep[thin circles][]{DeCoursey2025aApJ,DeCoursey2025bApJ,Coulter2026ApJ}.
The black dashed curve is a guideline of $M_{\rm UV}$ corresponding to the observed magnitude of 28 mag (without lensing).
}
\label{fig:context}
\end{figure*}

Similar metal-poor galaxy candidates have been recently reported even at $z\sim3-6$ \citep[][]{Cai2025ApJ,Vanzella2025arXiv,Nakajima2025arXiv,Morishita2025arXiv} based on the lack of [O{\sc iii}]4959,5007 lines.
These metal-poor galaxy candidates are found to have remarkably lower $R3$ values than other faint-end galaxies \citep[e.g.,][]{Chemerynska2024ApJ,Asada2026arXiv,Korber2026AAP}, and, these candidates are all LAEs.
The \eos\ host galaxy is a similarly faint LAE with a low $R3$ value (Figure~\ref{fig:context}A), and the \eos\ host could be a member of the extremely metal-poor galaxies.
We here emphasize that SN \eos\ itself is also found to have low stellar metallicity \citep[$Z_\star<10\ \%\ Z_\odot$; ][]{Coulter2026arXiv}, which is consistent with the low-metallicity interpretation of the host gas-phase metallicity\footnote{It should be noted that type IIP SN spectra at this phase are observing the recombination of the hydrogen envelope of the progenitor, which is only weakly affected by stellar nucleosynthesis, thus the chemical abundance observed in the late plateau phase spectrum should be similar to the gas out of which the star was formed, unless many metals are transported to the surface by exceptionally effective convection.}.
If this is the case, SN \eos\ should mark the formation and explosion of a metal-poor star in the extremely metal-poor environment, and SN \eos\ is contributing to the the initial steps of the chemical enrichment of the host galaxy.
The SN \eos\ spectrum indeed shows several metal absorption features (e.g., Na {\sc i}D, Fe {\sc ii} 5169, and the Ca {\sc ii} triplet around $\sim8500$ \AA), and the metals are being ejected back to the ISM via the CCSN.

The inferred stellar mass and the gas-phase metallicity place the \eos\ host at least $0.5$ dex below the mass-metallicity relation at this redshift, and the host galaxy seems to be experiencing a delayed metal enrichment process \citep[the ``undershoot'' process proposed in][]{Asada2026arXiv}.
Under this process, the metal enrichment process is suppressed and the ISM is kept very metal-poor until the galaxy experiences the first major burst of (Pop~II) star-formation at a later cosmic time.
This picture fits well with the observations of SN \eos\ and its host, and the prolonged extremely metal-poor period in this scenario indeed increases the probability of finding a metal-poor CCSN like SN \eos.
%We also note that this scenario favors the lower-limit case of the stellar mass calculation, and thus the stellar mass of \eos\ host can be as low as $\sim10^{6.0}\ M_\odot$.
Particularly, cosmological simulations predict that even a single Population~III star-formation episode can enrich the gas-phase metallicity up to $\sim1\ \%\ Z_\odot$ level \citep[e.g.,][]{Jaacks2018MNRAS}, and thus the hydrogen envelope of SN \eos\ could preserve the information of the chemical abundance pattern from the initial (previous) Pop~III SN activity.
A further detailed abundance analysis in the \eos\ spectrum such as [Fe/H] or [Na/Fe] would be illuminating not only to reveal the chemical abundance pattern in the very early phase of metal enrichment but also to infer the formation history of the host galaxy. %, otherwise impossible in normal star-forming galaxies.

An important caveat is that the low $R3$ value can be realized not only with a low metallicity. 
The [O{\sc iii}]4959,5007 lines are collisionally de-excited at $n_e\gtrsim10^5\ {\rm cm^{-3}}$ and the line ratio becomes lower regardless of the oxygen abundance \citep[e.g.,][]{Hsiao2026arXiv}.
Moreover, some of the metal-poor galaxy candidates including the SN \eos\ host do not detect the H$\beta$ line and indirectly estimate the $R3$ value from the low [O{\sc iii}]-to-H$\alpha$ ratio, but this can be replicated also with a significant dust attenuation \citep[see, e.g.,][]{Trussler2026MNRAS}.
The blue rest-UV color and the possible turnover around $\lambda_{\rm rest}\sim1400$ \AA\ seen in the \eos\ host spectrum (Figure~\ref{fig:nirspec}B) may disfavor these scenarios, but they are only tentative, and a deeper and cleaner NIRSpec observation of the host galaxy is required to properly estimate the metallicity: detecting the host H$\beta$ line would more securely constrain the dust attenuation, and confirming the UV continuum turnover around $\lambda_{\rm rest}\sim1400$ \AA\ due to the two-photon continuum emission would place a tight upper limit on the electron density.

\subsection{Why is SN Eos found in an ultra-faint galaxy?}\label{subsec:result_why_so_faint}

\begin{figure}[t]
\centering
\includegraphics[width=0.9\columnwidth]{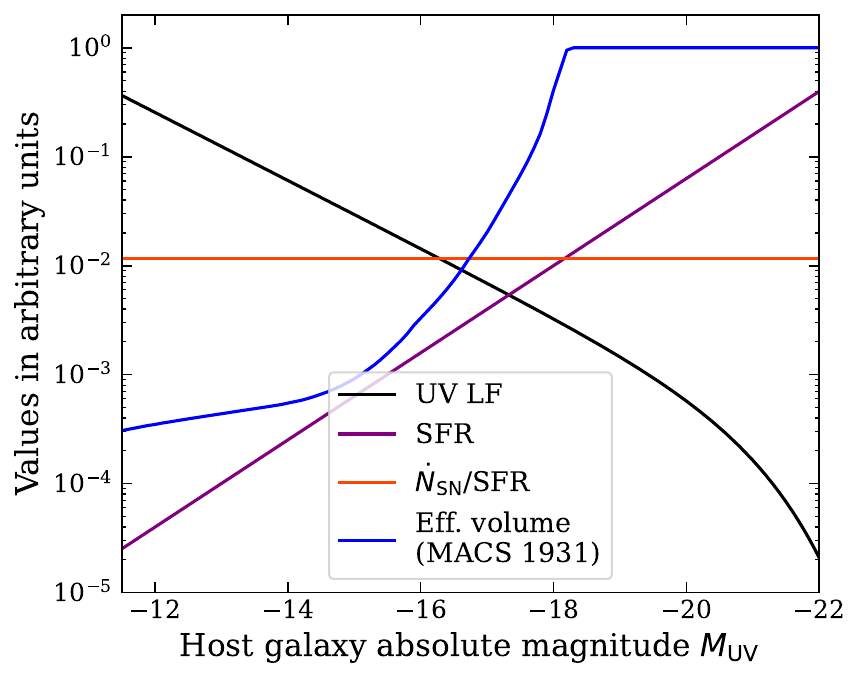}
\caption{Four key factors that determine the host $M_{\rm UV}$ probability distribution when a CCSN is found at high-$z$. The convolution of these four gives the shape of $P_{\rm host}(M_{\rm UV})$ (Figure \ref{fig:SN_finding} right), and thus adopting a different assumption on the CCSN rate per unit SFR (red curve in this plot; see also Figure \ref{fig:SN_finding} left) affects the overall shape of $P_{\rm host}(M_{\rm UV})$.
}
\label{fig:Prob_Muv_illust}
\end{figure}

\begin{figure*}[t]
\centering
\includegraphics[width=0.95\textwidth]{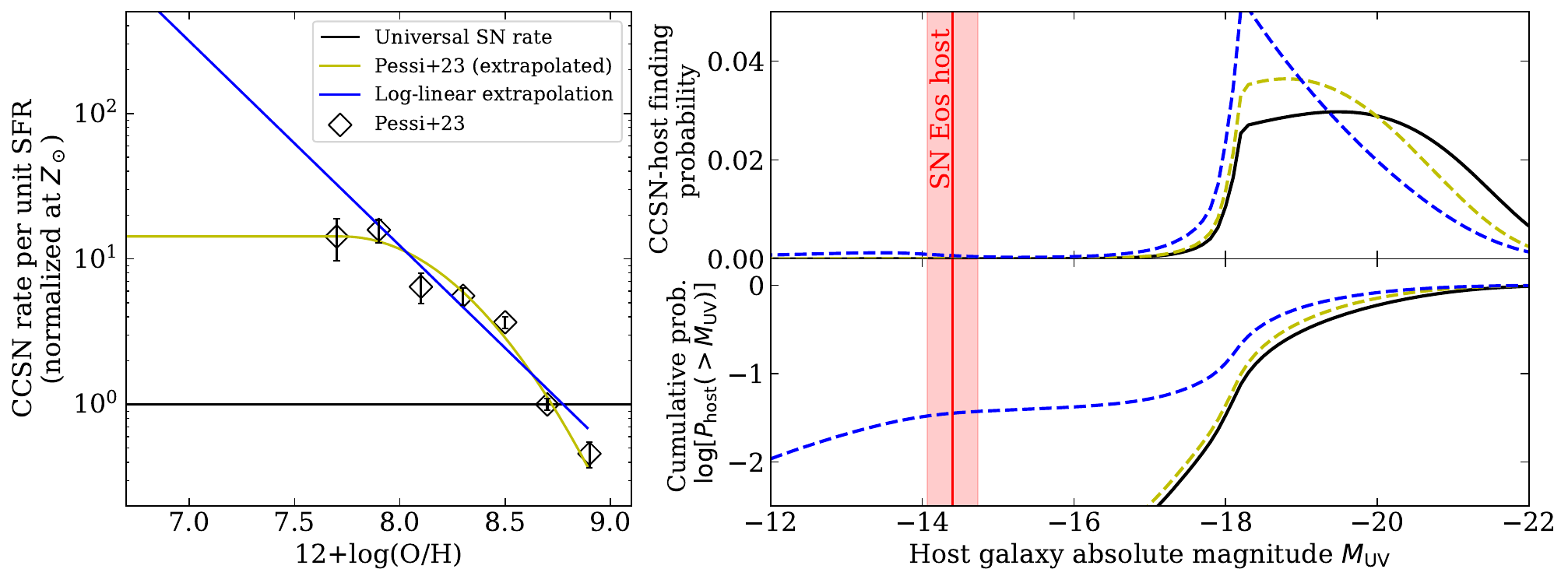}
\caption{Implication of finding of the CCSN in an ultra-faint galaxy. 
\textit{Left}: CCSN rate per unit SFR as a function of the metallicity, normalized at the solar metallicity. \citet{Pessi2023ApJ} found that the CCSN rate per unit SFR becomes higher in low-metallicity environments down to $\sim10$ \% $Z_\odot$ in the local universe. We explore three cases to infer the SN rate at even lower-metallicity: no metallicity-dependency (black), constant extrapolation of the 2nd order polynomial derived in \citet{Pessi2023ApJ} (yellow), and log-linear extrapolation (blue).
\textit{Right}: the probability distribution of the host-galaxy $M_{\rm UV}$ when a CCSN is found at $z\sim5$.
The probability function is computed by convolving the UV LF by \citet{Bouwens2022ApJ}, star-formation rate, and the $M_{\rm UV}$-dependent effective volume in the M1931 cluster field at $z\sim5.0$ (see also Figure \ref{fig:Prob_Muv_illust}). The top sub-panel shows the three probability distributions corresponding to the three assumptions of the CCSN rate's metallicity-dependency.
The bottom sub-panel displays the cumulative probability.
}
\label{fig:SN_finding}
\end{figure*}

Finding the first spectroscopically-confirmed CCSN at $z\sim5$ in such an ultra-faint galaxy has important implications for early galaxy evolution and/or stellar evolution in the high-$z$ universe.
Assuming a universal SN rate per unit SFR, the host $M_{\rm UV}$ probability distribution when a CCSN is found at $z\sim5$ can be calculated as $P_{\rm host}(M_{\rm UV})\propto\phi_{\rm LF} \cdot {\rm SFR} \cdot S_{\rm eff}$, where $\phi_{\rm LF}$, SFR, and $S_{\rm eff}$ are the UV luminosity function (LF), star-formation rate, and the effective survey area\footnote{In lensing clusters, the effective survey area varies depending on the magnification and so as on the intrinsic source luminosity.} at the given host $M_{\rm UV}$.
Figure \ref{fig:Prob_Muv_illust} illustrates these four factors that determine the host $M_{\rm UV}$ probability distribution, and the convolution of these four gives the shape of $P_{\rm host}(M_{\rm UV})$ (see Appendix \ref{apx:compute_prob} for further details).
Under the constant CCSN rate per unit SFR, the host $M_{\rm UV}$ probability distribution at $z\sim5$ should peak at the bright-end $M_{\rm UV}\sim-20$ mag (the black curve in Figure \ref{fig:SN_finding}), and the CCSN-hosting ultra-faint galaxies like the SN \eos\ host should be extremely rare -- the probability of the CCSN-host galaxy found being fainter than $-15$ mag is $\sim0.01$ \% (Figure \ref{fig:SN_finding}, right bottom).

The SN \eos\ host identification should thus indicate an elevated SN rate per unit SFR in low-metallicity environments at high-$z$. Indeed, \citet{Pessi2023ApJ} showed that the number of CCSNe per unit SFR is $\sim1$ dex higher in lower-metallicity environments even in the local universe.
Fainter, higher-$z$ galaxies are typically less metal-enriched, and our finding of the SN \eos\ host points to a metallicity-dependent SN rate in the high-$z$ universe.
In Figure \ref{fig:SN_finding} right, we show the CCSN-host finding probability ($P_{\rm host}$) under different assumptions on the metallicity-dependent CCSN rate.
\citet{Pessi2023ApJ} examined metallicity-dependent CCSN rates down to $10\ \%\ Z_\odot$, while faint-end galaxies at $z\sim5$ can be much metal-poorer and an extrapolation is needed.
We explore three scenarios: no metallicity dependency (black), constant extrapolation of the best-fit 2nd order polynomial obtained by \citet{Pessi2023ApJ} (yellow), and log-linear extrapolation of the CCSN rate (blue).
With the constant extrapolation of \citet{Pessi2023ApJ} relation (yellow), the peak of the $P_{\rm host}$ distribution shifts to fainter galaxies because of the higher SN rates, but the peak is still around $M_{\rm UV}\sim-19$ mag and the probability of finding an SN \eos\ host-like galaxy remains quite low.
Alternatively, if we adopt the log-linear extrapolation (blue), the $P_{\rm host}$ has non-negligible probability at the faint-end, and the probability of the CCSN-host being fainter than $-15$ mag increases to $\sim4$ \%.

The elevated CCSN rate in high-$z$, lower-metallicity environment could be attributed to several causes.
Recent high-resolution simulations have shown that the IMF becomes more top-heavy in high-$z$ low-metallicity environment \citep[e.g.,][]{Chon2024MNRAS,Mauerhofer2025AAP}, which leads to a higher fraction of massive stars formed at a given stellar mass.
The ``explodability'' of massive stars also depends on the stellar metallicity. At lower stellar metallicity, the line-driven stellar winds get weaker and the mass-loss during the main-sequence phase reduces \citep[e.g.,][]{Ibeling2013ApJ}. The stellar metallicity also affects the pre-SN profile structures of massive stars and differs the fate of the star \citep[i.e., whether a successful or failed explosion;][]{Maltsev2025AAP,Ono2026arXiv}.
Both minimum and maximum zero-age main sequence (ZAMS) stellar mass thresholds, determining whether the star can explode as a CCSN, therefore vary depending on the stellar metallicity.
%This is essentially because 
%\textcolor{blue}{Any other scenarios?}

Alternatively, the stellar collision may play a role in increasing the CCSN rate.
High-$z$ faint galaxies are typically found to be compact and have as high a stellar surface density as young star clusters \citep[e.g.,][]{Adamo2024Natur,Mowla2024Natur,Claeyssens2026arXiv}.
Numerical simulations have shown that such a dense environment can drive runaway stellar collisions \citep[e.g.,][]{PortegiesZwart2004Natur,Freitag2006MNRAS,Gurkan2006ApJ,Giersz2015MNRAS}, increasing the stellar mass and leading to the formation of very massive stars whose stellar masses exceed those allowed from single star evolution \citep[e.g.,][]{Kremer2020ApJ}.
Runaway collision channel may also increase the number of normal massive stars that end up with a standard CCSN explosion, although the effect of such a dynamical channel on the overall CCSN rates within dense star clusters has not yet been probed.

Although several scenarios are plausible, it remains largely uncertain which of them are most likely until we build a larger sample of CCSN-host galaxies at high-$z$.
The log-linear extrapolation requires an inflated CCSN rate by $\gtrsim1-2$ dex in the lowest-metallicity environments ($\sim1$ \%; see Figure \ref{fig:SN_finding}), and this is even more extreme than usually predicted by simulations -- e.g., just adopting a top-heavy IMF would increase the SN rate only by a factor of $\sim2-3$ \citep[e.g.,][]{Mauerhofer2025AAP}.
Moreover, when the IMF becomes top-heavy, the UV-to-SFR conversion factor gets lower, which acts to reduce the number of CCSNe for a given UV luminosity. These two factors are effectively canceled out, so only considering a $Z$-dependent IMF is not enough to enhance the CCSN rate significantly.
Nevertheless, if statistically confirmed, the high number density of ultra-faint CCSN-hosting galaxy at high-$z$ could be very robust observational evidence invoking extreme physical conditions of the star-formation or stellar evolution in earliest galaxy evolution.
Further statistical surveys of high-$z$ CCSNe-host galaxies in gravitational lensing fields are necessary to draw a conclusive picture.
%It is therefore crucial to build a larger sample of CCSN-host galaxies at high-$z$ and to statistically compare the $M_{\rm UV}$ distribution with theoretical predictions, to further study 

%the distribution of CCSN-hosting galaxies (as the function of the absolute UV magnitude of host galaxies) should be computed as the convolution of the UV LF and the UV luminosity ($\propto$ SFR).

\subsection{Origin of hostless supernovae}\label{subsec:result_hostless}
The boost in signal due to gravitational lensing enables the detection of the ultra-faint host galaxy of SN \eos, and it marks the highest-$z_{\rm spec}$ and faintest record of CCSN-host galaxies (the red star in Figure~\ref{fig:context}B).
Given its intrinsic faintness, the host  ($m\sim32$ mag) would not be detectable in any blank field \textit{JWST}/NIRCam surveys ever carried out, while SN \eos\ itself could be seen even without lensing ($m\sim28.5$ mag).
Although the number of host galaxies of spectroscopically confirmed transients at high-$z$ is limited (thick stars in Figure~\ref{fig:context}B; \citealt{Cooke2009Natur,Cooke2012Natur,Schulze2018MNRAS}), deep NIRCam imaging observations have identified substantial numbers of photometric CCSNe candidates in blank field surveys (thin circles in Figure~\ref{fig:context}B; \citealt{DeCoursey2025aApJ,DeCoursey2025bApJ,Coulter2026ApJ}).
There are also a large number of SN candidates without any associated host found in \textit{JWST}/NIRCam images \citep[{\it hostless} SNe; e.g.,][]{DeCoursey2025aApJ,Tee2025TNSTR}, and the origins of these hostless SNe are under debate \citep[e.g.,][]{Strolger2025ApJ}.
Our observation provides direct evidence supporting a population of SNe hosted in dwarf galaxies below the detection limit. %that the origin is a SN hosted in a dwarf galaxy below the detection limit.
Conversely, this also indicates that there could be many more faint-end galaxies like the SN \eos\ host galaxy below the detection limit experiencing CCSN feedback. %, given the large number of {\it hostless} SNe found in JWST blank field surveys.
Further detailed follow-up observations both on SN \eos\ itself and its host galaxy is necessary to connect findings in the SN \eos\ + host system with faint SN host galaxies at high-$z$ -- e.g., accurate estimations of the host stellar mass and metallicity and the progenitor ZAMS mass of SN \eos\ will be critical to estimate the number densities of similar galaxies.
%More detailed follow-up observations both on SN \eos\ itself and its host galaxy will allow us to further characterize SN host galaxies in general -- in particular, an accurate estimations of the host stellar mass and metallicity and the progenitor ZAMS mass of SN \eos\ will be critical to expand the finding to the general galaxy/CCSN population.

%having a more accurate measurement of the host stellar mass or SFH and the progenitor ZAMS mass of SN \eos\ would test the galaxy IMF, comparing the total CCSNe number counts found in JWST NIRCam observations and the prediction from the galaxy UV luminosity density would .

\section{Summary}
We report the physical properties of the host galaxy of a gravitationally lensed, type IIP SN at $z=5.13$ (SN \eos).
SN \eos\ and its host galaxy are located behind the lensing cluster MACS1931 and are predicted to be quintuply imaged.
The two most highly-magnified images (both $\mu\sim27$) near the critical line provide an unparalleled opportunity to study the role of CCSNe in early galaxy formation directly from observations.
We characterize the \eos\ host galaxy leveraging \textit{JWST}/NIRCam and NIRSpec observations, supplemented with archival data from VLT/MUSE.
Our key findings are:
\begin{enumerate}
    \item The SN \eos\ host galaxy is a low-mass ($M_\star\sim10^{6.5\pm0.4}\ M_\odot$), ultra-faint ($M_{\rm UV}=-14.4\pm0.3$ mag) LAE at $z=5.13$, with a very high Ly$\alpha$ EW (${\rm EW}_{ {\rm Ly}\alpha}^0=156\pm48$ \AA). The Ly$\alpha$ line profile is relatively symmetric and very close to the system redshift ($\Delta v_{{\rm Ly}\alpha}<100\ {\rm km\ s^{-1}}$), and the line is also spatially offset from the ionizing source (i.e., rest-frame UV) position, where SN \eos\ exploded. The observations should imply an anisotropic leakage of the Ly$\alpha$ photon from the SN \eos\ host galaxy ($f_{\rm esc}({\rm Ly}\alpha)=29 \pm 4\ \%$) or the presence of an even fainter companion with a higher Ly$\alpha$ EW.
    \item The NIRSpec PRISM spectrum finds the host-origin H$\alpha$ line and the lack of clear signal of the [O{\sc iii}]4959,5007 lines. The lack of [O{\sc iii}] lines results in a very low gas-phase metallicity ($R3<0.66$, $Z_{\rm gas}\lesssim1\ \%\ Z_\odot$), when an empirical $R3$-metallicity conversion is assumed. If this is the case, SN \eos\ possibly marks the formation and explosion of a metal-poor star in the extremely metal-poor environment, releasing the metals back to the ISM with the CCSN explosion. Alternatively, the low $R3$ may be due to significant dust obscuration and/or high electron density, although the blue UV color and the possible turnover at $\lambda_{\rm rest}\sim1400$ \AA\ in the \eos\ host spectrum could disfavor these scenarios.
    \item The $M_{\rm UV}$ probability of the host galaxy when a CCSN is found should be peak at a bright-end ($\sim-20$ mag) if a universal SN rate is assumed. Our finding of the first spectroscopically-confirmed CCSN at $z\sim5$ in an ultra-faint galaxy ($M_{\rm UV}=-14.4$ mag) could thus point to an elevated SN rate in fainter galaxies with lower-metallicity environments. Although it remains uncertain without a large statistical sample of CCSN-host galaxies at high-$z$, such a SN rate enhancement could be indicative of a $Z$-dependent IMF, $Z$-dependent massive star explodability, or massive star formation via runaway stellar collisions in dense stellar clusters at high-$z$. Our result thus signifies the importance of statistical search for high-$z$ SN host galaxies particularly in lensing cluster fields to unveil the astrophysics behind the early galaxy formation.
    \item The high lens magnification allows us to detect and characterize the SN \eos\ host galaxy, which would otherwise not be possible with \textit{JWST}/NIRCam sensitivity if it appeared in a blank field. On the other hand, the SN component can be detected with a medium-to-deep NIRCam survey strategy. This indicates that SN \eos\ would have been recognized as a {\it hostless} SN without the lensing effect, a type of which is frequently found in \textit{JWST} NIRCam surveys. The SN \eos\ host galaxy can be a representative of the origin of {\it hostless} SNe, SNe hosted in dwarf galaxies below the detection limit, and there could be far more faint-end galaxies like the SN \eos\ host.
\end{enumerate}

The discovery of SN \eos\ and the identification of its ultra-faint host galaxy at $z=5.13$ opens a new window to directly study the role of CCSNe in early galaxy formation. However, key information both about SN \eos\ and the host is still missing to fully characterize the system (e.g., ZAMS mass of the \eos\ progenitor, the abundance pattern in the \eos\ spectrum, more accurate measurements of the stellar mass and the SFH of the host, and a proper gas-phase metallicity measurement taking account for the degeneracy between dust attenuation/electron density).
Further follow-up observations of SN \eos\ and a deeper and cleaner observation of the host are required to push forward our understanding of early galaxy evolution with CCSNe.

%% Please use the acknowledgment and contribution environments. This will 
%% be anonomyized when the "anonymous" style option is used. 
\begin{acknowledgments}
This research used the Canadian Advanced Network For Astronomy Research (CANFAR) operated in partnership by the Canadian Astronomy Data Centre and The Digital Research Alliance of Canada with support from the National Research Council of Canada, the Canadian Space Agency, CANARIE and the Canadian Foundation for Innovation.
This work is supported by the Canadian Space Agency (CSA) through 25JWGO4A18.
YA and SF acknowledges support from the Dunlap Institute, funded through an endowment established by the David Dunlap family and the University of Toronto. CL acknowledges support from HST-GO-17474.
PD warmly acknowledges support from an NSERC discovery grant (RGPIN-2025-06182).
KM acknowledges support from JSPS KAKENHI grant (JP24KK0070, JP24K00682, and JP23H04894).
MB acknowledges support from the ERC Grant FIRSTLIGHT \# 101053208, Slovenian national research agency ARIS through grants N1-0238 and P1-0188, and ESA PRODEX Experiment Arrangement No.~4000149972.
FEB acknowledges support from ANID-Chile BASAL CATA FB210003 and FONDECYT Regular 1241005.
PAAL thanks the support from CNPq, grants 310260/2025-6 and 404160/2025-5.
MN acknowledges support from KAKENHI Grant Nos. 25KJ0828 through Japan Society for the Promotion of Science (JSPS).
GN acknowledges support by the Canadian Space Agency under a contract with NRC Herzberg Astronomy and Astrophysics.
MGO acknowledges financial support from the State Agency for Research of the Spanish MCIU through Center of Excellence Severo Ochoa award to the Instituto de Astrofísica de Andalucía CEX2021-001131-S funded by MCIN/AEI/10.13039/501100011033, and from the grant PID2022-136598NB-C32 “Estallidos8” funded by MCIN/AEI/10.13039/501100011033 and by “ERDF A way of making Europe”. MGO also acknowledges the support by the project ref. AST22\_00001\_Subp\_11 funded from the EU – NextGenerationEU, PPCC Junta de Andalucía.
KI acknowledges support from the National Natural Science Foundation of China (12573015, W2532003), the Beijing Natural Science Foundation (IS25003), and the China Manned Space Program (CMS-CSST-2025-A09).
Y.J-T. acknowledges financial support from the State Agency for Research of the Spanish MCIU through Center of Excellence Severo Ochoa award to the Instituto de Astrof\'isica de Andaluc\'ia CEX2021-001131-S funded by MCIN/AEI/10.13039/501100011033, and from the grant PID2022-136598NB-C32 Estallidos and project ref. AST22-00001-Subp-15 funded by the EU-NextGenerationEU.
KK acknowledges the support by JSPS KAKENHI Grant Numbers JP22H04939, JP23K20035, and JP24H00004.
GEM acknowledges the Villum Fonden research grant 13160 “Gas to stars, stars to dust: tracing star formation across cosmic time,” grant 37440, “The Hidden Cosmos,” and the Cosmic Dawn Center of Excellence funded by the Danish National Research Foundation under the grant No. 140.
EV acknowledges financial support through grants 
INAF GO Grant 2024 ``Mapping Star Cluster Feedback in a Galaxy 450 Myr after the Big Bang'' and the European Union – NextGenerationEU within PRIN 2022 project n.20229YBSAN - Globular clusters in cosmological simulations and lensed fields: from their birth to the present epoch.
RAW acknowledges support from NASA JWST Interdisciplinary Scientist grants
NAG5-12460, NNX14AN10G and 80NSSC18K0200 from GSFC.
AZ acknowledges support by the Israel Science Foundation Grant No. 864/23.
\end{acknowledgments}

\section*{Data Availability}

The JWST data used in this paper were obtained from the Mikulski Archive for Space Telescopes (MAST) at the Space Telescope Science Institute.
Data can be found at \dataset[doi: 10.17909/p2k3-3543]{http://dx.doi.org/10.17909/p2k3-3543}.

%\begin{contribution}
%%This section gives authors the space to recognize author contributions. The text inside this environment is NOT counted towards the total word quanta. At a minimum, manuscripts are expected to include this text:

%% But authors are expected to provide more specific details, e.g. 
%%
%%SC was responsible for writing and submitting the manuscript.
%%WWM came up with the initial research concept and edited the manuscript.
%%OTS obtained the funding and edited the manuscript.
%%EBF provided the formal analysis and validation. He also edited the manuscript.
%%GEH Supervised the undergraduates, wrote the software and administers the project github and Zenodo repositories.
%%
%% Authors can use the Contributor Role Taxonomy (CRediT) at
%% for ideas on how write a good statement tailored to their needs.

%\end{contribution}

\facilities{HST(ACS and WFC3), JWST(NIRCam and NIRSpec), VLT(MUSE)}

\software{astropy \citep{2013A&A...558A..33A,2018AJ....156..123A,2022ApJ...935..167A},  
          aperpy \citep{Weaver2024ApJS}, 
          emcee \citep{emcee},
          photutils \citep{photutils1.12.0}
          }


\appendix

\section{Point-source subtractions}\label{apx:pssub}
This appendix section presents the details of the point-source subtraction in each NIRCam image.
We first select point source candidates at $m_{277}=21-25$ mag in the NIRCam image based on the compactness, which is measured as the flux ratios of fixed-aperture photometry in $D=0.\!^{\prime\prime}4$ and $D=0.\!^{\prime\prime}1$ apertures. 
For each point source candidate, we use \texttt{EAzY} to do SED fitting with galaxy templates and stellar templates, and use only sources whose SED are better fit with stellar templates ($\Delta \chi^2 = \chi^2_{\rm star}-\chi^2_{\rm galaxy}<-20$).
We finally visually inspect all candidates and remove any sources with bright neighbor, close to the edge of the NIRCam footprint, or significant contamination by other sources (e.g., diffraction spikes).

We then build the PSFs in each NIRCam filter.
Although an empirical PSF built from actual point sources in NIRCam images is a proper model of the PSF of the image, it has high-frequency noise dominating at the outer wings, which could introduce an artificial pattern in the convolution kernel.
On the other hand, the simulated PSF given by the STPSF package is noise-free, but it is not necessarily a proper model of the actual observation and simulated PSFs are typically sharper than real empirical PSFs.
We thus generate a "smoothed" STPSF in each NIRCam filter, to use the noise-free simulated PSF while matching the PSF size to the empirical PSF size.
To this end, we first generate the empirical PSFs from the selected point sources utilizing the \texttt{aperpy} package \citep{Weaver2024ApJS}.
We also generate the on-flight simulated PSFs with the \texttt{STPSF} package, and then we search the optimal kernel size for smoothing to match the STPSF to the empirical PSF. We use a 2D Gaussian kernel for the smoothing.

\begin{figure*}[t]
\centering
\includegraphics[width=0.8\textwidth]{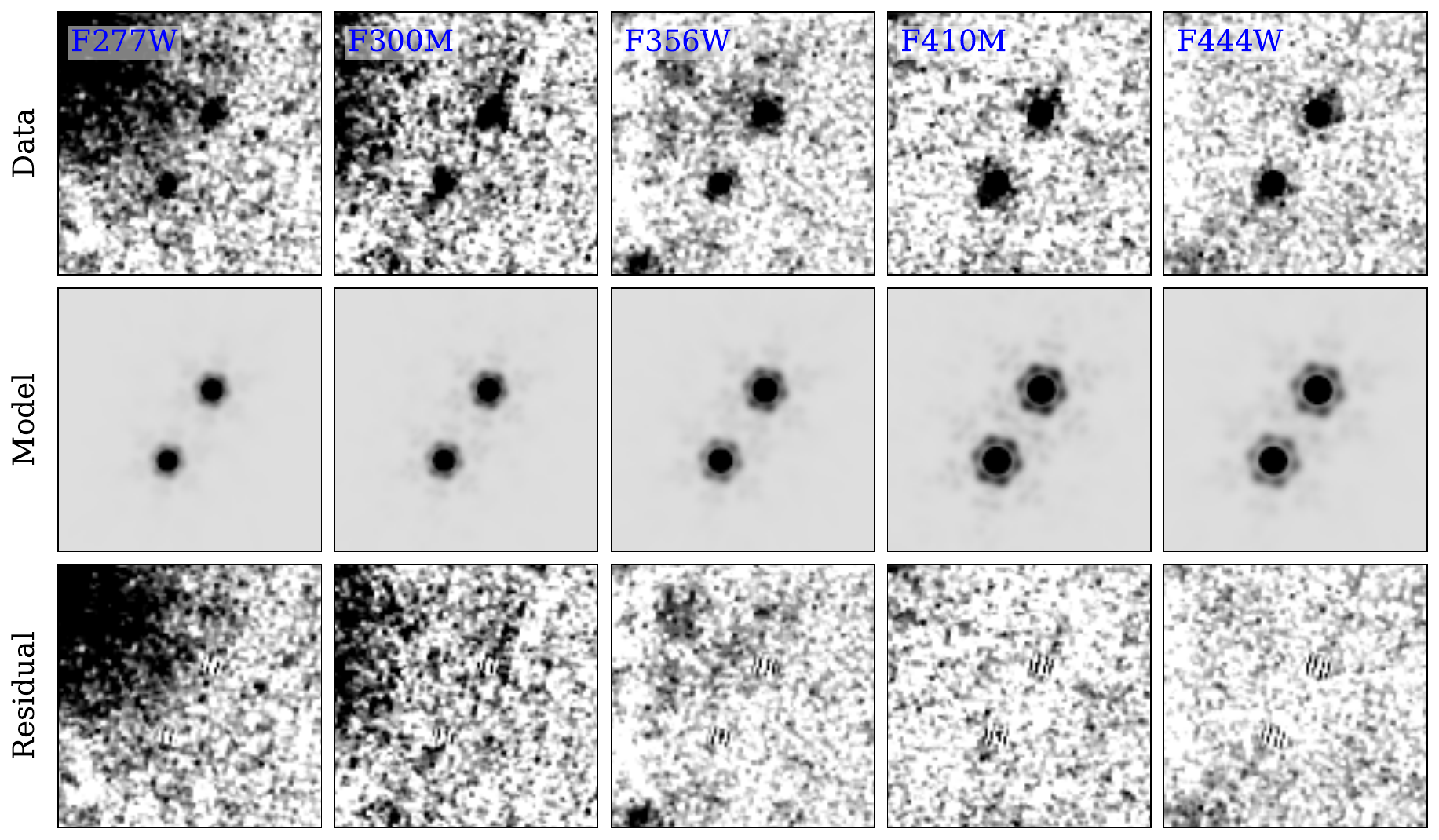}
\caption{Point-source subtractions in NIRCam F277W, F300M, F356W, F410M, and F444W images.
}
\label{fig:pssub}
\end{figure*}

With the PSF built in each NIRCam image, we perform point-source subtraction using the \texttt{Galfit} package \citep{Peng2010AJ}.
Figure~\ref{fig:pssub} shows the result.
The point-source component is well subtracted in the residual images (bottom row), and the elongated arc host component is barely seen in F300M, F410M, and F444W filters, where the strong emission lines from the host fall.
The lensed host galaxy component is clearly seen when these three images are combined (Figure~\ref{fig:fig1} right).

\section{Robustness of the Host-SN decomposition in NIRSpec spectra}\label{apx:NL_validation}

Although we find that the H$\alpha$ line profile of SN Eos can be significantly better fit with the double Gaussian model, the underlying P-Cyg profile of type IIP SNe can have more complex line profile than a simple broad Gaussian.
Given the low spectral resolution of NIRSpec/PRISM, it is not clear if the narrow + broad line decomposition indeed detects the host-origin H$\alpha$ line or it is just due to the complex H$\alpha$ line profile of SN Eos.
To validate this, we perform a statistical test on the robustness of the host-origin narrow H$\alpha$ line detection with an underlying broad P-Cyg profile of type IIP SNe.

\subsection{Fitting configurations}
\label{subsec:}
We first select type IIP SNe spectra in the local universe that have similar spectral features as SN Eos but are known to be free from host contamination, to use them as a reference template of the pure type IIP SN component.
We use SN~1992H spectrum \citep[+103 days after the explosion;][]{Clocchiatti1996AJ} and SN~2015bs \citep[+82 days;][]{Anderson2018NatAs} as references. SN~1992H is a type IIP SN occurred in NGC5377 with a long plateau phase, and the spectrum at +103 days is identified as the best match to the whole SN Eos spectrum in \citet{Coulter2026arXiv}.
SN~2015bs is found in a metal-poor ($\sim4\ \%\ Z_\odot$), ultra-faint ($M_{R}\sim-12$ mag) dwarf galaxy in the local universe, and is one of the lowest-metallicity type IIP SNe reported so far \citep[similar to SN Eos;][]{Coulter2026arXiv} and also has a long plateau phase.
The spectrum of SN~2015bs is found good match to that of the SN Eos as well, particularly around the metallicity-sensitivity absorption lines such as Fe {\sc ii} $\lambda$5169.
We then redshift the reference spectra to $z=5.13$ and degrade them to NIRSpec/PRISM resolution and wavelength sampling using the \texttt{SpectRes} package
\citep{Carnall2017}.

We then fit these spectra over a velocity window of $\pm 25{,}000$~km~s$^{-1}$ centred on the observed H$\alpha$ wavelength. Two models are compared:

$\bullet$ \textit{Pure SN model:} A P-Cygni profile consisting of a broad emission Gaussian, a blue-shifted absorption Gaussian, and a linear underlying continuum. This model only accounts for SN IIP emissions, and it has 8 free parameters in total.

$\bullet$ \textit{SN+host model:} The same pure SN model plus a narrow emission component representing host-galaxy H$\alpha$. The narrow component center and width are both free
parameters, yielding 11 free parameters in total.

We then compare the Pure SN model and SN+host model using the Bayesian Information Criterion,
\begin{equation}
    \mathrm{BIC} = \chi^2 + k \ln N,
\end{equation}
where $k$ is the number of free parameters and $N$ is the number of data points. We define $\Delta\mathrm{BIC} = \mathrm{BIC}_{\rm pureSN} -
\mathrm{BIC}_{\rm SN+host}$, which is positive when the data favour the SN+host model.

\subsection{Injection-Recovery Simulations}
\label{subsec:mc}

To evaluate the false-positive fraction and the detection sensitivity, we perform Monte Carlo (MC)
injection-recovery tests on each reference SN template. A narrow Gaussian
emission line is injected at the host-galaxy systemic velocity with width fixed at $\sigma_{\rm LSF}$, representing an unresolved host H$\alpha$ line. The injection amplitude is set so that the
ratio of the integrated narrow-line flux to the integrated broad H$\alpha$
emission flux equals a target value $f$; we explore
$f = [0,\, 0.02,\, 0.05,\, 0.10,\, 0.20,\, 0.30,\, 0.50]$.
For each SN template and each value of $f$, we generate $N = 1000$
realisations, by adding a Gaussian noise to the template spectrum assuming the same signal-to-noise ratio as our SN~Eos NIRSpec/PRISM observation.
The noise-added spectrum is then fitted with both the pure SN model and SN+host model, and
$\Delta\mathrm{BIC}$ and the recovered narrow-line width ($\hat{\sigma}_n$) are
recorded.
A mock spectrum is classified as a host-galaxy detection only when $\Delta\mathrm{BIC} > 20$ and $\hat{\sigma}_n \leq 2\,\sigma_{\rm LSF}$
%The threshold $\Delta\mathrm{BIC} > 20$ is chosen to limit the false-positive rate (FPR) — the detection fraction at $f = 0$ — to below 5\%, as verified from the $f = 0$ null-distribution MC realisations for each template.
From the injection-recovery simulations we derive, for each SN template,
(1) the false-positive rate (i.e., the detection fraction at $f = 0$),
(2) the detection completeness as a function of $f$, (3) the minimum flux ratio $f^*$ at which the completeness exceeds 95\%, and (4) the recovered-to-injected flux ratio as a measure of systematic bias.

\begin{figure*}[tb]
\centering
    \includegraphics[width=0.45\textwidth]{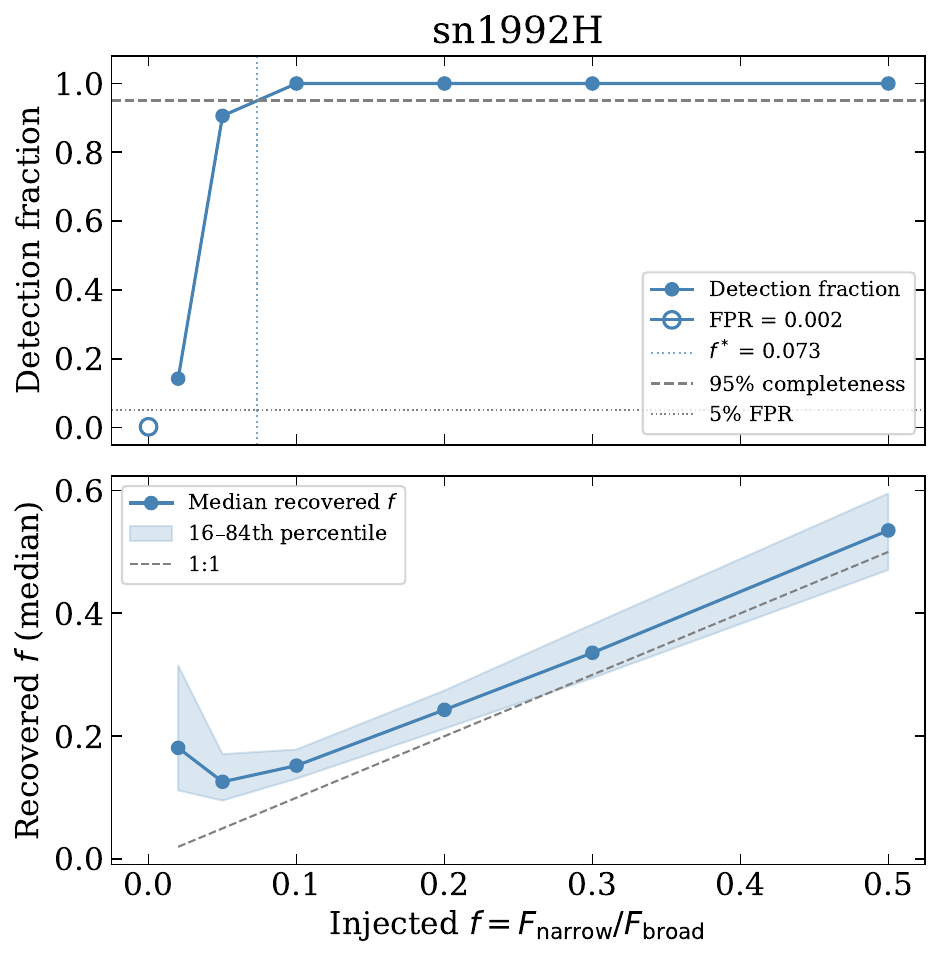}
    \includegraphics[width=0.45\textwidth]{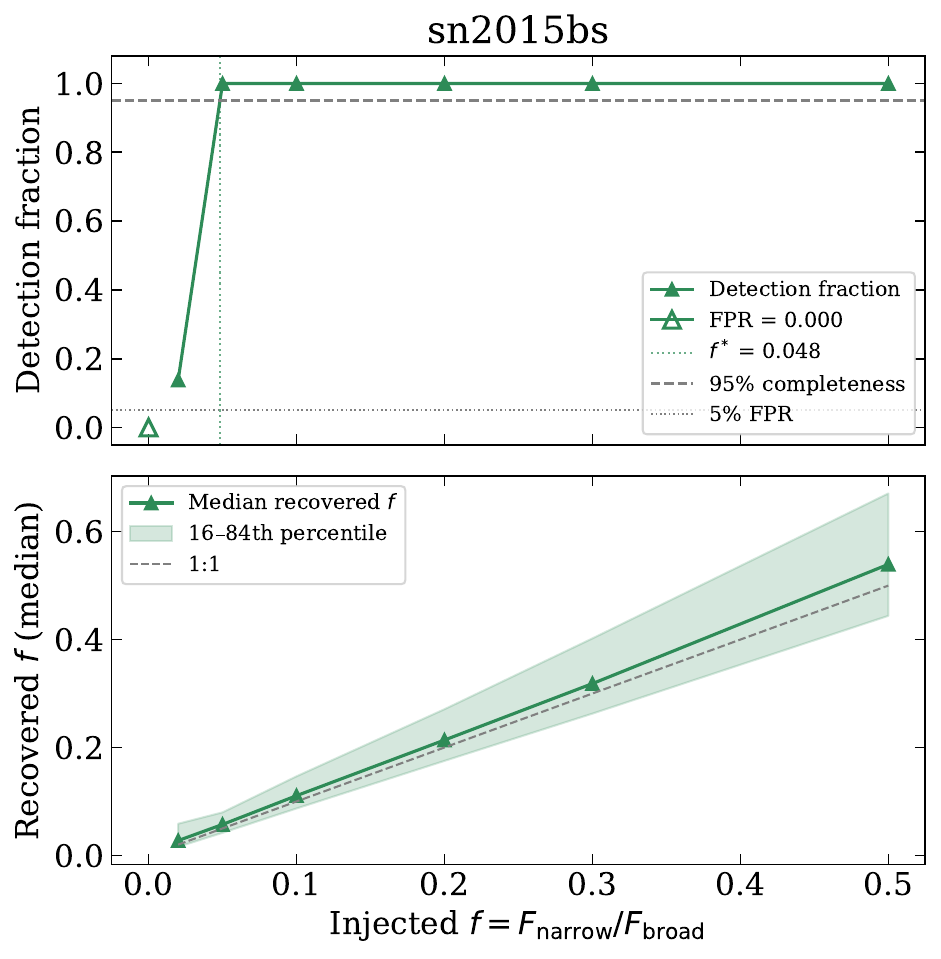}
\vspace{5mm}
\caption{Result of the injection-recovery simulation of the host-origin narrow H$\alpha$ line blended with late plateau-phase type IIP SN spectra. Results with SN~1992H and SN~2015bs spectra are displayed on the left and right, respectively.
{\it Top}: the recovery fraction of host-origin narrow line as the function of the injected narrow H$\alpha$ line flux. $N=1000$ mock spectra are generated at each grid of the injected narrow line fluxes, and the recovery fraction is computed by performing the host-origin H$\alpha$ identification based on $\Delta {\rm BIC}$ and the narrow component line width (see text). The recovery fraction at $f=0$ grid (no injection) corresponds to the false-positive rate.
{\it Bottom}: recovered narrow H$\alpha$ flux, compared against the injected line flux (both normalized with the underlying broad H$\alpha$ from the SN spectrum). The dashed line denotes the 1:1 relation. Data points with the shaded range shows the median and the 16-84th percentile of the recovered narrow line fluxes.}
\label{fig:injection_recovery}
\end{figure*}

Figure \ref{fig:injection_recovery} shows the result of each case of SN~1992H and SN~2015bs. It shows that the false positive fraction is $0.2\ \%$ with the SN~1992H template and $<0.1$ \% with the SN~2015bs template (i.e., no MCMC realization at the $f=0$ grid is identified as host-galaxy detection), while the narrow-line H$\alpha$ flux originated from the host can be well recovered at $F_{\rm narrow}/F_{\rm broad}>0.05$.
This indicates that the host-origin narrow H$\alpha$ detection criteria is rather conservative, and we can robustly measure the host H$\alpha$ line flux even with the NIRSpec/PRISM resolution when $F_{\rm narrow}/F_{\rm broad}\gtrsim0.1$, assuming that the H$\alpha$ line profile of the SN component is resemble to the later plateau phase of SN~1992H or SN~2015bs.
In case of the SN \eos\ spectra, all image 1, image 2, and the composite spectrum meet the host-H$\alpha$ detection criteria ($\Delta\mathrm{BIC}=88.4$, $26.7$, and $123.7$, respectively, and the narrow H$\alpha$ component are all $<2\sigma_{\rm LSF}$). This suggests the narrow component found in the SN \eos\ spectra is the host-origin H$\alpha$ emission line.
%The narrow-to-broad H$\alpha$ line flux ratios are $f=0.11$, $0.08$, and $0.10$, respectively.

%\iffalse
\section{Computing the host $M_{\rm UV}$ probability distribution}\label{apx:compute_prob}
The nominal computation of the CCSN-finding probability is considered extremely complicated and requires the full survey areas of all lensing clusters we have searched for, each with proper lens modeling.
Instead, we here focus on the host $M_{\rm UV}$ probability distribution given one CCSN is found.
Since the CCSN itself is bright enough to be detected with the NIRCam sensitivity without lensing, the host $M_{\rm UV}$ probability distribution should be computed as the number of detectable host galaxy in each $M_{\rm UV}$ bin weighted by their CCSN number rate.

The CCSN rate of a galaxy can be computed from SFR assuming the IMF and the explodable stellar mass range. Under a canonical assumption of the universal IMF and explodability, the conversion from SFR to the CCSN-rate becomes a constant factor.
Therefore, when a universal SN rate is adopted, the probability distribution can be computed as $P_{\rm host}(M_{\rm UV})\propto\phi_{\rm LF}\cdot {\rm SFR}\cdot S_{\rm eff}$.
We use the gravitational lens model for the MACS1931 cluster by \citet{Allingham2026arXiv} to compute the source-plane survey area at a given host $M_{\rm UV}$, assuming the nominal detection limit (in the image plane) is $-18.4$ mag. This corresponds to the VENUS survey depth at $z=5$. The result is shown by black in Figure \ref{fig:SN_finding}.

In contrast, when the CCSN rate per unit SFR is not constant but rather dependent on the metallicity, $P_{\rm host}(M_{\rm UV})$ needs to involve the additional correction factor to account for the metallicity-dependency.
We first obtain an empirical $M_{\rm UV}$-$Z$ relation by doing a simple regression on $z\sim5-8$ galaxies observed with JWST \citep{Nakajima2023ApJS,Chemerynska2024ApJ,Asada2026arXiv}.
Based on the empirical $M_{\rm UV}$-$Z$ relation, we estimate the typical metallicity at a given $M_{\rm UV}$, compute the metallicity-dependent SN rate per unit SFR $\nu_{\rm SN}(M_{\rm UV})$, and add this factor to the $P_{\rm host}(M_{\rm UV})$ computation.

We also take account for the subpopulation of extremely metal-poor galaxies (see Figure \ref{fig:context} left).
Although rare, small number of faint-end galaxies show significantly lower metallicity ($<1\ \%$ solar) than other faint galaxies. 
The fraction of these etremely-metal poor galaxies at a given $M_{\rm UV}$ is largely unknown at present, and thus we simply assume that the fraction monotonically increase from 0 at $M_{\rm UV}=-17$ mag to 1 at $M_{\rm UV}=-12$ mag, and fixed the metallicity of these galaxies at $10^{-2.5}\ Z_\odot$.
This effectively assigns a ``metallicity floor'' of $10^{-2.5}\ Z_\odot$ at the faint-end of the $M_{\rm UV}$-$Z$ relation.
The resulting $P_{\rm host}(M_{\rm UV})$ with the metallicity-dependent SN rates are shown by blue and yellow in Figure \ref{fig:SN_finding}.
%\fi

%% For this sample we use BibTeX plus aasjournalv7.bst to generate the

\bibliography{sample701}{}

@ARTICLE{Jaacks2018MNRAS,
       author = {{Jaacks}, Jason and {Thompson}, Robert and {Finkelstein}, Steven L. and {Bromm}, Volker},
        title = "{Baseline metal enrichment from Population III star formation in cosmological volume simulations}",
      journal = {\mnras},
         year = 2018,
        month = apr,
       volume = {475},
       number = {4},
        pages = {4396-4410},
          doi = {10.1093/mnras/sty062},
archivePrefix = {arXiv},
       eprint = {1705.08059},
 primaryClass = {astro-ph.GA},
       adsurl = {https://ui.adsabs.harvard.edu/abs/2018MNRAS.475.4396J}
}

@ARTICLE{Smith2019MNRAS,
       author = {{Smith}, Aaron and {Ma}, Xiangcheng and {Bromm}, Volker and {Finkelstein}, Steven L. and {Hopkins}, Philip F. and {Faucher-Gigu{\`e}re}, Claude-Andr{\'e} and {Kere{\v{s}}}, Du{\v{s}}an},
        title = "{The physics of Lyman {\ensuremath{\alpha}} escape from high-redshift galaxies}",
      journal = {\mnras},
         year = 2019,
        month = mar,
       volume = {484},
       number = {1},
        pages = {39-59},
          doi = {10.1093/mnras/sty3483},
archivePrefix = {arXiv},
       eprint = {1810.08185},
 primaryClass = {astro-ph.GA},
       adsurl = {https://ui.adsabs.harvard.edu/abs/2019MNRAS.484...39S}
}

@ARTICLE{Karlsson2013RvMP,
       author = {{Karlsson}, Torgny and {Bromm}, Volker and {Bland-Hawthorn}, Joss},
        title = "{Pregalactic metal enrichment: The chemical signatures of the first stars}",
      journal = {Reviews of Modern Physics},
         year = 2013,
        month = apr,
       volume = {85},
       number = {2},
        pages = {809-848},
          doi = {10.1103/RevModPhys.85.809},
archivePrefix = {arXiv},
       eprint = {1101.4024},
 primaryClass = {astro-ph.CO},
       adsurl = {https://ui.adsabs.harvard.edu/abs/2013RvMP...85..809K}
}

@ARTICLE{Dayal2018PhR,
       author = {{Dayal}, Pratika and {Ferrara}, Andrea},
        title = "{Early galaxy formation and its large-scale effects}",
      journal = {\physrep},
         year = 2018,
        month = dec,
       volume = {780},
        pages = {1-64},
          doi = {10.1016/j.physrep.2018.10.002},
archivePrefix = {arXiv},
       eprint = {1809.09136},
 primaryClass = {astro-ph.GA},
       adsurl = {https://ui.adsabs.harvard.edu/abs/2018PhR...780....1D}
}

@ARTICLE{Bromm2011ARAA,
       author = {{Bromm}, Volker and {Yoshida}, Naoki},
        title = "{The First Galaxies}",
      journal = {\araa},
         year = 2011,
        month = sep,
       volume = {49},
       number = {1},
        pages = {373-407},
          doi = {10.1146/annurev-astro-081710-102608},
archivePrefix = {arXiv},
       eprint = {1102.4638},
 primaryClass = {astro-ph.CO},
       adsurl = {https://ui.adsabs.harvard.edu/abs/2011ARA&A..49..373B}
}

@ARTICLE{Kremer2020ApJ,
       author = {{Kremer}, Kyle and {Spera}, Mario and {Becker}, Devin and {Chatterjee}, Sourav and {Di Carlo}, Ugo N. and {Fragione}, Giacomo and {Rodriguez}, Carl L. and {Ye}, Claire S. and {Rasio}, Frederic A.},
        title = "{Populating the Upper Black Hole Mass Gap through Stellar Collisions in Young Star Clusters}",
      journal = {\apj},
         year = 2020,
        month = nov,
       volume = {903},
       number = {1},
          eid = {45},
        pages = {45},
          doi = {10.3847/1538-4357/abb945},
archivePrefix = {arXiv},
       eprint = {2006.10771},
 primaryClass = {astro-ph.HE},
       adsurl = {https://ui.adsabs.harvard.edu/abs/2020ApJ...903...45K}
}

@ARTICLE{Freitag2006MNRAS,
       author = {{Freitag}, Marc and {G{\"u}rkan}, M. Atakan and {Rasio}, Frederic A.},
        title = "{Runaway collisions in young star clusters - II. Numerical results}",
      journal = {\mnras},
         year = 2006,
        month = may,
       volume = {368},
       number = {1},
        pages = {141-161},
          doi = {10.1111/j.1365-2966.2006.10096.x},
archivePrefix = {arXiv},
       eprint = {astro-ph/0503130},
 primaryClass = {astro-ph},
       adsurl = {https://ui.adsabs.harvard.edu/abs/2006MNRAS.368..141F}
}

@ARTICLE{Gurkan2006ApJ,
       author = {{G{\"u}rkan}, M. Atakan and {Fregeau}, John M. and {Rasio}, Frederic A.},
        title = "{Massive Black Hole Binaries from Collisional Runaways}",
      journal = {\apjl},
         year = 2006,
        month = mar,
       volume = {640},
       number = {1},
        pages = {L39-L42},
          doi = {10.1086/503295},
archivePrefix = {arXiv},
       eprint = {astro-ph/0512642},
 primaryClass = {astro-ph},
       adsurl = {https://ui.adsabs.harvard.edu/abs/2006ApJ...640L..39G}
}

@ARTICLE{PortegiesZwart2004Natur,
       author = {{Portegies Zwart}, Simon F. and {Baumgardt}, Holger and {Hut}, Piet and {Makino}, Junichiro and {McMillan}, Stephen L.~W.},
        title = "{Formation of massive black holes through runaway collisions in dense young star clusters}",
      journal = {\nat},
         year = 2004,
        month = apr,
       volume = {428},
       number = {6984},
        pages = {724-726},
          doi = {10.1038/nature02448},
archivePrefix = {arXiv},
       eprint = {astro-ph/0402622},
 primaryClass = {astro-ph},
       adsurl = {https://ui.adsabs.harvard.edu/abs/2004Natur.428..724P}
}

@ARTICLE{Giersz2015MNRAS,
       author = {{Giersz}, Mirek and {Leigh}, Nathan and {Hypki}, Arkadiusz and {L{\"u}tzgendorf}, Nora and {Askar}, Abbas},
        title = "{MOCCA code for star cluster simulations - IV. A new scenario for intermediate mass black hole formation in globular clusters}",
      journal = {\mnras},
         year = 2015,
        month = dec,
       volume = {454},
       number = {3},
        pages = {3150-3165},
          doi = {10.1093/mnras/stv2162},
archivePrefix = {arXiv},
       eprint = {1506.05234},
 primaryClass = {astro-ph.GA},
       adsurl = {https://ui.adsabs.harvard.edu/abs/2015MNRAS.454.3150G}
}

@ARTICLE{Claeyssens2026arXiv,
       author = {{Claeyssens}, Ad{\'e}la{\"\i}de and {Adamo}, Angela and {Kokorev}, Vasily and {Furtak}, Lukas and {Richard}, Johan and {Beauchesne}, Benjamin and {Dessauges-Zavadsky}, Miroslava and {Atek}, Hakim and {Chisholm}, John and {Endsley}, Ryan and {Fujimoto}, Seiji and {Korber}, Damien and {Pan}, Richard and {Saldana-Lopez}, Alberto and {Schaerer}, Daniel},
        title = "{A first GLIMPSE into star clusters populations across cosmic time}",
      journal = {arXiv e-prints},
         year = 2026,
        month = jan,
          eid = {arXiv:2601.16281},
        pages = {arXiv:2601.16281},
          doi = {10.48550/arXiv.2601.16281},
archivePrefix = {arXiv},
       eprint = {2601.16281},
 primaryClass = {astro-ph.GA},
       adsurl = {https://ui.adsabs.harvard.edu/abs/2026arXiv260116281C}
}

@ARTICLE{Adamo2024Natur,
       author = {{Adamo}, Angela and {Bradley}, Larry D. and {Vanzella}, Eros and {Claeyssens}, Ad{\'e}la{\"\i}de and {Welch}, Brian and {Diego}, Jose M. and {Mahler}, Guillaume and {Oguri}, Masamune and {Sharon}, Keren and {Abdurro'uf} and {Hsiao}, Tiger Yu-Yang and {Xu}, Xinfeng and {Messa}, Matteo and {Lassen}, Augusto E. and {Zackrisson}, Erik and {Brammer}, Gabriel and {Coe}, Dan and {Kokorev}, Vasily and {Ricotti}, Massimo and {Zitrin}, Adi and {Fujimoto}, Seiji and {Inoue}, Akio K. and {Resseguier}, Tom and {Rigby}, Jane R. and {Jim{\'e}nez-Teja}, Yolanda and {Windhorst}, Rogier A. and {Hashimoto}, Takuya and {Tamura}, Yoichi},
        title = "{Bound star clusters observed in a lensed galaxy 460 Myr after the Big Bang}",
      journal = {\nat},
         year = 2024,
        month = aug,
       volume = {632},
       number = {8025},
        pages = {513-516},
          doi = {10.1038/s41586-024-07703-7},
archivePrefix = {arXiv},
       eprint = {2401.03224},
 primaryClass = {astro-ph.GA},
       adsurl = {https://ui.adsabs.harvard.edu/abs/2024Natur.632..513A}
}

@ARTICLE{Mowla2024Natur,
       author = {{Mowla}, Lamiya and {Iyer}, Kartheik and {Asada}, Yoshihisa and {Desprez}, Guillaume and {Tan}, Vivian Yun Yan and {Martis}, Nicholas and {Sarrouh}, Ghassan and {Strait}, Victoria and {Abraham}, Roberto and {Brada{\v{c}}}, Maru{\v{s}}a and {Brammer}, Gabriel and {Muzzin}, Adam and {Pacifici}, Camilla and {Ravindranath}, Swara and {Sawicki}, Marcin and {Willott}, Chris and {Estrada-Carpenter}, Vince and {Jahan}, Nusrath and {Noirot}, Ga{\"e}l and {Matharu}, Jasleen and {Rihtar{\v{s}}i{\v{c}}}, Gregor and {Zabl}, Johannes},
        title = "{Formation of a low-mass galaxy from star clusters in a 600-million-year-old Universe}",
      journal = {\nat},
         year = 2024,
        month = dec,
       volume = {636},
       number = {8042},
        pages = {332-336},
          doi = {10.1038/s41586-024-08293-0},
archivePrefix = {arXiv},
       eprint = {2402.08696},
 primaryClass = {astro-ph.GA},
       adsurl = {https://ui.adsabs.harvard.edu/abs/2024Natur.636..332M}
}

@ARTICLE{Tee2025TNSTR,
       author = {{Tee}, W.~L.},
        title = "{TSST\_COSMOS\_3D Transient Discovery Report for 2025-07-14}",
      journal = {Transient Name Server Discovery Report},
         year = 2025,
        month = jul,
       volume = {2025-2691},
        pages = {1},
       adsurl = {https://ui.adsabs.harvard.edu/abs/2025TNSTR2691....1T}
}

@ARTICLE{Markov2026AAP,
       author = {{Markov}, V. and {Brada{\v{c}}}, M. and {Estrada-Carpenter}, V. and {Desprez}, G. and {Rihtar{\v{s}}i{\v{c}}}, G. and {Jude{\v{z}}}, J. and {Tripodi}, R. and {Sawicki}, M. and {Noirot}, G. and {Martis}, N. and {Willott}, C. and {Abraham}, R. and {Asada}, Y. and {Brammer}, G. and {Matharu}, J. and {Muzzin}, A. and {Sarrouh}, G.~T.~E. and {Withers}, S. and {Ferrara}, A. and {Fujimoto}, S. and {Gallerani}, S. and {Goovaerts}, I. and {Harshan}, A.},
        title = "{Resolving dust and Lyman {\ensuremath{\alpha}} emission in a lensed galaxy at the epoch of reionization with JWST/CANUCS}",
      journal = {\aap},
         year = 2026,
        month = apr,
       volume = {708},
          eid = {A236},
        pages = {A236},
          doi = {10.1051/0004-6361/202558580},
archivePrefix = {arXiv},
       eprint = {2512.13778},
 primaryClass = {astro-ph.GA},
       adsurl = {https://ui.adsabs.harvard.edu/abs/2026A&A...708A.236M}
}

@ARTICLE{Vanzella2023AAP,
       author = {{Vanzella}, E. and {Loiacono}, F. and {Bergamini}, P. and {Me{\v{s}}tri{\'c}}, U. and {Castellano}, M. and {Rosati}, P. and {Meneghetti}, M. and {Grillo}, C. and {Calura}, F. and {Mignoli}, M. and {Brada{\v{c}}}, M. and {Adamo}, A. and {Rihtar{\v{s}}i{\v{c}}}, G. and {Dickinson}, M. and {Gronke}, M. and {Zanella}, A. and {Annibali}, F. and {Willott}, C. and {Messa}, M. and {Sani}, E. and {Acebron}, A. and {Bolamperti}, A. and {Comastri}, A. and {Gilli}, R. and {Caputi}, K.~I. and {Ricotti}, M. and {Gruppioni}, C. and {Ravindranath}, S. and {Mercurio}, A. and {Strait}, V. and {Martis}, N. and {Pascale}, R. and {Caminha}, G.~B. and {Annunziatella}, M. and {Nonino}, M.},
        title = "{An extremely metal-poor star complex in the reionization era: Approaching Population III stars with JWST}",
      journal = {\aap},
         year = 2023,
        month = oct,
       volume = {678},
          eid = {A173},
        pages = {A173},
          doi = {10.1051/0004-6361/202346981},
archivePrefix = {arXiv},
       eprint = {2305.14413},
 primaryClass = {astro-ph.GA},
       adsurl = {https://ui.adsabs.harvard.edu/abs/2023A&A...678A.173V}
}

@ARTICLE{Maltsev2025AAP,
       author = {{Maltsev}, K. and {Schneider}, F.~R.~N. and {Mandel}, I. and {M{\"u}ller}, B. and {Heger}, A. and {R{\"o}pke}, F.~K. and {Laplace}, E.},
        title = "{Explodability criteria for the neutrino-driven supernova mechanism}",
      journal = {\aap},
         year = 2025,
        month = aug,
       volume = {700},
          eid = {A20},
        pages = {A20},
          doi = {10.1051/0004-6361/202554931},
archivePrefix = {arXiv},
       eprint = {2503.23856},
 primaryClass = {astro-ph.SR},
       adsurl = {https://ui.adsabs.harvard.edu/abs/2025A&A...700A..20M}
}

@ARTICLE{Zitrin2026arXiv,
       author = {{Zitrin}, Adi},
        title = "{Strong Gravitational Lensing with the James Webb Space Telescope}",
      journal = {arXiv e-prints},
         year = 2026,
        month = may,
          eid = {arXiv:2605.15189},
        pages = {arXiv:2605.15189},
          doi = {10.48550/arXiv.2605.15189},
archivePrefix = {arXiv},
       eprint = {2605.15189},
 primaryClass = {astro-ph.CO},
       adsurl = {https://ui.adsabs.harvard.edu/abs/2026arXiv260515189Z}
}

@ARTICLE{Rosati2014Msngr,
       author = {{Rosati}, P. and {Balestra}, I. and {Grillo}, C. and {Mercurio}, A. and {Nonino}, M. and {Biviano}, A. and {Girardi}, M. and {Vanzella}, E. and {Clash-VLT Team}},
        title = "{CLASH-VLT: A VIMOS Large Programme to Map the Dark Matter Mass Distribution in Galaxy Clusters and Probe Distant Lensed Galaxies}",
      journal = {The Messenger},
         year = 2014,
        month = dec,
       volume = {158},
        pages = {48-53},
       adsurl = {https://ui.adsabs.harvard.edu/abs/2014Msngr.158...48R}
}

@software{Brammer2022zndo,
       author = {{Brammer}, Gabe},
        title = "{gbrammer/msaexp: Full working version with 2d drizzling and extraction}",
         year = 2022,
        month = nov,
          eid = {10.5281/zenodo.7299501},
          doi = {10.5281/zenodo.7299501},
      version = {0.3},
    publisher = {Zenodo},
       adsurl = {https://ui.adsabs.harvard.edu/abs/2022zndo...7299501B}
}

@software{photutils1.12.0,
       author = {{Bradley}, Larry and {Sip{\H{o}}cz}, Brigitta and {Robitaille}, Thomas and {Tollerud}, Erik and {Vin{\'\i}cius}, Z{\'e} and {Deil}, Christoph and {Barbary}, Kyle and {Wilson}, Tom J and {Busko}, Ivo and {Donath}, Axel and {G{\"u}nther}, Hans Moritz and {Cara}, Mihai and {Lim}, P.~L. and {Me{\ss}linger}, Sebastian and {Burnett}, Zach and {Conseil}, Simon and {Droettboom}, Michael and {Bostroem}, Azalee and {Bray}, E.~M. and {Andersen Bratholm}, Lars and {Jamieson}, William and {Ginsburg}, Adam and {Barentsen}, Geert and {Craig}, Matt and {Pascual}, Sergio and {Rathi}, Shivangee and {Perrin}, Marshall and {Morris}, Brett M. and {Perren}, Gabriel},
        title = "{astropy/photutils: 1.12.0}",
         year = 2024,
        month = apr,
          eid = {10.5281/zenodo.10967176},
          doi = {10.5281/zenodo.10967176},
      version = {1.12.0},
    publisher = {Zenodo},
       adsurl = {https://ui.adsabs.harvard.edu/abs/2024zndo..10967176B}
}

@ARTICLE{Shahbandeh2025ApJ,
       author = {{Shahbandeh}, Melissa and {Fox}, Ori D. and {Temim}, Tea and {Dwek}, Eli and {Sarangi}, Arkaprabha and {Smith}, Nathan and {Dessart}, Luc and {Nickson}, Bryony and {Engesser}, Michael and {Filippenko}, Alexei V. and {Brink}, Thomas G. and {Zheng}, WeiKang and {Szalai}, Tam{\'a}s and {Johansson}, Joel and {Rest}, Armin and {Van Dyk}, Schuyler D. and {Andrews}, Jennifer and {Ashall}, Chris and {Clayton}, Geoffrey C. and {De Looze}, Ilse and {DerKacy}, James M. and {Dulude}, Michael and {Foley}, Ryan J. and {Gezari}, Suvi and {Gomez}, Sebastian and {Gonzaga}, Shireen and {Indukuri}, Siva and {Jencson}, Jacob and {Kasliwal}, Mansi and {Lane}, Zachary G. and {Lau}, Ryan and {Law}, David and {Marston}, Anthony and {Milisavljevic}, Dan and {O'Steen}, Richard and {Pierel}, Justin and {Siebert}, Matthew and {Skrutskie}, Michael and {Strolger}, Lou and {Tinyanont}, Samaporn and {Wang}, Qinan and {Williams}, Brian and {Xiao}, Lin and {Yang}, Yi and {Zs{\'\i}ros}, Szanna},
        title = "{JWST/MIRI Observations of Newly Formed Dust in the Cold, Dense Shell of the Type IIn SN 2005ip}",
      journal = {\apj},
         year = 2025,
        month = jun,
       volume = {985},
       number = {2},
          eid = {262},
        pages = {262},
          doi = {10.3847/1538-4357/adce77},
archivePrefix = {arXiv},
       eprint = {2410.09142},
 primaryClass = {astro-ph.HE},
       adsurl = {https://ui.adsabs.harvard.edu/abs/2025ApJ...985..262S}
}

@ARTICLE{Bouwens2022ApJ,
       author = {{Bouwens}, R.~J. and {Illingworth}, G. and {Ellis}, R.~S. and {Oesch}, P. and {Stefanon}, M.},
        title = "{z   2-9 Galaxies Magnified by the Hubble Frontier Field Clusters. II. Luminosity Functions and Constraints on a Faint-end Turnover}",
      journal = {\apj},
         year = 2022,
        month = nov,
       volume = {940},
       number = {1},
          eid = {55},
        pages = {55},
          doi = {10.3847/1538-4357/ac86d1},
archivePrefix = {arXiv},
       eprint = {2205.11526},
 primaryClass = {astro-ph.GA},
       adsurl = {https://ui.adsabs.harvard.edu/abs/2022ApJ...940...55B}
}

@ARTICLE{Ibeling2013ApJ,
       author = {{Ibeling}, Duligur and {Heger}, Alexander},
        title = "{The Metallicity Dependence of the Minimum Mass for Core-collapse Supernovae}",
      journal = {\apjl},
         year = 2013,
        month = mar,
       volume = {765},
       number = {2},
          eid = {L43},
        pages = {L43},
          doi = {10.1088/2041-8205/765/2/L43},
archivePrefix = {arXiv},
       eprint = {1301.5783},
 primaryClass = {astro-ph.SR},
       adsurl = {https://ui.adsabs.harvard.edu/abs/2013ApJ...765L..43I}
}

@ARTICLE{Ono2026arXiv,
       author = {{Ono}, Sojun and {Maeda}, Keiichi and {Suzuki}, Akihiro},
        title = "{Constraints on the Metallicity-dependent Explodability of Massive Stars from Galactic Chemical Evolution: Toward Alleviating the Red Supergiant Problem}",
      journal = {arXiv e-prints},
         year = 2026,
        month = may,
          eid = {arXiv:2605.15462},
        pages = {arXiv:2605.15462},
          doi = {10.48550/arXiv.2605.15462},
archivePrefix = {arXiv},
       eprint = {2605.15462},
 primaryClass = {astro-ph.HE},
       adsurl = {https://ui.adsabs.harvard.edu/abs/2026arXiv260515462O}
}

@ARTICLE{Chon2024MNRAS,
       author = {{Chon}, Sunmyon and {Hosokawa}, Takashi and {Omukai}, Kazuyuki and {Schneider}, Raffaella},
        title = "{Impact of radiative feedback on the initial mass function of metal-poor stars}",
      journal = {\mnras},
         year = 2024,
        month = may,
       volume = {530},
       number = {3},
        pages = {2453-2474},
          doi = {10.1093/mnras/stae1027},
archivePrefix = {arXiv},
       eprint = {2312.13339},
 primaryClass = {astro-ph.GA},
       adsurl = {https://ui.adsabs.harvard.edu/abs/2024MNRAS.530.2453C}
}

@ARTICLE{Mauerhofer2025AAP,
       author = {{Mauerhofer}, Valentin and {Dayal}, Pratika and {Haehnelt}, Martin G. and {Kimm}, Taysun and {Rosdahl}, Joakim and {Teyssier}, Romain},
        title = "{Synergising semi-analytical models and hydrodynamical simulations to interpret JWST data from the first billion years}",
      journal = {\aap},
         year = 2025,
        month = apr,
       volume = {696},
          eid = {A157},
        pages = {A157},
          doi = {10.1051/0004-6361/202554042},
archivePrefix = {arXiv},
       eprint = {2502.02647},
 primaryClass = {astro-ph.GA},
       adsurl = {https://ui.adsabs.harvard.edu/abs/2025A&A...696A.157M}
}

@ARTICLE{Pessi2023ApJ,
       author = {{Pessi}, Thallis and {Anderson}, Joseph P. and {Lyman}, Joseph D. and {Prieto}, Jose L. and {Galbany}, Llu{\'\i}s and {Kochanek}, Christopher S. and {S{\'a}nchez}, Sebastian F. and {Kuncarayakti}, Hanindyo},
        title = "{A Metallicity Dependence on the Occurrence of Core-collapse Supernovae}",
      journal = {\apjl},
         year = 2023,
        month = oct,
       volume = {955},
       number = {2},
          eid = {L29},
        pages = {L29},
          doi = {10.3847/2041-8213/acf7c6},
archivePrefix = {arXiv},
       eprint = {2306.11962},
 primaryClass = {astro-ph.SR},
       adsurl = {https://ui.adsabs.harvard.edu/abs/2023ApJ...955L..29P}
}

@ARTICLE{Hillier2019AAP,
       author = {{Hillier}, Desmond John and {Dessart}, Luc},
        title = "{Photometric and spectroscopic diversity of Type II supernovae}",
      journal = {\aap},
         year = 2019,
        month = nov,
       volume = {631},
          eid = {A8},
        pages = {A8},
          doi = {10.1051/0004-6361/201935100},
archivePrefix = {arXiv},
       eprint = {1908.02973},
 primaryClass = {astro-ph.SR},
       adsurl = {https://ui.adsabs.harvard.edu/abs/2019A&A...631A...8H}
}

@ARTICLE{Faran2018MNRAS,
       author = {{Faran}, T. and {Nakar}, E. and {Poznanski}, D.},
        title = "{The evolution of temperature and bolometric luminosity in Type II supernovae}",
      journal = {\mnras},
         year = 2018,
        month = jan,
       volume = {473},
       number = {1},
        pages = {513-537},
          doi = {10.1093/mnras/stx2288},
archivePrefix = {arXiv},
       eprint = {1707.07695},
 primaryClass = {astro-ph.HE},
       adsurl = {https://ui.adsabs.harvard.edu/abs/2018MNRAS.473..513F}
}

@ARTICLE{Dastidar2018MNRAS,
       author = {{Dastidar}, Raya and {Misra}, Kuntal and {Hosseinzadeh}, G. and {Pastorello}, A. and {Pumo}, M.~L. and {Valenti}, S. and {McCully}, C. and {Tomasella}, L. and {Arcavi}, I. and {Elias-Rosa}, N. and {Singh}, Mridweeka and {Gangopadhyay}, Anjasha and {Howell}, D.~A. and {Morales-Garoffolo}, Antonia and {Zampieri}, L. and {Kumar}, Brijesh and {Turatto}, M. and {Benetti}, S. and {Tartaglia}, L. and {Ochner}, P. and {Sahu}, D.~K. and {Anupama}, G.~C. and {Pandey}, S.~B.},
        title = "{SN 2015ba: a Type IIP supernova with a long plateau}",
      journal = {\mnras},
         year = 2018,
        month = sep,
       volume = {479},
       number = {2},
        pages = {2421-2442},
          doi = {10.1093/mnras/sty1634},
archivePrefix = {arXiv},
       eprint = {1806.05470},
 primaryClass = {astro-ph.HE},
       adsurl = {https://ui.adsabs.harvard.edu/abs/2018MNRAS.479.2421D}
}

@ARTICLE{Clocchiatti1996AJ,
       author = {{Clocchiatti}, Alejandro and {Benetti}, Stefano and {Wheeler}, J. Craig and {Wren}, William and {Boisseau}, J. and {Cappellaro}, Enrico and {Turatto}, Massimo and {Patat}, Ferdinando and {Swartz}, Douglas A. and {Harkness}, Robert P. and {Brotherton}, Michael S. and {Wills}, Beverly and {Hemenway}, Paul and {Cornell}, Mark and {Frueh}, Marian and {Kaiser}, Mary B.},
        title = "{A Study of SN 1992H in NGC 5377}",
      journal = {\aj},
         year = 1996,
        month = mar,
       volume = {111},
        pages = {1286},
          doi = {10.1086/117874},
       adsurl = {https://ui.adsabs.harvard.edu/abs/1996AJ....111.1286C}
}

@ARTICLE{Anderson2018NatAs,
       author = {{Anderson}, J.~P. and {Dessart}, L. and {Guti{\'e}rrez}, C.~P. and {Kr{\"u}hler}, T. and {Galbany}, L. and {Jerkstrand}, A. and {Smartt}, S.~J. and {Contreras}, C. and {Morrell}, N. and {Phillips}, M.~M. and {Stritzinger}, M.~D. and {Hsiao}, E.~Y. and {Gonz{\'a}lez-Gait{\'a}n}, S. and {Agliozzo}, C. and {Castell{\'o}n}, S. and {Chambers}, K.~C. and {Chen}, T.-W. and {Flewelling}, H. and {Gonzalez}, C. and {Hosseinzadeh}, G. and {Huber}, M. and {Fraser}, M. and {Inserra}, C. and {Kankare}, E. and {Mattila}, S. and {Magnier}, E. and {Maguire}, K. and {Lowe}, T.~B. and {Sollerman}, J. and {Sullivan}, M. and {Young}, D.~R. and {Valenti}, S.},
        title = "{The lowest-metallicity type II supernova from the highest-mass red supergiant progenitor}",
      journal = {Nature Astronomy},
         year = 2018,
        month = may,
       volume = {2},
        pages = {574-579},
          doi = {10.1038/s41550-018-0458-4},
archivePrefix = {arXiv},
       eprint = {1805.04434},
 primaryClass = {astro-ph.HE},
       adsurl = {https://ui.adsabs.harvard.edu/abs/2018NatAs...2..574A}
}

@ARTICLE{Maseda2018ApJ,
       author = {{Maseda}, Michael V. and {Bacon}, Roland and {Franx}, Marijn and {Brinchmann}, Jarle and {Schaye}, Joop and {Boogaard}, Leindert A. and {Bouch{\'e}}, Nicolas and {Bouwens}, Rychard J. and {Cantalupo}, Sebastiano and {Contini}, Thierry and {Hashimoto}, Takuya and {Inami}, Hanae and {Marino}, Raffaella A. and {Muzahid}, Sowgat and {Nanayakkara}, Themiya and {Richard}, Johan and {Schmidt}, Kasper B. and {Verhamme}, Anne and {Wisotzki}, Lutz},
        title = "{MUSE Spectroscopic Identifications of Ultra-faint Emission Line Galaxies with M $_{UV}$ {\ensuremath{\sim}} -15}",
      journal = {\apjl},
         year = 2018,
        month = sep,
       volume = {865},
       number = {1},
          eid = {L1},
        pages = {L1},
          doi = {10.3847/2041-8213/aade4b},
archivePrefix = {arXiv},
       eprint = {1809.01142},
 primaryClass = {astro-ph.GA},
       adsurl = {https://ui.adsabs.harvard.edu/abs/2018ApJ...865L...1M}
}

@ARTICLE{Carnall2017,
       author = {{Carnall}, A.~C.},
        title = "{SpectRes: A Fast Spectral Resampling Tool in Python}",
      journal = {arXiv e-prints},
         year = 2017,
        month = may,
          eid = {arXiv:1705.05165},
        pages = {arXiv:1705.05165},
          doi = {10.48550/arXiv.1705.05165},
archivePrefix = {arXiv},
       eprint = {1705.05165},
 primaryClass = {astro-ph.IM},
       adsurl = {https://ui.adsabs.harvard.edu/abs/2017arXiv170505165C}
}

@ARTICLE{Hsiao2026arXiv,
       author = {{Hsiao}, Tiger Yu-Yang and {Chisholm}, John and {Berg}, Danielle A. and {Finkelstein}, Steven L. and {Kokorev}, Vasily and {Atek}, Hakim and {Naidu}, Rohan P. and {Fujimoto}, Seiji and {Furtak}, Lukas J. and {Adamo}, Angela and {Aravindan}, Archana and {Asada}, Yoshihisa and {Basu}, Arghyadeep and {Blaizot}, Jeremy and {Choustikov}, Nicholas and {Dessauges-Zavadsky}, Miroslava and {Fei}, Qinyue and {Katz}, Harley and {Korber}, Damien and {McQuinn}, Kristen. B.~W. and {Mun}, Marcie and {Munoz}, Julian B. and {Natarajan}, Priyamvada and {Stephenson}, Mabel G. and {Schaerer}, Daniel},
        title = "{A Glimpse of the Low-Mass End of the Direct Mass-Metallicity Relation at $z\sim6-8$}",
      journal = {arXiv e-prints},
         year = 2026,
        month = may,
          eid = {arXiv:2605.06770},
        pages = {arXiv:2605.06770},
archivePrefix = {arXiv},
       eprint = {2605.06770},
 primaryClass = {astro-ph.GA},
       adsurl = {https://ui.adsabs.harvard.edu/abs/2026arXiv260506770H}
}

@ARTICLE{Strolger2025ApJ,
       author = {{Strolger}, Louis-Gregory and {Bovill}, Mia Sauda and {Perlman}, Eric and {Kolobow}, Craig and {Larison}, Conor and {Lane}, Zachary G.},
        title = "{On the Origins of ``Hostless'' Supernovae: Testing the Faint-end Galaxy Luminosity Function and Supernova Progenitors with Events in Dwarf Galaxies}",
      journal = {\apj},
         year = 2025,
        month = aug,
       volume = {988},
       number = {2},
          eid = {278},
        pages = {278},
          doi = {10.3847/1538-4357/ade43a},
archivePrefix = {arXiv},
       eprint = {2504.05117},
 primaryClass = {astro-ph.GA},
       adsurl = {https://ui.adsabs.harvard.edu/abs/2025ApJ...988..278S}
}

@ARTICLE{Schulze2018MNRAS,
       author = {{Schulze}, S. and {Kr{\"u}hler}, T. and {Leloudas}, G. and {Gorosabel}, J. and {Mehner}, A. and {Buchner}, J. and {Kim}, S. and {Ibar}, E. and {Amor{\'\i}n}, R. and {Herrero-Illana}, R. and {Anderson}, J.~P. and {Bauer}, F.~E. and {Christensen}, L. and {de Pasquale}, M. and {de Ugarte Postigo}, A. and {Gallazzi}, A. and {Hjorth}, J. and {Morrell}, N. and {Malesani}, D. and {Sparre}, M. and {Stalder}, B. and {Stark}, A.~A. and {Th{\"o}ne}, C.~C. and {Wheeler}, J.~C.},
        title = "{Cosmic evolution and metal aversion in superluminous supernova host galaxies}",
      journal = {\mnras},
         year = 2018,
        month = jan,
       volume = {473},
       number = {1},
        pages = {1258-1285},
          doi = {10.1093/mnras/stx2352},
archivePrefix = {arXiv},
       eprint = {1612.05978},
 primaryClass = {astro-ph.GA},
       adsurl = {https://ui.adsabs.harvard.edu/abs/2018MNRAS.473.1258S}
}

@ARTICLE{Cooke2009Natur,
       author = {{Cooke}, Jeff and {Sullivan}, Mark and {Barton}, Elizabeth J. and {Bullock}, James S. and {Carlberg}, Ray G. and {Gal-Yam}, Avishay and {Tollerud}, Erik},
        title = "{Type IIn supernovae at redshift z\raisebox{-0.5ex}\textasciitilde2 from archival data}",
      journal = {\nat},
         year = 2009,
        month = jul,
       volume = {460},
       number = {7252},
        pages = {237-239},
          doi = {10.1038/nature08082},
archivePrefix = {arXiv},
       eprint = {0907.1928},
 primaryClass = {astro-ph.CO},
       adsurl = {https://ui.adsabs.harvard.edu/abs/2009Natur.460..237C}
}

@ARTICLE{Willott2025ApJ,
       author = {{Willott}, Chris J. and {Asada}, Yoshihisa and {Iyer}, Kartheik G. and {Jude{\v{z}}}, Jon and {Rihtar{\v{s}}i{\v{c}}}, Gregor and {Martis}, Nicholas S. and {Sarrouh}, Ghassan T.~E. and {Desprez}, Guillaume and {Harshan}, Anishya and {Mowla}, Lamiya and {Noirot}, Ga{\"e}l and {Felicioni}, Giordano and {Brada{\v{c}}}, Maru{\v{s}}a and {Brammer}, Gabe and {Muzzin}, Adam and {Sawicki}, Marcin and {Antwi-Danso}, Jacqueline and {Markov}, Vladan and {Tripodi}, Roberta},
        title = "{In Search of the First Stars: An Ultra-compact and Very-low-metallicity Ly{\ensuremath{\alpha}} Emitter Deep within the Epoch of Reionization}",
      journal = {\apj},
         year = 2025,
        month = jul,
       volume = {988},
       number = {1},
          eid = {26},
        pages = {26},
          doi = {10.3847/1538-4357/addf49},
archivePrefix = {arXiv},
       eprint = {2502.07733},
 primaryClass = {astro-ph.GA},
       adsurl = {https://ui.adsabs.harvard.edu/abs/2025ApJ...988...26W}
}

@ARTICLE{Hsiao2025arXiv,
       author = {{Hsiao}, Tiger Yu-Yang and {Sun}, Fengwu and {Lin}, Xiaojing and {Coe}, Dan and {Egami}, Eiichi and {Eisenstein}, Daniel J. and {Fudamoto}, Yoshinobu and {Bunker}, Andrew J. and {Fan}, Xiaohui and {Harikane}, Yuichi and {Helton}, Jakob M. and {Kakiichi}, Koki and {Liu}, Yichen and {Liu}, Weizhe and {Maiolino}, Roberto and {Ouchi}, Masami and {Tee}, Wei Leong and {Wang}, Feige and {Wu}, Yunjing and {Xu}, Yi and {Yang}, Jinyi and {Zhu}, Yongda},
        title = "{SAPPHIRES: Extremely Metal-Poor Galaxy Candidates with $12+{\rm log(O/H)}<7.0$ at $z\sim5-7$ from Deep JWST/NIRCam Grism Observations}",
      journal = {arXiv e-prints},
         year = 2025,
        month = may,
          eid = {arXiv:2505.03873},
        pages = {arXiv:2505.03873},
          doi = {10.48550/arXiv.2505.03873},
archivePrefix = {arXiv},
       eprint = {2505.03873},
 primaryClass = {astro-ph.GA},
       adsurl = {https://ui.adsabs.harvard.edu/abs/2025arXiv250503873H}
}

@ARTICLE{Trussler2026MNRAS,
       author = {{Trussler}, James A.~A. and {Cameron}, Alex J. and {Eisenstein}, Daniel J. and {Katz}, Harley and {Adams}, Nathan J. and {Austin}, Duncan and {Bunker}, Andrew J. and {Carniani}, Stefano and {Conselice}, Christopher J. and {Curti}, Mirko and {Curtis-Lake}, Emma and {Hainline}, Kevin and {Harvey}, Thomas and {Johnson}, Benjamin D. and {Li}, Qiong and {Looser}, Tobias J. and {Rinaldi}, Pierluigi and {Robertson}, Brant and {Sun}, Fengwu and {Tacchella}, Sandro and {Williams}, Christina C. and {Willmer}, Christopher N.~A. and {Willott}, Chris and {Wu}, Zihao},
        title = "{Possible photometric signatures of nebular-dominated emission in 1.5 < z < 8.5 JADES galaxies}",
      journal = {\mnras},
         year = 2026,
        month = apr,
          doi = {10.1093/mnras/stag788},
archivePrefix = {arXiv},
       eprint = {2510.12622},
 primaryClass = {astro-ph.GA},
       adsurl = {https://ui.adsabs.harvard.edu/abs/2026MNRAS.tmp..751T}
}

@ARTICLE{Korber2026AAP,
       author = {{Korber}, Damien and {Chemerynska}, Iryna and {Furtak}, Lukas J. and {Atek}, Hakim and {Endsley}, Ryan and {Schaerer}, Daniel and {Chisholm}, John and {Kokorev}, Vasily and {Saldana-Lopez}, Alberto and {Adamo}, Angela and {Mu{\~n}oz}, Julian B. and {Oesch}, Pascal A. and {Meyer}, Romain and {Marques-Chaves}, Rui and {Fujimoto}, Seiji},
        title = "{A GLIMPSE into the very faint end of the H{\ensuremath{\beta}}+[O III]{\ensuremath{\lambda}}{\ensuremath{\lambda}}4960,5008 luminosity function at z {\ensuremath{\sim}} 7 ─ 9 behind Abell S1063}",
      journal = {\aap},
         year = 2026,
        month = mar,
       volume = {708},
          eid = {A43},
        pages = {A43},
          doi = {10.1051/0004-6361/202556868},
archivePrefix = {arXiv},
       eprint = {2510.04771},
 primaryClass = {astro-ph.GA},
       adsurl = {https://ui.adsabs.harvard.edu/abs/2026A&A...708A..43K}
}

@ARTICLE{Cai2025ApJ,
       author = {{Cai}, Sijia and {Li}, Mingyu and {Cai}, Zheng and {Wu}, Yunjing and {Yu}, Fujiang and {Dickinson}, Mark and {Sun}, Fengwu and {Fan}, Xiaohui and {Wang}, Ben and {Cullen}, Fergus and {Bian}, Fuyan and {Lin}, Xiaojing and {Zou}, Jiaqi},
        title = "{A Metal-free Galaxy at z = 3.19? Evidence of Late Population III Star Formation at Cosmic Noon}",
      journal = {\apjl},
         year = 2025,
        month = nov,
       volume = {993},
       number = {2},
          eid = {L52},
        pages = {L52},
          doi = {10.3847/2041-8213/ae1608},
archivePrefix = {arXiv},
       eprint = {2507.17820},
 primaryClass = {astro-ph.GA},
       adsurl = {https://ui.adsabs.harvard.edu/abs/2025ApJ...993L..52C}
}

@ARTICLE{Sanders2025arXiv,
       author = {{Sanders}, Ryan L. and {Shapley}, Alice E. and {Topping}, Michael W. and {Reddy}, Naveen A. and {Berg}, Danielle A. and {Khostovan}, Ali Ahmad and {Bouwens}, Rychard J. and {Brammer}, Gabriel and {Carnall}, Adam C. and {Cullen}, Fergus and {Dav{\'e}}, Romeel and {Dunlop}, James S. and {Ellis}, Richard S. and {F{\"o}rster Schreiber}, N.~M. and {Furlanetto}, Steven R. and {Glazebrook}, Karl and {Illingworth}, Garth D. and {Jones}, Tucker and {Kriek}, Mariska and {McLeod}, Derek J. and {McLure}, Ross J. and {Narayanan}, Desika and {Oesch}, Pascal A. and {Pahl}, Anthony J. and {Pettini}, Max and {Schaerer}, Daniel and {Stark}, Daniel P. and {Steidel}, Charles C. and {Tang}, Mengtao and {Clarke}, Leonardo and {Donnan}, Callum T. and {Kehoe}, Emily},
        title = "{The AURORA Survey: High-Redshift Empirical Metallicity Calibrations from Electron Temperature Measurements at z=2-10}",
      journal = {arXiv e-prints},
         year = 2025,
        month = aug,
          eid = {arXiv:2508.10099},
        pages = {arXiv:2508.10099},
          doi = {10.48550/arXiv.2508.10099},
archivePrefix = {arXiv},
       eprint = {2508.10099},
 primaryClass = {astro-ph.GA},
       adsurl = {https://ui.adsabs.harvard.edu/abs/2025arXiv250810099S}
}

@ARTICLE{Sanders2024ApJ,
       author = {{Sanders}, Ryan L. and {Shapley}, Alice E. and {Topping}, Michael W. and {Reddy}, Naveen A. and {Brammer}, Gabriel B.},
        title = "{Direct T $_{e}$-based Metallicities of z = 2─9 Galaxies with JWST/NIRSpec: Empirical Metallicity Calibrations Applicable from Reionization to Cosmic Noon}",
      journal = {\apj},
         year = 2024,
        month = feb,
       volume = {962},
       number = {1},
          eid = {24},
        pages = {24},
          doi = {10.3847/1538-4357/ad15fc},
archivePrefix = {arXiv},
       eprint = {2303.08149},
 primaryClass = {astro-ph.GA},
       adsurl = {https://ui.adsabs.harvard.edu/abs/2024ApJ...962...24S}
}

@ARTICLE{Vanzella2025arXiv,
       author = {{Vanzella}, E. and {Messa}, M. and {Zanella}, A. and {Bolamperti}, A. and {Castellano}, M. and {Loiacono}, F. and {Bergamini}, P. and {Roberts-Borsani}, G. and {Adamo}, A. and {Fontana}, A. and {Treu}, T. and {Calura}, F. and {Grillo}, C. and {Lombardi}, M. and {Rosati}, P. and {Gilli}, R. and {Meneghetti}, M.},
        title = "{A Pristine Star-Forming Complex at z=4.19}",
      journal = {arXiv e-prints},
         year = 2025,
        month = sep,
          eid = {arXiv:2509.07073},
        pages = {arXiv:2509.07073},
          doi = {10.48550/arXiv.2509.07073},
archivePrefix = {arXiv},
       eprint = {2509.07073},
 primaryClass = {astro-ph.GA},
       adsurl = {https://ui.adsabs.harvard.edu/abs/2025arXiv250907073V}
}

@ARTICLE{Nakajima2025arXiv,
       author = {{Nakajima}, Kimihiko and {Ouchi}, Masami and {Harikane}, Yuichi and {Vanzella}, Eros and {Ono}, Yoshiaki and {Isobe}, Yuki and {Nishigaki}, Moka and {Tsujimoto}, Takuji and {Nakamura}, Fumitaka and {Xu}, Yi and {Umeda}, Hiroya and {Zhang}, Yechi},
        title = "{An Ultra-Faint, Chemically Primitive Galaxy Forming at the Epoch of Reionization}",
      journal = {arXiv e-prints},
         year = 2025,
        month = jun,
          eid = {arXiv:2506.11846},
        pages = {arXiv:2506.11846},
          doi = {10.48550/arXiv.2506.11846},
archivePrefix = {arXiv},
       eprint = {2506.11846},
 primaryClass = {astro-ph.GA},
       adsurl = {https://ui.adsabs.harvard.edu/abs/2025arXiv250611846N}
}

@ARTICLE{Morishita2025arXiv,
       author = {{Morishita}, Takahiro and {Liu}, Zhaoran and {Stiavelli}, Massimo and {Treu}, Tommaso and {Bergamini}, Pietro and {Zhang}, Yechi},
        title = "{Pristine Massive Star Formation Caught at the Break of Cosmic Dawn}",
      journal = {arXiv e-prints},
         year = 2025,
        month = jul,
          eid = {arXiv:2507.10521},
        pages = {arXiv:2507.10521},
          doi = {10.48550/arXiv.2507.10521},
archivePrefix = {arXiv},
       eprint = {2507.10521},
 primaryClass = {astro-ph.CO},
       adsurl = {https://ui.adsabs.harvard.edu/abs/2025arXiv250710521M}
}

@ARTICLE{Chemerynska2024ApJ,
       author = {{Chemerynska}, Iryna and {Atek}, Hakim and {Dayal}, Pratika and {Furtak}, Lukas J. and {Feldmann}, Robert and {Greene}, Jenny E. and {Maseda}, Michael V. and {Nanayakkara}, Themiya and {Oesch}, Pascal A. and {Fujimoto}, Seiji and {Labb{\'e}}, Ivo and {Bezanson}, Rachel and {Brammer}, Gabriel and {Cutler}, Sam E. and {Leja}, Joel and {Pan}, Richard and {Price}, Sedona H. and {Wang}, Bingjie and {Weaver}, John R. and {Whitaker}, Katherine E.},
        title = "{The Extreme Low-mass End of the Mass{\textendash}Metallicity Relation at z {\ensuremath{\sim}} 7}",
      journal = {\apjl},
         year = 2024,
        month = nov,
       volume = {976},
       number = {1},
          eid = {L15},
        pages = {L15},
          doi = {10.3847/2041-8213/ad8dc9},
archivePrefix = {arXiv},
       eprint = {2407.17110},
 primaryClass = {astro-ph.GA},
       adsurl = {https://ui.adsabs.harvard.edu/abs/2024ApJ...976L..15C}
}

@ARTICLE{Nakajima2022ApJS,
       author = {{Nakajima}, Kimihiko and {Ouchi}, Masami and {Xu}, Yi and {Rauch}, Michael and {Harikane}, Yuichi and {Nishigaki}, Moka and {Isobe}, Yuki and {Kusakabe}, Haruka and {Nagao}, Tohru and {Ono}, Yoshiaki and {Onodera}, Masato and {Sugahara}, Yuma and {Kim}, Ji Hoon and {Komiyama}, Yutaka and {Lee}, Chien-Hsiu and {Zahedy}, Fakhri S.},
        title = "{EMPRESS. V. Metallicity Diagnostics of Galaxies over 12 + log(O/H) ≃ 6.9-8.9 Established by a Local Galaxy Census: Preparing for JWST Spectroscopy}",
      journal = {\apjs},
         year = 2022,
        month = sep,
       volume = {262},
       number = {1},
          eid = {3},
        pages = {3},
          doi = {10.3847/1538-4365/ac7710},
archivePrefix = {arXiv},
       eprint = {2206.02824},
 primaryClass = {astro-ph.GA},
       adsurl = {https://ui.adsabs.harvard.edu/abs/2022ApJS..262....3N}
}

@ARTICLE{Nakajima2023ApJS,
       author = {{Nakajima}, Kimihiko and {Ouchi}, Masami and {Isobe}, Yuki and {Harikane}, Yuichi and {Zhang}, Yechi and {Ono}, Yoshiaki and {Umeda}, Hiroya and {Oguri}, Masamune},
        title = "{JWST Census for the Mass-Metallicity Star Formation Relations at z = 4-10 with Self-consistent Flux Calibration and Proper Metallicity Calibrators}",
      journal = {\apjs},
         year = 2023,
        month = dec,
       volume = {269},
       number = {2},
          eid = {33},
        pages = {33},
          doi = {10.3847/1538-4365/acd556},
archivePrefix = {arXiv},
       eprint = {2301.12825},
 primaryClass = {astro-ph.GA},
       adsurl = {https://ui.adsabs.harvard.edu/abs/2023ApJS..269...33N}
}

@ARTICLE{Eldridge2017PASA,
       author = {{Eldridge}, J.~J. and {Stanway}, E.~R. and {Xiao}, L. and {McClelland}, L.~A.~S. and {Taylor}, G. and {Ng}, M. and {Greis}, S.~M.~L. and {Bray}, J.~C.},
        title = "{Binary Population and Spectral Synthesis Version 2.1: Construction, Observational Verification, and New Results}",
      journal = {\pasa},
         year = 2017,
        month = nov,
       volume = {34},
          eid = {e058},
        pages = {e058},
          doi = {10.1017/pasa.2017.51},
archivePrefix = {arXiv},
       eprint = {1710.02154},
 primaryClass = {astro-ph.SR},
       adsurl = {https://ui.adsabs.harvard.edu/abs/2017PASA...34...58E}
}

@ARTICLE{Chabrier2003PASP,
       author = {{Chabrier}, Gilles},
        title = "{Galactic Stellar and Substellar Initial Mass Function}",
      journal = {\pasp},
         year = 2003,
        month = jul,
       volume = {115},
       number = {809},
        pages = {763-795},
          doi = {10.1086/376392},
archivePrefix = {arXiv},
       eprint = {astro-ph/0304382},
 primaryClass = {astro-ph},
       adsurl = {https://ui.adsabs.harvard.edu/abs/2003PASP..115..763C}
}

@ARTICLE{Kennicutt1998ARAA,
       author = {{Kennicutt}, Jr., Robert C.},
        title = "{Star Formation in Galaxies Along the Hubble Sequence}",
      journal = {\araa},
         year = 1998,
        month = jan,
       volume = {36},
        pages = {189-232},
          doi = {10.1146/annurev.astro.36.1.189},
archivePrefix = {arXiv},
       eprint = {astro-ph/9807187},
 primaryClass = {astro-ph},
       adsurl = {https://ui.adsabs.harvard.edu/abs/1998ARA&A..36..189K}
}

@ARTICLE{Heckman2011ApJ,
       author = {{Heckman}, Timothy M. and {Borthakur}, Sanchayeeta and {Overzier}, Roderik and {Kauffmann}, Guinevere and {Basu-Zych}, Antara and {Leitherer}, Claus and {Sembach}, Ken and {Martin}, D. Chris and {Rich}, R. Michael and {Schiminovich}, David and {Seibert}, Mark},
        title = "{Extreme Feedback and the Epoch of Reionization: Clues in the Local Universe}",
      journal = {\apj},
         year = 2011,
        month = mar,
       volume = {730},
       number = {1},
          eid = {5},
        pages = {5},
          doi = {10.1088/0004-637X/730/1/5},
archivePrefix = {arXiv},
       eprint = {1101.4219},
 primaryClass = {astro-ph.CO},
       adsurl = {https://ui.adsabs.harvard.edu/abs/2011ApJ...730....5H}
}

@ARTICLE{Kimm2019MNRAS,
       author = {{Kimm}, Taysun and {Blaizot}, J{\'e}r{\'e}my and {Garel}, Thibault and {Michel-Dansac}, L{\'e}o and {Katz}, Harley and {Rosdahl}, Joakim and {Verhamme}, Anne and {Haehnelt}, Martin},
        title = "{Understanding the escape of LyC and Ly{\ensuremath{\alpha}} photons from turbulent clouds}",
      journal = {\mnras},
         year = 2019,
        month = jun,
       volume = {486},
       number = {2},
        pages = {2215-2237},
          doi = {10.1093/mnras/stz989},
archivePrefix = {arXiv},
       eprint = {1901.05990},
 primaryClass = {astro-ph.GA},
       adsurl = {https://ui.adsabs.harvard.edu/abs/2019MNRAS.486.2215K}
}

@ARTICLE{Gronke2016ApJ,
       author = {{Gronke}, Max and {Dijkstra}, Mark and {McCourt}, Michael and {Oh}, S. Peng},
        title = "{From Mirrors to Windows: Lyman-alpha Radiative Transfer in a Very Clumpy Medium}",
      journal = {\apjl},
         year = 2016,
        month = dec,
       volume = {833},
       number = {2},
          eid = {L26},
        pages = {L26},
          doi = {10.3847/2041-8213/833/2/L26},
archivePrefix = {arXiv},
       eprint = {1611.01161},
 primaryClass = {astro-ph.GA},
       adsurl = {https://ui.adsabs.harvard.edu/abs/2016ApJ...833L..26G}
}

@ARTICLE{Gazagnes2020AAP,
       author = {{Gazagnes}, S. and {Chisholm}, J. and {Schaerer}, D. and {Verhamme}, A. and {Izotov}, Y.},
        title = "{The origin of the escape of Lyman {\ensuremath{\alpha}} and ionizing photons in Lyman continuum emitters}",
      journal = {\aap},
         year = 2020,
        month = jul,
       volume = {639},
          eid = {A85},
        pages = {A85},
          doi = {10.1051/0004-6361/202038096},
archivePrefix = {arXiv},
       eprint = {2005.07215},
 primaryClass = {astro-ph.GA},
       adsurl = {https://ui.adsabs.harvard.edu/abs/2020A&A...639A..85G}
}

@ARTICLE{Jaskot2019ApJ,
       author = {{Jaskot}, Anne E. and {Dowd}, Tara and {Oey}, M.~S. and {Scarlata}, Claudia and {McKinney}, Jed},
        title = "{New Insights on Ly{\ensuremath{\alpha}} and Lyman Continuum Radiative Transfer in the Greenest Peas}",
      journal = {\apj},
         year = 2019,
        month = nov,
       volume = {885},
       number = {1},
          eid = {96},
        pages = {96},
          doi = {10.3847/1538-4357/ab3d3b},
archivePrefix = {arXiv},
       eprint = {1908.09763},
 primaryClass = {astro-ph.GA},
       adsurl = {https://ui.adsabs.harvard.edu/abs/2019ApJ...885...96J}
}

@ARTICLE{Verhamme2015AAP,
       author = {{Verhamme}, Anne and {Orlitov{\'a}}, Ivana and {Schaerer}, Daniel and {Hayes}, Matthew},
        title = "{Using Lyman-{\ensuremath{\alpha}} to detect galaxies that leak Lyman continuum}",
      journal = {\aap},
         year = 2015,
        month = jun,
       volume = {578},
          eid = {A7},
        pages = {A7},
          doi = {10.1051/0004-6361/201423978},
archivePrefix = {arXiv},
       eprint = {1404.2958},
 primaryClass = {astro-ph.GA},
       adsurl = {https://ui.adsabs.harvard.edu/abs/2015A&A...578A...7V}
}

@ARTICLE{Zackrisson2013ApJ,
       author = {{Zackrisson}, Erik and {Inoue}, Akio K. and {Jensen}, Hannes},
        title = "{The Spectral Evolution of the First Galaxies. II. Spectral Signatures of Lyman Continuum Leakage from Galaxies in the Reionization Epoch}",
      journal = {\apj},
         year = 2013,
        month = nov,
       volume = {777},
       number = {1},
          eid = {39},
        pages = {39},
          doi = {10.1088/0004-637X/777/1/39},
archivePrefix = {arXiv},
       eprint = {1304.6404},
 primaryClass = {astro-ph.CO},
       adsurl = {https://ui.adsabs.harvard.edu/abs/2013ApJ...777...39Z}
}

@ARTICLE{Rauch2011MN,
       author = {{Rauch}, Michael and {Becker}, George D. and {Haehnelt}, Martin G. and {Gauthier}, Jean-Rene and {Ravindranath}, Swara and {Sargent}, Wallace L.~W.},
        title = "{Filamentary infall of cold gas and escape of Ly{\ensuremath{\alpha}} and hydrogen ionizing radiation from an interacting high-redshift galaxy}",
      journal = {\mnras},
         year = 2011,
        month = dec,
       volume = {418},
       number = {2},
        pages = {1115-1126},
          doi = {10.1111/j.1365-2966.2011.19556.x},
archivePrefix = {arXiv},
       eprint = {1105.4876},
 primaryClass = {astro-ph.CO},
       adsurl = {https://ui.adsabs.harvard.edu/abs/2011MNRAS.418.1115R}
}

@ARTICLE{Lemaux2021MN,
       author = {{Lemaux}, B.~C. and {Fuller}, S. and {Brada{\v{c}}}, M. and {Pentericci}, L. and {Hoag}, A. and {Strait}, V. and {Treu}, T. and {Alvarez}, C. and {Bolan}, P. and {Gandhi}, P.~J. and {Huang}, K.-H. and {Jones}, T. and {Mason}, C. and {Pelliccia}, D. and {Ribeiro}, B. and {Ryan}, R.~E. and {Schmidt}, K.~B. and {Vanzella}, E. and {Khusanova}, Y. and {Le F{\`e}vre}, O. and {Guaita}, L. and {Hathi}, N.~P. and {Koekemoer}, A. and {Pforr}, J.},
        title = "{The size and pervasiveness of Ly {\ensuremath{\alpha}}-UV spatial offsets in star-forming galaxies at z {\ensuremath{\sim}} 6}",
      journal = {\mnras},
         year = 2021,
        month = jul,
       volume = {504},
       number = {3},
        pages = {3662-3681},
          doi = {10.1093/mnras/stab924},
archivePrefix = {arXiv},
       eprint = {2007.01310},
 primaryClass = {astro-ph.GA},
       adsurl = {https://ui.adsabs.harvard.edu/abs/2021MNRAS.504.3662L}
}

@ARTICLE{Claeyssens2022AAP,
       author = {{Claeyssens}, A. and {Richard}, J. and {Blaizot}, J. and {Garel}, T. and {Kusakabe}, H. and {Bacon}, R. and {Bauer}, F.~E. and {Guaita}, L. and {Jeanneau}, A. and {Lagattuta}, D. and {Leclercq}, F. and {Maseda}, M. and {Matthee}, J. and {Nanayakkara}, T. and {Pello}, R. and {Thai}, T.~T. and {Tuan-Anh}, P. and {Verhamme}, A. and {Vitte}, E. and {Wisotzki}, L.},
        title = "{The Lensed Lyman-Alpha MUSE Arcs Sample (LLAMAS). I. Characterisation of extended Lyman-alpha halos and spatial offsets}",
      journal = {\aap},
         year = 2022,
        month = oct,
       volume = {666},
          eid = {A78},
        pages = {A78},
          doi = {10.1051/0004-6361/202142320},
archivePrefix = {arXiv},
       eprint = {2201.04674},
 primaryClass = {astro-ph.GA},
       adsurl = {https://ui.adsabs.harvard.edu/abs/2022A&A...666A..78C}
}

@ARTICLE{emcee,
   author = {{Foreman-Mackey}, D. and {Hogg}, D.~W. and {Lang}, D. and {Goodman}, J.},
    title = {emcee: The MCMC Hammer},
  journal = {PASP},
     year = 2013,
   volume = 125,
    pages = {306-312},
   eprint = {1202.3665},
      doi = {10.1086/670067}
}

@ARTICLE{Roberts-Borsani2024ApJ,
       author = {{Roberts-Borsani}, Guido and {Treu}, Tommaso and {Shapley}, Alice and {Fontana}, Adriano and {Pentericci}, Laura and {Castellano}, Marco and {Morishita}, Takahiro and {Bergamini}, Pietro and {Rosati}, Piero},
        title = "{Between the Extremes: A JWST Spectroscopic Benchmark for High-redshift Galaxies Using {\ensuremath{\sim}}500 Confirmed Sources at z {\ensuremath{\geq}} 5}",
      journal = {\apj},
         year = 2024,
        month = dec,
       volume = {976},
       number = {2},
          eid = {193},
        pages = {193},
          doi = {10.3847/1538-4357/ad85d3},
archivePrefix = {arXiv},
       eprint = {2403.07103},
 primaryClass = {astro-ph.GA},
       adsurl = {https://ui.adsabs.harvard.edu/abs/2024ApJ...976..193R}
}

@ARTICLE{Saxena2024AAP,
       author = {{Saxena}, Aayush and {Bunker}, Andrew J. and {Jones}, Gareth C. and {Stark}, Daniel P. and {Cameron}, Alex J. and {Witstok}, Joris and {Arribas}, Santiago and {Baker}, William M. and {Baum}, Stefi and {Bhatawdekar}, Rachana and {Bowler}, Rebecca and {Boyett}, Kristan and {Carniani}, Stefano and {Charlot}, Stephane and {Chevallard}, Jacopo and {Curti}, Mirko and {Curtis-Lake}, Emma and {Eisenstein}, Daniel J. and {Endsley}, Ryan and {Hainline}, Kevin and {Helton}, Jakob M. and {Johnson}, Benjamin D. and {Kumari}, Nimisha and {Looser}, Tobias J. and {Maiolino}, Roberto and {Rieke}, Marcia and {Rix}, Hans-Walter and {Robertson}, Brant E. and {Sandles}, Lester and {Simmonds}, Charlotte and {Smit}, Renske and {Tacchella}, Sandro and {Williams}, Christina C. and {Willmer}, Christopher N.~A. and {Willott}, Chris},
        title = "{JADES: The production and escape of ionizing photons from faint Lyman-alpha emitters in the epoch of reionization}",
      journal = {\aap},
         year = 2024,
        month = apr,
       volume = {684},
          eid = {A84},
        pages = {A84},
          doi = {10.1051/0004-6361/202347132},
archivePrefix = {arXiv},
       eprint = {2306.04536},
 primaryClass = {astro-ph.GA},
       adsurl = {https://ui.adsabs.harvard.edu/abs/2024A&A...684A..84S}
}

@ARTICLE{Peng2010AJ,
       author = {{Peng}, Chien Y. and {Ho}, Luis C. and {Impey}, Chris D. and {Rix}, Hans-Walter},
        title = "{Detailed Decomposition of Galaxy Images. II. Beyond Axisymmetric Models}",
      journal = {\aj},
         year = 2010,
        month = jun,
       volume = {139},
       number = {6},
        pages = {2097-2129},
          doi = {10.1088/0004-6256/139/6/2097},
archivePrefix = {arXiv},
       eprint = {0912.0731},
 primaryClass = {astro-ph.CO},
       adsurl = {https://ui.adsabs.harvard.edu/abs/2010AJ....139.2097P}
}

@ARTICLE{Weaver2024ApJS,
       author = {{Weaver}, John R. and {Cutler}, Sam E. and {Pan}, Richard and {Whitaker}, Katherine E. and {Labb{\'e}}, Ivo and {Price}, Sedona H. and {Bezanson}, Rachel and {Brammer}, Gabriel and {Marchesini}, Danilo and {Leja}, Joel and {Wang}, Bingjie and {Furtak}, Lukas J. and {Zitrin}, Adi and {Atek}, Hakim and {Chemerynska}, Iryna and {Coe}, Dan and {Dayal}, Pratika and {van Dokkum}, Pieter and {Feldmann}, Robert and {F{\"o}rster Schreiber}, Natascha M. and {Franx}, Marijn and {Fujimoto}, Seiji and {Fudamoto}, Yoshinobu and {Glazebrook}, Karl and {de Graaff}, Anna and {Greene}, Jenny E. and {Juneau}, St{\'e}phanie and {Kassin}, Susan and {Kriek}, Mariska and {Khullar}, Gourav and {Maseda}, Michael V. and {Mowla}, Lamiya A. and {Muzzin}, Adam and {Nanayakkara}, Themiya and {Nelson}, Erica J. and {Oesch}, Pascal A. and {Pacifici}, Camilla and {Papovich}, Casey and {Setton}, David J. and {Shapley}, Alice E. and {Shipley}, Heath V. and {Smit}, Renske and {Stefanon}, Mauro and {Taylor}, Edward N. and {Weibel}, Andrea and {Williams}, Christina C.},
        title = "{The UNCOVER Survey: A First-look HST + JWST Catalog of 60,000 Galaxies near A2744 and beyond}",
      journal = {\apjs},
         year = 2024,
        month = jan,
       volume = {270},
       number = {1},
          eid = {7},
        pages = {7},
          doi = {10.3847/1538-4365/ad07e0},
archivePrefix = {arXiv},
       eprint = {2301.02671},
 primaryClass = {astro-ph.GA},
       adsurl = {https://ui.adsabs.harvard.edu/abs/2024ApJS..270....7W}
}

@ARTICLE{Allingham2026arXiv,
       author = {{Allingham}, Joseph F.~V. and {Zitrin}, Adi and {Kokorev}, Vasily and {Yanagisawa}, Hiroto and {Diego}, Jose M. and {Furtak}, Lukas J. and {Asada}, Yoshihisa and {Coe}, Dan and {Coulter}, David A. and {Fujimoto}, Seiji and {Larison}, Conor and {Oguri}, Masamune and {Pierel}, Justin D.~R. and {Sun}, Fengwu and {Bradac}, Marusa and {Dayal}, Pratika and {Lopes}, Paulo A.~A. and {Meena}, Ashish K. and {Pascale}, Massimo and {Akins}, Hollis B. and {Bauer}, Franz E. and {Bradley}, Larry D. and {Brammer}, Gabriel and {Chisholm}, John and {Desprez}, Guillaume and {Fei}, Qinyue and {Ferguson}, Henry C. and {Finkelstein}, Steven L. and {Frye}, Brenda and {Golubchik}, Miriam and {Inayoshi}, Kohei and {Koekemoer}, Anton M. and {Lucas}, Ray A. and {Magdis}, Georgios E. and {Martis}, Nicholas S. and {Pan}, Richard and {Richard}, Johan and {Ricotti}, Massimo and {Rihtarsic}, Gregor and {Robbins}, Luke and {Sheu}, William and {Welch}, Brian and {Willott}, Chris and {Windhorst}, Rogier A.},
        title = "{VENUS: Strong-lensing model of MACS J1931.8-2635 -- revealing the farthest multiply imaged supernova}",
      journal = {arXiv e-prints},
         year = 2026,
        month = feb,
          eid = {arXiv:2602.14074},
        pages = {arXiv:2602.14074},
          doi = {10.48550/arXiv.2602.14074},
archivePrefix = {arXiv},
       eprint = {2602.14074},
 primaryClass = {astro-ph.CO},
       adsurl = {https://ui.adsabs.harvard.edu/abs/2026arXiv260214074A}
}

@ARTICLE{Ciocan2021AAP,
       author = {{Ciocan}, B.~I. and {Ziegler}, B.~L. and {Verdugo}, M. and {Papaderos}, P. and {Fogarty}, K. and {Donahue}, M. and {Postman}, M.},
        title = "{The VLT-MUSE and ALMA view of the MACS 1931.8-2635 brightest cluster galaxy}",
      journal = {\aap},
         year = 2021,
        month = may,
       volume = {649},
          eid = {A23},
        pages = {A23},
          doi = {10.1051/0004-6361/202040010},
archivePrefix = {arXiv},
       eprint = {2101.10718},
 primaryClass = {astro-ph.GA},
       adsurl = {https://ui.adsabs.harvard.edu/abs/2021A&A...649A..23C}
}

@ARTICLE{Oke1983ApJ,
       author = {{Oke}, J.~B. and {Gunn}, J.~E.},
        title = "{Secondary standard stars for absolute spectrophotometry.}",
      journal = {\apj},
         year = 1983,
        month = mar,
       volume = {266},
        pages = {713-717},
          doi = {10.1086/160817},
       adsurl = {https://ui.adsabs.harvard.edu/abs/1983ApJ...266..713O}
}

@MISC{Fujimoto_Go6882,
       author = {{Fujimoto}, Seiji and {Coe}, Dan and {Abdurro'uf}, Abdurro'uf and {Abraham}, Roberto G. and {Adamo}, Angela and {Akins}, Hollis and {Amorin}, Ricardo and {Arrabal Haro}, Pablo and {Asada}, Yoshihisa and {Atek}, Hakim and {Bagley}, Micaela and {Bhatawdekar}, Rachana and {Bradac}, Marusa and {Bradley}, Larry and {Brammer}, Gabriel and {Bromm}, Volker and {Casey}, Caitlin M. and {Chisholm}, John and {Conselice}, Christopher and {Dai}, Liang and {Dayal}, Pratika and {Desprez}, Guillaume and {Dessauges-Zavadsky}, Miroslava and {Dickinson}, Mark and {Diego}, Jose M. and {Egami}, Eiichi and {Eisenstein}, Daniel J. and {Ferguson}, Henry C. and {Finkelstein}, Steven L. and {Furtak}, Lukas Jonathan and {Hamilton}, Timothy S. and {Harikane}, Yuichi and {Hashimoto}, Takuya and {Hathi}, Nimish P. and {Hsiao}, Tiger and {Inayoshi}, Kohei and {Jimenez-Teja}, Yolanda and {Jogee}, Shardha and {Kartaltepe}, Jeyhan and {Koekemoer}, Anton M. and {Kohno}, Kotaro and {Kokorev}, Vasily and {Kumari}, Nimisha and {Labbe}, Ivo and {Larson}, Rebecca L. and {Lucas}, Ray A. and {Magdis}, Georgios and {Marchesini}, Danilo and {Markov}, Vladan and {Martis}, Nicholas and {Matthee}, Jorryt and {Meena}, Ashish Kumar and {Messa}, Matteo and {Mowla}, Lamiya and {Munoz}, Julian B. and {Naidu}, Rohan and {Nakajima}, Kimihiko and {Nakane}, Minami and {Noirot}, Gael and {Oesch}, Pascal and {Oguri}, Masamune and {Ono}, Yoshiaki and {Ouchi}, Masami and {Pan}, Richard and {Papovich}, Casey and {Pascale}, Massimo and {Pierel}, Justin and {Postman}, Marc and {Resseguier}, Tom and {Rest}, Armin and {Richard}, Johan Pierre and {Ricotti}, Massimo and {Rigby}, Jane R. and {Sawicki}, Marcin and {Schneider}, Raffaella and {Shimasaku}, Kazuhiro and {Strolger}, Louis-Gregory and {Sun}, Fengwu and {Toft}, Sune and {Tripodi}, Roberta and {Trussler}, James and {Tsujita}, Akiyoshi and {Umeda}, Hiroya and {Valentino}, Francesco Maria and {Vanzella}, Eros and {Venditti}, Alessandra and {Watson}, Darach and {Weaver}, John R. and {Welch}, Brian and {Willott}, Chris J. and {Windhorst}, Rogier A. and {Xu}, Yi and {Yanagisawa}, Hiroto and {Zackrisson}, Erik and {Zitrin}, Adi},
        title = "{Vast Exploration for Nascent, Unexplored Sources (VENUS)}",
 howpublished = {JWST Proposal. Cycle 4, ID. \#6882},
         year = 2025,
        month = mar,
        pages = {6882},
       adsurl = {https://ui.adsabs.harvard.edu/abs/2025jwst.prop.6882F}
}

@ARTICLE{DeCoursey2025aApJ,
       author = {{DeCoursey}, Christa and {Egami}, Eiichi and {Pierel}, Justin D.~R. and {Sun}, Fengwu and {Rest}, Armin and {Coulter}, David A. and {Engesser}, Michael and {Siebert}, Matthew R. and {Hainline}, Kevin N. and {Johnson}, Benjamin D. and {Bunker}, Andrew J. and {Cargile}, Phillip A. and {Charlot}, Stephane and {Chen}, Wenlei and {Curti}, Mirko and {DeFour-Remy}, Shea and {Eisenstein}, Daniel J. and {Fox}, Ori D. and {Gezari}, Suvi and {Gomez}, Sebastian and {Jencson}, Jacob and {Joshi}, Bhavin A. and {Khairnar}, Sanvi and {Lyu}, Jianwei and {Maiolino}, Roberto and {Moriya}, Takashi J. and {Quimby}, Robert M. and {Rieke}, George H. and {Rieke}, Marcia J. and {Robertson}, Brant and {Shahbandeh}, Melissa and {Strolger}, Louis-Gregory and {Tacchella}, Sandro and {Wang}, Qinan and {Williams}, Christina C. and {Willmer}, Christopher N.~A. and {Willott}, Chris and {Zenati}, Yossef},
        title = "{The JADES Transient Survey: Discovery and Classification of Supernovae in the JADES Deep Field}",
      journal = {\apj},
         year = 2025,
        month = feb,
       volume = {979},
       number = {2},
          eid = {250},
        pages = {250},
          doi = {10.3847/1538-4357/ad8fab},
archivePrefix = {arXiv},
       eprint = {2406.05060},
 primaryClass = {astro-ph.HE},
       adsurl = {https://ui.adsabs.harvard.edu/abs/2025ApJ...979..250D}
}

@ARTICLE{Coulter2026ApJ,
       author = {{Coulter}, D.~A. and {Pierel}, J.~D.~R. and {DeCoursey}, C. and {Moriya}, T.~J. and {Siebert}, M.~R. and {Joshi}, B.~A. and {Engesser}, M. and {Rest}, A. and {Egami}, E. and {Shahbandeh}, M. and {Chen}, W. and {Fox}, O.~D. and {Strolger}, L.~G. and {Zenati}, Y. and {Bunker}, A.~J. and {Cargile}, P.~A. and {Curti}, M. and {Eisenstein}, D.~J. and {Gezari}, S. and {Gomez}, S. and {Guolo}, M. and {Hainline}, K. and {Jencson}, J. and {Johnson}, B.~D. and {Karmen}, M. and {Maiolino}, R. and {Quimby}, R.~M. and {Rinaldi}, P. and {Robertson}, B. and {Tacchella}, S. and {Sun}, F. and {Wang}, Q. and {Wevers}, T.},
        title = "{Discovery of a Likely Type II Supernova at z = 3.6 with JWST}",
      journal = {\apj},
         year = 2026,
        month = may,
       volume = {1002},
       number = {1},
          eid = {83},
        pages = {83},
          doi = {10.3847/1538-4357/ae595d},
archivePrefix = {arXiv},
       eprint = {2501.05513},
 primaryClass = {astro-ph.HE},
       adsurl = {https://ui.adsabs.harvard.edu/abs/2026ApJ..1002...83C}
}

@ARTICLE{DeCoursey2025bApJ,
       author = {{DeCoursey}, Christa and {Egami}, Eiichi and {Sun}, Fengwu and {Akhtarkavan}, Arshia and {Bhatawdekar}, Rachana and {Bunker}, Andrew J. and {Coulter}, David A. and {Engesser}, Michael and {Fox}, Ori D. and {Gomez}, Sebastian and {Inayoshi}, Kohei and {Johnson}, Benjamin D. and {Karmen}, Mitchell and {Larison}, Conor and {Lin}, Xiaojing and {Lyu}, Jianwei and {Mattila}, Seppo and {Moriya}, Takashi J. and {Pierel}, Justin D.~R. and {Pusk{\'a}s}, D{\'a}vid and {Rest}, Armin and {Rieke}, George H. and {Robertson}, Brant and {Salamat}, Sepehr and {Strolger}, Louis-Gregory and {Tacchella}, Sandro and {Vassallo}, Christian and {Williams}, Christina C. and {Zenati}, Yossef and {Zhang}, Junyu},
        title = "{The First Photometric Evidence of a Transient/Variable Source at z > 5 with JWST}",
      journal = {\apj},
         year = 2025,
        month = sep,
       volume = {990},
       number = {1},
          eid = {31},
        pages = {31},
          doi = {10.3847/1538-4357/ade78c},
archivePrefix = {arXiv},
       eprint = {2504.17007},
 primaryClass = {astro-ph.HE},
       adsurl = {https://ui.adsabs.harvard.edu/abs/2025ApJ...990...31D}
}

@ARTICLE{Cooke2012Natur,
       author = {{Cooke}, Jeff and {Sullivan}, Mark and {Gal-Yam}, Avishay and {Barton}, Elizabeth J. and {Carlberg}, Raymond G. and {Ryan-Weber}, Emma V. and {Horst}, Chuck and {Omori}, Yuuki and {D{\'\i}az}, C. Gonzalo},
        title = "{Superluminous supernovae at redshifts of 2.05 and 3.90}",
      journal = {\nat},
         year = 2012,
        month = nov,
       volume = {491},
       number = {7423},
        pages = {228-231},
          doi = {10.1038/nature11521},
archivePrefix = {arXiv},
       eprint = {1211.2003},
 primaryClass = {astro-ph.CO},
       adsurl = {https://ui.adsabs.harvard.edu/abs/2012Natur.491..228C}
}

@ARTICLE{Silk2012RAA,
       author = {{Silk}, Joseph and {Mamon}, Gary A.},
        title = "{The current status of galaxy formation}",
      journal = {Research in Astronomy and Astrophysics},
         year = 2012,
        month = aug,
       volume = {12},
       number = {8},
        pages = {917-946},
          doi = {10.1088/1674-4527/12/8/004},
archivePrefix = {arXiv},
       eprint = {1207.3080},
 primaryClass = {astro-ph.CO},
       adsurl = {https://ui.adsabs.harvard.edu/abs/2012RAA....12..917S}
}

@ARTICLE{Nelson2019MNRAS,
       author = {{Nelson}, Dylan and {Pillepich}, Annalisa and {Springel}, Volker and {Pakmor}, R{\"u}diger and {Weinberger}, Rainer and {Genel}, Shy and {Torrey}, Paul and {Vogelsberger}, Mark and {Marinacci}, Federico and {Hernquist}, Lars},
        title = "{First results from the TNG50 simulation: galactic outflows driven by supernovae and black hole feedback}",
      journal = {\mnras},
         year = 2019,
        month = dec,
       volume = {490},
       number = {3},
        pages = {3234-3261},
          doi = {10.1093/mnras/stz2306},
archivePrefix = {arXiv},
       eprint = {1902.05554},
 primaryClass = {astro-ph.GA},
       adsurl = {https://ui.adsabs.harvard.edu/abs/2019MNRAS.490.3234N}
}

@ARTICLE{Katz2023MNRAS,
       author = {{Katz}, Harley and {Saxena}, Aayush and {Rosdahl}, Joki and {Kimm}, Taysun and {Blaizot}, Jeremy and {Garel}, Thibault and {Michel-Dansac}, Leo and {Haehnelt}, Martin and {Ellis}, Richard S. and {Pentericci}, Laura and {Devriendt}, Julien and {Slyz}, Adrianne},
        title = "{Two modes of LyC escape from bursty star formation: implications for [C II] deficits and the sources of reionization}",
      journal = {\mnras},
         year = 2023,
        month = jan,
       volume = {518},
       number = {1},
        pages = {270-285},
          doi = {10.1093/mnras/stac3019},
archivePrefix = {arXiv},
       eprint = {2210.09156},
 primaryClass = {astro-ph.GA},
       adsurl = {https://ui.adsabs.harvard.edu/abs/2023MNRAS.518..270K}
}

@ARTICLE{Tollet2019MNRAS,
       author = {{Tollet}, {\'E}douard and {Cattaneo}, Andrea and {Macci{\`o}}, Andrea V. and {Dutton}, Aaron A. and {Kang}, Xi},
        title = "{NIHAO XIX: how supernova feedback shapes the galaxy baryon cycle}",
      journal = {\mnras},
         year = 2019,
        month = may,
       volume = {485},
       number = {2},
        pages = {2511-2531},
          doi = {10.1093/mnras/stz545},
archivePrefix = {arXiv},
       eprint = {1902.03888},
 primaryClass = {astro-ph.GA},
       adsurl = {https://ui.adsabs.harvard.edu/abs/2019MNRAS.485.2511T}
}

@ARTICLE{Zhang2018Galax,
       author = {{Zhang}, Dong},
        title = "{A Review of the Theory of Galactic Winds Driven by Stellar Feedback}",
      journal = {Galaxies},
         year = 2018,
        month = nov,
       volume = {6},
       number = {4},
          eid = {114},
        pages = {114},
          doi = {10.3390/galaxies6040114},
archivePrefix = {arXiv},
       eprint = {1811.00558},
 primaryClass = {astro-ph.GA},
       adsurl = {https://ui.adsabs.harvard.edu/abs/2018Galax...6..114Z}
}

@ARTICLE{Schneider2024AARv,
       author = {{Schneider}, Raffaella and {Maiolino}, Roberto},
        title = "{The formation and cosmic evolution of dust in the early Universe: I. Dust sources}",
      journal = {\aapr},
         year = 2024,
        month = apr,
       volume = {32},
       number = {1},
          eid = {2},
        pages = {2},
          doi = {10.1007/s00159-024-00151-2},
archivePrefix = {arXiv},
       eprint = {2310.00053},
 primaryClass = {astro-ph.GA},
       adsurl = {https://ui.adsabs.harvard.edu/abs/2024A&ARv..32....2S}
}

@ARTICLE{Kobayashi2020ApJ,
       author = {{Kobayashi}, Chiaki and {Karakas}, Amanda I. and {Lugaro}, Maria},
        title = "{The Origin of Elements from Carbon to Uranium}",
      journal = {\apj},
         year = 2020,
        month = sep,
       volume = {900},
       number = {2},
          eid = {179},
        pages = {179},
          doi = {10.3847/1538-4357/abae65},
archivePrefix = {arXiv},
       eprint = {2008.04660},
 primaryClass = {astro-ph.GA},
       adsurl = {https://ui.adsabs.harvard.edu/abs/2020ApJ...900..179K}
}

@ARTICLE{Fujimoto2025NatAs,
       author = {{Fujimoto}, S. and {Ouchi}, M. and {Kohno}, K. and {Valentino}, F. and {Gim{\'e}nez-Arteaga}, C. and {Brammer}, G.~B. and {Furtak}, L.~J. and {Kohandel}, M. and {Oguri}, M. and {Pallottini}, A. and {Richard}, J. and {Zitrin}, A. and {Bauer}, F.~E. and {Boylan-Kolchin}, M. and {Dessauges-Zavadsky}, M. and {Egami}, E. and {Finkelstein}, S.~L. and {Ma}, Z. and {Smail}, I. and {Watson}, D. and {Hutchison}, T.~A. and {Rigby}, J.~R. and {Welch}, B.~D. and {Ao}, Y. and {Bradley}, L.~D. and {Caminha}, G.~B. and {Caputi}, K.~I. and {Espada}, D. and {Endsley}, R. and {Fudamoto}, Y. and {Gonz{\'a}lez-L{\'o}pez}, J. and {Hatsukade}, B. and {Koekemoer}, A.~M. and {Kokorev}, V. and {Laporte}, N. and {Lee}, M. and {Magdis}, G.~E. and {Ono}, Y. and {Rizzo}, F. and {Shibuya}, T. and {Shimasaku}, K. and {Sun}, F. and {Toft}, S. and {Umehata}, H. and {Wang}, T. and {Yajima}, H.},
        title = "{Primordial rotating disk composed of at least 15 dense star-forming clumps at cosmic dawn}",
      journal = {Nature Astronomy},
         year = 2025,
        month = aug,
          doi = {10.1038/s41550-025-02592-w},
archivePrefix = {arXiv},
       eprint = {2402.18543},
 primaryClass = {astro-ph.GA},
       adsurl = {https://ui.adsabs.harvard.edu/abs/2025NatAs.tmp..174F}
}

@ARTICLE{Berg2025arXiv,
       author = {{Berg}, Danielle A. and {Naidu}, Rohan P. and {Chisholm}, John and {Atek}, Hakim and {Fujimoto}, Seiji and {Kokorev}, Vasily and {Furtak}, Lukas J. and {Kobayashi}, Chiaki and {Schaerer}, Daniel and {Adamo}, Angela and {Fei}, Qinyue and {Korber}, Damien and {Matthee}, Jorryt and {Marques-Chaves}, Rui and {Martinez}, Zorayda and {Mcquinn}, Kristen B.~W. and {Mu{\~n}oz}, Julian B. and {Oesch}, Pascal A. and {Stark}, Daniel P. and {Stephenson}, Mabel G. and {Hsiao}, Tiger Yu-Yang},
        title = "{A Fleeting GLIMPSE of N/O Enrichment at Cosmic Dawn: Evidence for Wolf Rayet N Stars in a z = 6.1 Galaxy}",
      journal = {arXiv e-prints},
         year = 2025,
        month = nov,
          eid = {arXiv:2511.13591},
        pages = {arXiv:2511.13591},
          doi = {10.48550/arXiv.2511.13591},
archivePrefix = {arXiv},
       eprint = {2511.13591},
 primaryClass = {astro-ph.GA},
       adsurl = {https://ui.adsabs.harvard.edu/abs/2025arXiv251113591B}
}

@ARTICLE{Asada2026arXiv,
       author = {{Asada}, Yoshihisa and {Fujimoto}, Seiji and {Chisholm}, John and {Naidu}, Rohan P. and {Atek}, Hakim and {Brammer}, Gabriel and {Furtak}, Lukas J. and {Kokorev}, Vasily and {Pan}, Richard and {Basu}, Arghyadeep and {Bromm}, Volker and {Dessauges-Zavadsky}, Miroslava and {Hsiao}, Tiger Yu-Yang and {Jecmen}, Michelle and {Korber}, Damien and {Liu}, Boyuan and {McKinney}, Jed and {McQuinn}, Kristen B.~W. and {Schaerer}, Daniel},
        title = "{GLIMPSE-DDT spectroscopic properties of faint-end galaxies at $z\sim6$: Towards first metal enrichment, dust production, and ionizing photon production}",
      journal = {arXiv e-prints},
         year = 2026,
        month = jan,
          eid = {arXiv:2601.20045},
        pages = {arXiv:2601.20045},
          doi = {10.48550/arXiv.2601.20045},
archivePrefix = {arXiv},
       eprint = {2601.20045},
 primaryClass = {astro-ph.GA},
       adsurl = {https://ui.adsabs.harvard.edu/abs/2026arXiv260120045A}
}

@ARTICLE{Asada2024MNRAS,
       author = {{Asada}, Yoshihisa and {Sawicki}, Marcin and {Abraham}, Roberto and {Brada{\v{c}}}, Maru{\v{s}}a and {Brammer}, Gabriel and {Desprez}, Guillaume and {Estrada-Carpenter}, Vince and {Iyer}, Kartheik and {Martis}, Nicholas and {Matharu}, Jasleen and {Mowla}, Lamiya and {Muzzin}, Adam and {Noirot}, Ga{\"e}l and {Sarrouh}, Ghassan T.~E. and {Strait}, Victoria and {Willott}, Chris J. and {Harshan}, Anishya},
        title = "{Bursty star formation and galaxy-galaxy interactions in low-mass galaxies 1 Gyr after the Big Bang}",
      journal = {\mnras},
         year = 2024,
        month = feb,
       volume = {527},
       number = {4},
        pages = {11372-11392},
          doi = {10.1093/mnras/stad3902},
archivePrefix = {arXiv},
       eprint = {2310.02314},
 primaryClass = {astro-ph.GA},
       adsurl = {https://ui.adsabs.harvard.edu/abs/2024MNRAS.52711372A}
}

@ARTICLE{Harikane2025ApJ,
       author = {{Harikane}, Yuichi and {Sanders}, Ryan L. and {Ellis}, Richard and {Jones}, Tucker and {Ouchi}, Masami and {Laporte}, Nicolas and {Roberts-Borsani}, Guido and {Katz}, Harley and {Nakajima}, Kimihiko and {Ono}, Yoshiaki and {Gupta}, Mansi},
        title = "{JWST and ALMA Joint Analysis with [O II] {\ensuremath{\lambda}}{\ensuremath{\lambda}}3726, 3729, [O III] {\ensuremath{\lambda}}4363, [O III] 88 {\ensuremath{\mu}}m, and [O III] 52 {\ensuremath{\mu}}m: Multizone Evolution of Electron Densities at z {\ensuremath{\sim}} 0─14 and its Impact on Metallicity Measurements}",
      journal = {\apj},
         year = 2025,
        month = nov,
       volume = {993},
       number = {2},
          eid = {204},
        pages = {204},
          doi = {10.3847/1538-4357/ae0e53},
archivePrefix = {arXiv},
       eprint = {2505.09186},
 primaryClass = {astro-ph.GA},
       adsurl = {https://ui.adsabs.harvard.edu/abs/2025ApJ...993..204H}
}

@ARTICLE{Curti2024AAP,
       author = {{Curti}, Mirko and {Maiolino}, Roberto and {Curtis-Lake}, Emma and {Chevallard}, Jacopo and {Carniani}, Stefano and {D'Eugenio}, Francesco and {Looser}, Tobias J. and {Scholtz}, Jan and {Charlot}, Stephane and {Cameron}, Alex and {{\"U}bler}, Hannah and {Witstok}, Joris and {Boyett}, Kristian and {Laseter}, Isaac and {Sandles}, Lester and {Arribas}, Santiago and {Bunker}, Andrew and {Giardino}, Giovanna and {Maseda}, Michael V. and {Rawle}, Tim and {Rodr{\'\i}guez Del Pino}, Bruno and {Smit}, Renske and {Willott}, Chris J. and {Eisenstein}, Daniel J. and {Hausen}, Ryan and {Johnson}, Benjamin and {Rieke}, Marcia and {Robertson}, Brant and {Tacchella}, Sandro and {Williams}, Christina C. and {Willmer}, Christopher and {Baker}, William M. and {Bhatawdekar}, Rachana and {Egami}, Eiichi and {Helton}, Jakob M. and {Ji}, Zhiyuan and {Kumari}, Nimisha and {Perna}, Michele and {Shivaei}, Irene and {Sun}, Fengwu},
        title = "{JADES: Insights into the low-mass end of the mass-metallicity-SFR relation at 3 < z < 10 from deep JWST/NIRSpec spectroscopy}",
      journal = {\aap},
         year = 2024,
        month = apr,
       volume = {684},
          eid = {A75},
        pages = {A75},
          doi = {10.1051/0004-6361/202346698},
archivePrefix = {arXiv},
       eprint = {2304.08516},
 primaryClass = {astro-ph.GA},
       adsurl = {https://ui.adsabs.harvard.edu/abs/2024A&A...684A..75C}
}

@ARTICLE{Endsley2024MNRAS,
       author = {{Endsley}, Ryan and {Stark}, Daniel P. and {Whitler}, Lily and {Topping}, Michael W. and {Johnson}, Benjamin D. and {Robertson}, Brant and {Tacchella}, Sandro and {Alberts}, Stacey and {Baker}, William M. and {Bhatawdekar}, Rachana and {Boyett}, Kristan and {Bunker}, Andrew J. and {Cameron}, Alex J. and {Carniani}, Stefano and {Charlot}, Stephane and {Chen}, Zuyi and {Chevallard}, Jacopo and {Curtis-Lake}, Emma and {Danhaive}, A. Lola and {Egami}, Eiichi and {Eisenstein}, Daniel J. and {Hainline}, Kevin and {Helton}, Jakob M. and {Ji}, Zhiyuan and {Looser}, Tobias J. and {Maiolino}, Roberto and {Nelson}, Erica and {Pusk{\'a}s}, D{\'a}vid and {Rieke}, George and {Rieke}, Marcia and {Rix}, Hans-Walter and {Sandles}, Lester and {Saxena}, Aayush and {Simmonds}, Charlotte and {Smit}, Renske and {Sun}, Fengwu and {Williams}, Christina C. and {Willmer}, Christopher N.~A. and {Willott}, Chris and {Witstok}, Joris},
        title = "{The star-forming and ionizing properties of dwarf z 6-9 galaxies in JADES: insights on bursty star formation and ionized bubble growth}",
      journal = {\mnras},
         year = 2024,
        month = sep,
       volume = {533},
       number = {1},
        pages = {1111-1142},
          doi = {10.1093/mnras/stae1857},
archivePrefix = {arXiv},
       eprint = {2306.05295},
 primaryClass = {astro-ph.GA},
       adsurl = {https://ui.adsabs.harvard.edu/abs/2024MNRAS.533.1111E}
}

@ARTICLE{Coulter2026arXiv,
       author = {{Coulter}, David A. and {Larison}, Conor and {Pierel}, Justin D.~R. and {Fujimoto}, Seiji and {Kokorev}, Vasily and {Allingham}, Joseph F.~V. and {Moriya}, Takashi J. and {Siebert}, Matthew and {Asada}, Yoshihisa and {Bezanson}, Rachel and {Brada{\v{c}}}, Maru{\v{s}}a and {Brammer}, Gabriel and {Chisholm}, John and {Coe}, Dan and {Dayal}, Pratika and {Engesser}, Michael and {Finkelstein}, Steven L. and {Fox}, Ori D. and {Furtak}, Lukas J. and {Koekemoer}, Anton M. and {Moore}, Thomas and {Nakane}, Minami and {Ouchi}, Masami and {Pan}, Richard and {Quimby}, Robert and {Rest}, Armin and {Richard}, Johan and {Robbins}, Luke and {Strolger}, Louis-Gregory and {Sun}, Fengwu and {Treu}, Tommaso and {Yanagisawa}, Hiroto and {Abdurro'uf} and {Agrawal}, Aadya and {Amor{\'\i}n}, Ricardo and {Anderson}, Joseph P. and {Angulo}, Rodrigo and {Atek}, Hakim and {Bauer}, Franz E. and {Bradley}, Larry D. and {Bromm}, Volker and {Bronikowski}, Mateusz and {Conselice}, Christopher J. and {DeCoursey}, Christa and {DerKacy}, James M. and {Desprez}, Guillaume and {Dhawan}, Suhail and {Diego}, Jose M. and {Egami}, Eiichi and {Faisst}, Andreas and {Frye}, Brenda and {Gomez}, Sebastian and {Gonz{\'a}lez-Otero}, Mauro and {Griggio}, Massimo and {Harikane}, Yuichi and {Inayoshi}, Kohei and {Jha}, Saurabh W. and {Jim{\'e}nez-Teja}, Yolanda and {Kartaltepe}, Jeyhan S. and {Kelly}, Patrick L. and {Kwok}, Lindsey A. and {Lane}, Zachary G. and {Li}, Xiaolong and {Lobbe}, Ivo and {Lopes}, Paulo A.~A. and {Lucas}, Ray A. and {Magdis}, Georgios E. and {Martis}, Nicholas S. and {Matthee}, Jorryt and {Meena}, Ashish K. and {Naidu}, Rohan P. and {Noirot}, Ga{\"e}l and {Oguri}, Masamune and {Padilla Gonzalez}, Estefania and {Pascale}, Massimo and {Petrushevska}, Tanja and {Ricotti}, Massimo and {Schaerer}, Daniel and {Schuldt}, Stefan and {Shahbandeh}, Melissa and {Sheu}, William and {Shukawa}, Koji and {Tsujita}, Akiyoshi and {Vanzella}, Eros and {Wang}, Qinan and {Weaver}, John and {Williams}, Robert and {Windhorst}, Rogier and {Xu}, Yi and {Zenati}, Yossef and {Zitrin}, Adi},
        title = "{A spectroscopically confirmed, strongly lensed, metal-poor Type II supernova at z = 5.13}",
      journal = {arXiv e-prints},
         year = 2026,
        month = jan,
          eid = {arXiv:2601.04156},
        pages = {arXiv:2601.04156},
          doi = {10.48550/arXiv.2601.04156},
archivePrefix = {arXiv},
       eprint = {2601.04156},
 primaryClass = {astro-ph.HE},
       adsurl = {https://ui.adsabs.harvard.edu/abs/2026arXiv260104156C}
}

@ARTICLE{2022ApJ...935..167A,
       author = {{Astropy Collaboration} and {Price-Whelan}, Adrian M. and {Lim}, Pey Lian and {Earl}, Nicholas and {Starkman}, Nathaniel and {Bradley}, Larry and {Shupe}, David L. and {Patil}, Aarya A. and {Corrales}, Lia and {Brasseur}, C.~E. and {N{\"o}the}, Maximilian and {Donath}, Axel and {Tollerud}, Erik and {Morris}, Brett M. and {Ginsburg}, Adam and {Vaher}, Eero and {Weaver}, Benjamin A. and {Tocknell}, James and {Jamieson}, William and {van Kerkwijk}, Marten H. and {Robitaille}, Thomas P. and {Merry}, Bruce and {Bachetti}, Matteo and {G{\"u}nther}, H. Moritz and {Aldcroft}, Thomas L. and {Alvarado-Montes}, Jaime A. and {Archibald}, Anne M. and {B{\'o}di}, Attila and {Bapat}, Shreyas and {Barentsen}, Geert and {Baz{\'a}n}, Juanjo and {Biswas}, Manish and {Boquien}, M{\'e}d{\'e}ric and {Burke}, D.~J. and {Cara}, Daria and {Cara}, Mihai and {Conroy}, Kyle E. and {Conseil}, Simon and {Craig}, Matthew W. and {Cross}, Robert M. and {Cruz}, Kelle L. and {D'Eugenio}, Francesco and {Dencheva}, Nadia and {Devillepoix}, Hadrien A.~R. and {Dietrich}, J{\"o}rg P. and {Eigenbrot}, Arthur Davis and {Erben}, Thomas and {Ferreira}, Leonardo and {Foreman-Mackey}, Daniel and {Fox}, Ryan and {Freij}, Nabil and {Garg}, Suyog and {Geda}, Robel and {Glattly}, Lauren and {Gondhalekar}, Yash and {Gordon}, Karl D. and {Grant}, David and {Greenfield}, Perry and {Groener}, Austen M. and {Guest}, Steve and {Gurovich}, Sebastian and {Handberg}, Rasmus and {Hart}, Akeem and {Hatfield-Dodds}, Zac and {Homeier}, Derek and {Hosseinzadeh}, Griffin and {Jenness}, Tim and {Jones}, Craig K. and {Joseph}, Prajwel and {Kalmbach}, J. Bryce and {Karamehmetoglu}, Emir and {Ka{\l}uszy{\'n}ski}, Miko{\l}aj and {Kelley}, Michael S.~P. and {Kern}, Nicholas and {Kerzendorf}, Wolfgang E. and {Koch}, Eric W. and {Kulumani}, Shankar and {Lee}, Antony and {Ly}, Chun and {Ma}, Zhiyuan and {MacBride}, Conor and {Maljaars}, Jakob M. and {Muna}, Demitri and {Murphy}, N.~A. and {Norman}, Henrik and {O'Steen}, Richard and {Oman}, Kyle A. and {Pacifici}, Camilla and {Pascual}, Sergio and {Pascual-Granado}, J. and {Patil}, Rohit R. and {Perren}, Gabriel I. and {Pickering}, Timothy E. and {Rastogi}, Tanuj and {Roulston}, Benjamin R. and {Ryan}, Daniel F. and {Rykoff}, Eli S. and {Sabater}, Jose and {Sakurikar}, Parikshit and {Salgado}, Jes{\'u}s and {Sanghi}, Aniket and {Saunders}, Nicholas and {Savchenko}, Volodymyr and {Schwardt}, Ludwig and {Seifert-Eckert}, Michael and {Shih}, Albert Y. and {Jain}, Anany Shrey and {Shukla}, Gyanendra and {Sick}, Jonathan and {Simpson}, Chris and {Singanamalla}, Sudheesh and {Singer}, Leo P. and {Singhal}, Jaladh and {Sinha}, Manodeep and {Sip{\H{o}}cz}, Brigitta M. and {Spitler}, Lee R. and {Stansby}, David and {Streicher}, Ole and {{\v{S}}umak}, Jani and {Swinbank}, John D. and {Taranu}, Dan S. and {Tewary}, Nikita and {Tremblay}, Grant R. and {de Val-Borro}, Miguel and {Van Kooten}, Samuel J. and {Vasovi{\'c}}, Zlatan and {Verma}, Shresth and {de Miranda Cardoso}, Jos{\'e} Vin{\'\i}cius and {Williams}, Peter K.~G. and {Wilson}, Tom J. and {Winkel}, Benjamin and {Wood-Vasey}, W.~M. and {Xue}, Rui and {Yoachim}, Peter and {Zhang}, Chen and {Zonca}, Andrea and {Astropy Project Contributors}},
        title = "{The Astropy Project: Sustaining and Growing a Community-oriented Open-source Project and the Latest Major Release (v5.0) of the Core Package}",
      journal = {\apj},
         year = 2022,
        month = aug,
       volume = {935},
       number = {2},
          eid = {167},
        pages = {167},
          doi = {10.3847/1538-4357/ac7c74},
archivePrefix = {arXiv},
       eprint = {2206.14220},
 primaryClass = {astro-ph.IM},
       adsurl = {https://ui.adsabs.harvard.edu/abs/2022ApJ...935..167A}
}

@ARTICLE{2018AJ....156..123A,
       author = {{Astropy Collaboration} and {Price-Whelan}, A.~M. and {Sip{\H{o}}cz}, B.~M. and {G{\"u}nther}, H.~M. and {Lim}, P.~L. and {Crawford}, S.~M. and {Conseil}, S. and {Shupe}, D.~L. and {Craig}, M.~W. and {Dencheva}, N. and {Ginsburg}, A. and {VanderPlas}, J.~T. and {Bradley}, L.~D. and {P{\'e}rez-Su{\'a}rez}, D. and {de Val-Borro}, M. and {Aldcroft}, T.~L. and {Cruz}, K.~L. and {Robitaille}, T.~P. and {Tollerud}, E.~J. and {Ardelean}, C. and {Babej}, T. and {Bach}, Y.~P. and {Bachetti}, M. and {Bakanov}, A.~V. and {Bamford}, S.~P. and {Barentsen}, G. and {Barmby}, P. and {Baumbach}, A. and {Berry}, K.~L. and {Biscani}, F. and {Boquien}, M. and {Bostroem}, K.~A. and {Bouma}, L.~G. and {Brammer}, G.~B. and {Bray}, E.~M. and {Breytenbach}, H. and {Buddelmeijer}, H. and {Burke}, D.~J. and {Calderone}, G. and {Cano Rodr{\'\i}guez}, J.~L. and {Cara}, M. and {Cardoso}, J.~V.~M. and {Cheedella}, S. and {Copin}, Y. and {Corrales}, L. and {Crichton}, D. and {D'Avella}, D. and {Deil}, C. and {Depagne}, {\'E}. and {Dietrich}, J.~P. and {Donath}, A. and {Droettboom}, M. and {Earl}, N. and {Erben}, T. and {Fabbro}, S. and {Ferreira}, L.~A. and {Finethy}, T. and {Fox}, R.~T. and {Garrison}, L.~H. and {Gibbons}, S.~L.~J. and {Goldstein}, D.~A. and {Gommers}, R. and {Greco}, J.~P. and {Greenfield}, P. and {Groener}, A.~M. and {Grollier}, F. and {Hagen}, A. and {Hirst}, P. and {Homeier}, D. and {Horton}, A.~J. and {Hosseinzadeh}, G. and {Hu}, L. and {Hunkeler}, J.~S. and {Ivezi{\'c}}, {\v{Z}}. and {Jain}, A. and {Jenness}, T. and {Kanarek}, G. and {Kendrew}, S. and {Kern}, N.~S. and {Kerzendorf}, W.~E. and {Khvalko}, A. and {King}, J. and {Kirkby}, D. and {Kulkarni}, A.~M. and {Kumar}, A. and {Lee}, A. and {Lenz}, D. and {Littlefair}, S.~P. and {Ma}, Z. and {Macleod}, D.~M. and {Mastropietro}, M. and {McCully}, C. and {Montagnac}, S. and {Morris}, B.~M. and {Mueller}, M. and {Mumford}, S.~J. and {Muna}, D. and {Murphy}, N.~A. and {Nelson}, S. and {Nguyen}, G.~H. and {Ninan}, J.~P. and {N{\"o}the}, M. and {Ogaz}, S. and {Oh}, S. and {Parejko}, J.~K. and {Parley}, N. and {Pascual}, S. and {Patil}, R. and {Patil}, A.~A. and {Plunkett}, A.~L. and {Prochaska}, J.~X. and {Rastogi}, T. and {Reddy Janga}, V. and {Sabater}, J. and {Sakurikar}, P. and {Seifert}, M. and {Sherbert}, L.~E. and {Sherwood-Taylor}, H. and {Shih}, A.~Y. and {Sick}, J. and {Silbiger}, M.~T. and {Singanamalla}, S. and {Singer}, L.~P. and {Sladen}, P.~H. and {Sooley}, K.~A. and {Sornarajah}, S. and {Streicher}, O. and {Teuben}, P. and {Thomas}, S.~W. and {Tremblay}, G.~R. and {Turner}, J.~E.~H. and {Terr{\'o}n}, V. and {van Kerkwijk}, M.~H. and {de la Vega}, A. and {Watkins}, L.~L. and {Weaver}, B.~A. and {Whitmore}, J.~B. and {Woillez}, J. and {Zabalza}, V. and {Astropy Contributors}},
        title = "{The Astropy Project: Building an Open-science Project and Status of the v2.0 Core Package}",
      journal = {\aj},
         year = 2018,
        month = sep,
       volume = {156},
       number = {3},
          eid = {123},
        pages = {123},
          doi = {10.3847/1538-3881/aabc4f},
archivePrefix = {arXiv},
       eprint = {1801.02634},
 primaryClass = {astro-ph.IM},
       adsurl = {https://ui.adsabs.harvard.edu/abs/2018AJ....156..123A}
}

@ARTICLE{2013A&A...558A..33A,
       author = {{Astropy Collaboration} and {Robitaille}, Thomas P. and
         {Tollerud}, Erik J. and {Greenfield}, Perry and {Droettboom}, Michael and
         {Bray}, Erik and {Aldcroft}, Tom and {Davis}, Matt and
         {Ginsburg}, Adam and {Price-Whelan}, Adrian M. and
         {Kerzendorf}, Wolfgang E. and {Conley}, Alexander and {Crighton}, Neil and
         {Barbary}, Kyle and {Muna}, Demitri and {Ferguson}, Henry and
         {Grollier}, Fr{\'e}d{\'e}ric and {Parikh}, Madhura M. and
         {Nair}, Prasanth H. and {Unther}, Hans M. and {Deil}, Christoph and
         {Woillez}, Julien and {Conseil}, Simon and {Kramer}, Roban and
         {Turner}, James E.~H. and {Singer}, Leo and {Fox}, Ryan and
         {Weaver}, Benjamin A. and {Zabalza}, Victor and {Edwards}, Zachary I. and
         {Azalee Bostroem}, K. and {Burke}, D.~J. and {Casey}, Andrew R. and
         {Crawford}, Steven M. and {Dencheva}, Nadia and {Ely}, Justin and
         {Jenness}, Tim and {Labrie}, Kathleen and {Lim}, Pey Lian and
         {Pierfederici}, Francesco and {Pontzen}, Andrew and {Ptak}, Andy and
         {Refsdal}, Brian and {Servillat}, Mathieu and {Streicher}, Ole},
        title = "{Astropy: A community Python package for astronomy}",
      journal = {\aap},
         year = "2013",
        month = "Oct",
       volume = {558},
          eid = {A33},
        pages = {A33},
          doi = {10.1051/0004-6361/201322068},
archivePrefix = {arXiv},
       eprint = {1307.6212},
 primaryClass = {astro-ph.IM},
       adsurl = {https://ui.adsabs.harvard.edu/abs/2013A&A...558A..33A}
}
\bibliographystyle{aasjournalv7}



\end{document}